\documentclass[a4paper,11pt]{article}
\usepackage[latin1]{inputenc}
\usepackage[T1]{fontenc}
\usepackage[english]{babel}
\usepackage[hmargin={32mm,32mm},vmargin={32mm,35mm}]{geometry}
\usepackage{amsmath}
\usepackage{amsfonts}
\usepackage{amssymb}
\usepackage{enumerate}
\usepackage{graphicx}
\usepackage{tocloft}
\usepackage{bbm}
\usepackage{feynmf}
\usepackage{epsfig}
\usepackage{color}
\usepackage{slashed}
\usepackage{xspace}
\usepackage{url}
\usepackage{undertilde}
\usepackage{accents}
\usepackage[makeroom]{cancel}
\usepackage[hidelinks]{hyperref}
\usepackage{afterpage}
\usepackage{cite}
\usepackage{listings}
\usepackage{caption}
\usepackage{epstopdf}
\usepackage{enumitem}

\newcommand{\bra}[1]{\langle #1 |}
\newcommand{\ket}[1]{|#1\rangle}
\newcommand{\braket}[2]{\langle #1 | #2 \rangle}
\newcommand{\bbraket}[2]{\bigl\langle #1 \big| #2 \bigr\rangle}

\newcommand{\bket}[1]{\bigl|#1\bigr\rangle}

\newcommand{\be}{\begin{equation}}
\newcommand{\ee}{\end{equation}}
\newcommand{\bea}{\begin{eqnarray}}
\newcommand{\eea}{\end{eqnarray}}
\newcommand{\bal}{\begin{align}}
\newcommand{\eal}{\end{align}}

\newcommand{\eg}{e.g.\@\xspace}
\newcommand{\ie}{i.e.\@\xspace}
\newcommand{\Eq}[1]{Eq.\@\xspace\eqref{#1}}
\newcommand{\Eqs}[1]{Eqs.\@\xspace\eqref{#1}}

\newcommand{\updown}[2]{^{#1}_{\phantom{#1}#2}}
\newcommand{\downup}[2]{_{#1}^{\phantom{#1}#2}}

\DeclareMathOperator{\Tr}{{\rm Tr}}
\newcommand{\half}{\tfrac{1}{2}}

\newcommand{\Id}{\mathbbm{1}}
\newcommand{\R}{\mathbbm{R}}
\newcommand{\C}{\mathbbm{C}}

\newcommand{\makeSymbol}[1]{\mathord{\vcenter{\hbox{#1}}}}
\newcommand{\RealSymb}[2]{\makeSymbol{\includegraphics[scale=#2]{#1}}}

\numberwithin{equation}{section}

\newsavebox{\mybox}

\newcommand{\D}[4]{{D^{(#1)#2}}_{#3}(#4)}

\newcommand{\CG}[4]{C^{(#1#2)}{}\downup{#3}{#4}}
\newcommand{\CGi}[4]{C^{(#1#2)}{}\updown{#3}{#4}}

\newcommand{\Tau}[4]{({\tau^{(#1)}_{#2}}){}\updown{#3}{#4}}
\newcommand{\threej}[6]{\begin{pmatrix} #1&#2&#3 \\ #4&#5&#6 \end{pmatrix}}
\newcommand{\sixj}[6]{\begin{Bmatrix} #1&#2&#3 \\ #4&#5&#6 \end{Bmatrix}}
\newcommand{\ninej}[9]{\begin{Bmatrix} #1&#2&#3 \\ #4&#5&#6 \\ #7&#8&#9 \end{Bmatrix}}

\addto\captionsenglish{\renewcommand{\contentsname}{{\LARGE Contents}}}

\newcounter{nopage}
\newenvironment{nopage}
 {\clearpage\stepcounter{nopage}
  
  \thispagestyle{empty}}
 {\clearpage\addtocounter{page}{-1}}

\begin{document}

\begin{nopage}

\begin{center}

\textbf{\LARGE Introduction to $SU(2)$ recoupling theory and} \\

\vspace{4pt}

\textbf{\LARGE graphical methods for loop quantum gravity} \\

\vspace{20pt}

{\large Ilkka Mäkinen} \\

\vspace{8pt}

Faculty of Physics, University of Warsaw \\
Pasteura 5, 02-093 Warsaw, Poland \\

\vspace{12pt}

\end{center}

\begin{small}

\noindent {\bf Abstract.} We present an introduction to $SU(2)$ recoupling theory, focusing on those aspects of the topic which are useful for practical calculations in loop quantum gravity. In particular, we give a self-contained presentation of the powerful graphical formalism, which is an indispensable tool for performing computations in the spin network basis of loop quantum gravity. The use of the graphical techniques in loop quantum gravity is illustrated by several detailed example calculations. Plenty of exercises are included for the benefit of the ambitious student.

\end{small}

\renewcommand{\contentsname}{{\Large Contents}}
\tableofcontents

\end{nopage}

\section*{Introduction}
\addcontentsline{toc}{section}{Introduction}

The quantum degrees of freedom of the discrete, quantized spatial geometries of loop quantum gravity are encoded in intertwiners, or invariant tensors of $SU(2)$. Since these intertwiners are likely to be prominently featured in virtually any calculation in loop quantum gravity, a good practical knowledge of $SU(2)$ recoupling theory is essential for being able to perform calculations in the theory. An indispensable technical tool for carrying out such calculations quickly and efficiently is provided by the graphical formalism of $SU(2)$ recoupling theory. This formalism consists of a diagrammatic notation for the basic elements of $SU(2)$ recoupling theory, together with a set of simple rules according to which diagrams appearing in a graphical calculation can be manipulated.

For a student of loop quantum gravity wishing to become familiar with the graphical techniques of $SU(2)$ recoupling theory (such as the author in the beginning of his PhD studies), it can be difficult to find suitable references from which the subject could be efficiently learned. Information about the topic tends to be scattered around in the literature of quantum angular momentum \cite{BrinkSatchler, Varshalovich, YLV}, and in research articles where graphical methods are applied to calculations in loop quantum gravity \cite{AlesciLiegenerZipfel, AlesciRovelli, AlesciThiemannZipfel, YangMa1, YangMa2, paper1}, instead of being available in a single, comprehensive reference -- especially one written from the perspective of loop quantum gravity.

The goal of these notes is therefore to give an accessible and self-contained presentation of those aspects of $SU(2)$ recoupling theory which are relevant to practical calculations in loop quantum gravity, focusing particularly on the graphical techniques which provide an invaluable tool for dealing with calculations involving the spin network states of the theory. We hope that our presentation can be useful both for students interested in developing a thorough understanding of the graphical calculus of $SU(2)$ as a powerful computational technique in loop quantum gravity, and for researchers looking for a reference in which the definitions and the central results of the graphical formalism are collected in one place, under a single, consistent set of conventions.

A set of notes having the same premise, undoubtedly motivated by similar considerations as the present manuscript, has recently been published by Martin-Dussaud \cite{Martin-Dussaud}. His notes are written from the point of view of covariant loop quantum gravity, and therefore discuss also the group $SL(2,\C)$. In contrast, the document at hand features only the group $SU(2)$, reflecting the author's background in canonical loop quantum gravity.

Our presentation in these notes will be focused entirely on introducing the graphical formalism of $SU(2)$ recoupling theory from the perspective of loop quantum gravity. While we highlight the use of the graphical techniques as an efficient tool for calculations in loop quantum gravity, essentially no discussion of the framework of loop quantum gravity itself will be given. The student of loop quantum gravity should therefore study these notes in connection with learning (or after having learned) the foundations of loop quantum gravity from any of the standard references on the subject. The conceptual and mathematical aspects of the theory are very comprehensively discussed respectively in the books by Rovelli \cite{Rovelli} and Thiemann \cite{Thiemann}. The elements of loop quantum gravity are also explained in the books \cite{Bojowald, Cianfrani, GambiniPullin}, the introductory articles \cite{Ashtekar, status, Norbert, DonaSpeziale, GieselSahlmann, Han, Mercuri, Thiemann-lect}, and in the author's PhD thesis \cite{thesis}, from which most of the material in these notes has been taken. While focusing on different aspects of the topic, all of these references emphasize the canonical formulation of loop quantum gravity. Introductory treatments of covariant (or spin foam) formulation of the theory include the book \cite{RovelliVidotto} and the articles \cite{Bianchi, Perez, Zakopane}.

\newpage

The material in this manuscript is organized as follows. In the first chapter we review some basic facts of $SU(2)$ representation theory, the presentation of this elementary material serving to establish our notation and to make our treatment fully self-contained. In the second chapter, we discuss the theory of intertwiners, or $SU(2)$ invariant tensors. This discussion provides the essential foundations underlying the powerful graphical formalism for $SU(2)$ recoupling theory. The graphical framework itself is introduced in the third chapter, in which we describe the basic elements of the formalism, and derive the rules according to which graphical diagrams can be manipulated, and which therefore elevate the formalism from a mere graphical notation into a graphical calculus. In the fourth chapter, we illustrate the use of the graphical techniques in loop quantum gravity by going through three detailed example calculations. Each chapter includes several exercises, which will help the determined student to test and deepen their understanding of the material. At the end of the manuscript, a collection of useful graphical formulas is given.

The graphical formalism, which constitutes the main subject of these notes, is merely one aspect (even if a very useful one) of the theory of $SU(2)$ as it pertains to loop quantum gravity. Below we give an incomplete list of topics related to $SU(2)$, which are not covered by our presentation, but which nevertheless occupy an important place in the technical toolbox of loop quantum gravity. The list of references accompanying each topic is also likely to be incomplete, due to the author's imperfect knowledge of the literature.
\begin{itemize}
\item The different coherent states associated with $SU(2)$: Angular momentum coherent states \cite{Perelomov, Radcliffe}, Livine--Speziale coherent intertwiners \cite{LivineSpeziale}, and the heat kernel coherent states introduced into loop quantum gravity by Thiemann \cite{BahrThiemann, BMP, GCS1, GCS2, GCS3, GCS4}.
\item The formalism of spinors and twistors \cite{spinors1, spinors2, spinors3, spinors4}.
\item Various lines of research where the machinery of spinors, twistors and coherent states is used to describe and analyze $SU(2)$ intertwiners \cite{applications1, applications2, applications3, applications4, applications5, applications6}.
\item Schwinger's oscillator formulation of angular momentum \cite{Schwinger} and its uses in loop quantum gravity \cite{oscillator1, oscillator2, oscillator3, oscillator4}.
\item Asymptotic properties of the Wigner $nj$-symbols \cite{asymptotics2, asymptotics3, asymptotics4, asymptotics5} and other objects of $SU(2)$ recoupling theory \cite{asymptotics1}.
\end{itemize}
The ambitious student can consider this list as a suggestion of topics for further study, after the material presented in these notes has been mastered.

While a significant effort has been made to ensure that the contents of these notes are free from mistakes, the author thinks it is not very probable that his attempts have been perfectly successful. We therefore encourage the reader -- especially the reader who wants to apply the formulas given here to their own calculations -- to keep an eye out for any typos, incorrect signs, and other mistakes which may have escaped the author's attention.

\newpage

\section{Representations of $SU(2)$}

The first chapter of these notes consists of a review of the necessary elements of the representation theory of $SU(2)$. While the material in this chapter is elementary and most of it can be found in any good textbook on quantum mechanics or Lie groups, the chapter nevertheless serves a number of purposes: to establish notation, to make our treatment fully self-contained, and to put forward the author's personal bias for the quantum-mechanical theory of angular momentum, as opposed to the formal machinery of mathematical group theory, as the natural framework for discussing the theory of $SU(2)$ in the context of physics.

\subsection{Fundamental representation}

A general element of $SU(2)$, the group of unitary $2\times 2$ -matrices with determinant $+1$, has the form
\be\label{gAB}
g\updown{A}{B} = \begin{pmatrix} \alpha&\beta \\ -\bar\beta &\bar\alpha \end{pmatrix}\qquad \text{with} \qquad |\alpha|^2+|\beta|^2=1.
\ee
The fundamental representation of the group is realized by the action of the matrices $g\updown{A}{B}$ on two-component vectors of the form
\be
v^A = \begin{pmatrix} v^0 \\ v^1 \end{pmatrix}.
\ee
We denote the vector space consisting of such vectors as ${\cal H}_{1/2}$. The natural scalar product on ${\cal H}_{1/2}$, defined by
\be\label{Hprod}
\braket{u}{v} = \bar u^0 v^0 + \bar u^1 v^1,
\ee
is invariant under the action of $SU(2)$. The antisymmetric tensors
\be
\epsilon_{AB} = \begin{pmatrix} 0&1 \\ -1&0 \end{pmatrix}, \qquad \epsilon^{AB} = \begin{pmatrix} 0&1 \\ -1&0 \end{pmatrix}
\ee
are also $SU(2)$-invariant:
\be
\epsilon_{AB}g\updown{A}{C}g\updown{B}{D} = \epsilon_{CD}
\ee
and similarly for $\epsilon^{AB}$. By manipulating this relation, one can show that the matrix elements of the inverse matrix $g^{-1}$ are related to those of $g$ by
\be
(g^{-1})\updown{A}{B} = \epsilon^{AC}\epsilon_{BD}g\updown{D}{C}.
\ee
Since we define both $\epsilon_{AB}$ and $\epsilon^{AB}$ to have the same numerical value, their contraction gives
\be
\epsilon_{AB}\epsilon^{BC} = -\delta_A^C.
\ee
The epsilon tensor can be used for raising and lowering of $SU(2)$ indices. We adopt the convention
\be\label{raise-1/2}
v^A = \epsilon^{AB}v_B, \qquad v_A = v^B\epsilon_{BA}.
\ee
Using $\epsilon_{AB}$, one may also define the invariant antisymmetric product
\be\label{eprod}
(u|v) = \epsilon_{AB}u^Av^B = u^0v^1 - u^1v^0
\ee
between two vectors in ${\cal H}_{1/2}$.

By introducing the Pauli matrices
\be
\sigma_x = \begin{pmatrix} 0&1 \\ 1&0 \end{pmatrix}, \qquad \sigma_y = \begin{pmatrix} 0&-i \\ i&0 \end{pmatrix}, \qquad \sigma_z = \begin{pmatrix} 1&0 \\ 0&-1 \end{pmatrix},
\ee
a general $SU(2)$-element can be expressed in terms of an angle $\alpha$ and a unit vector $\vec n$ as
\be\label{g_n(a)}
g(\alpha,\vec n) = e^{-i\alpha\vec n\cdot\vec\sigma/2} = \cos\frac{\alpha}{2} - i\sin\frac{\alpha}{2}(\vec n\cdot\vec\sigma),
\ee
where the second equality follows from expanding the exponential and observing that $(\vec n\cdot\vec\sigma)^2 = 1$. The parametrization \eqref{g_n(a)} makes it particularly clear that elements of $SU(2)$ can be viewed as representing rotations in three-dimensional space. Let us give a geometrical argument to justify this interpretation. Consider the following sequence of infinitesimal rotations:
\begin{itemize}
\item Around the $x$-axis by an infinitesimal angle $\epsilon$;
\item Around the $y$-axis by another infinitesimal angle $\epsilon'$;
\item Around the $x$-axis by the angle $-\epsilon$;
\item Around the $y$-axis by the angle $-\epsilon'$.
\end{itemize}
By elementary geometry it is possible to convince oneself that, at lowest nontrivial order in the angles, the sequence is equivalent to a single rotation around the $z$-axis by the angle $-\epsilon\epsilon'$.

Suppose that rotations around the coordinate axes are generated by the operators $(J_x,J_y,J_z)$, so that the operator
\be
R_i(\alpha) = e^{-i\alpha J_i}
\ee
gives a rotation by an angle $\alpha$ around the $i$-axis. Then the rotations involved in the above sequence are given by
\begin{align}
R_x(\epsilon) &= 1 - i\epsilon J_x - \frac{\epsilon^2}{2}J_x^2 + {\cal O}(\epsilon^3), \\
R_y(\epsilon') &= 1 - i\epsilon' J_y - \frac{\epsilon'{}^2}{2}J_y^2 + {\cal O}(\epsilon'{}^3).
\end{align}
Now a direct calculation shows that the entire sequence is represented by the operator
\be
R_y(-\epsilon')R_x(-\epsilon)R_y(\epsilon')R_x(\epsilon) = 1 + \epsilon\epsilon'[J_x,J_y] + \dots
\ee
But this operator is supposed to be $R_z(-\epsilon\epsilon') = 1+i\epsilon\epsilon'J_z + \dots$, so we conclude that the generators must satisfy $[J_x,J_y] = iJ_z$. By considering cyclic permutations of the coordinate axes, we find the complete commutation relation
\be
[J_i,J_j] = i\epsilon\downup{ij}{k}J_k.
\ee
Hence we see that this commutation relation has a direct geometrical significance: it encodes the way in which successive rotations in three-dimensional space are combined.

The Pauli matrices satisfy the commutation relation
\be
[\sigma_i,\sigma_j] = 2i\epsilon\downup{ij}{k}\sigma_k,
\ee
which shows that they can be interpreted as generators of rotations by making the identification $J_i=\sigma_i/2$. Under this interpretation, $SU(2)$ elements of the form
\be
g_i(\alpha) = e^{-i\alpha\sigma_i/2}
\ee
describe rotations around the coordinate axes. The general element of \Eq{g_n(a)} corresponds to a rotation by the angle $\alpha$ around the direction given by the vector $\vec n$.

\subsection{The angular momentum operator}

In quantum mechanics, any (Hermitian) vector operator $\vec J$ whose components satisfy the commutation relation
\be
[J_i,J_j] = i\epsilon\downup{ij}{k}J_k
\ee
is called an angular momentum operator. All components of $\vec J$ commute with the squared angular momentum
\be
J^2 = J_x^2 + J_y^2 + J_z^2.
\ee
Therefore one can simultaneously diagonalize $J^2$ and one of the components, conventionally chosen as $J_z$. Let us write the eigenvalue equations as
\begin{align}
J^2\ket{\lambda,\mu} &= \lambda\ket{\lambda,\mu}, \label{J2ab} \\
J_z\ket{\lambda,\mu} &= \mu\ket{\lambda,\mu}. \label{Jzab}
\end{align}
To derive the solution to the eigenvalue problem, it is useful to define the raising and lowering operators
\be
J_\pm = J_x\pm iJ_y.
\ee
Their commutators with $J^2$ and $J_z$ are given by
\begin{align}
[J^2,J_\pm] = 0, \qquad [J_z,J_\pm] = \pm J_\pm.
\end{align}
These relations imply that the raising and lowering operators indeed raise and lower the eigenvalue of $J_z$ by one, while leaving the eigenvalue of $J^2$ unchanged: 
\begin{align}
J^2(J_\pm\ket{\lambda,\mu}) &= \lambda J_\pm\ket{\lambda,\mu}, \\
J_z(J_\pm\ket{\lambda,\mu}) &= (\mu\pm 1)J_\pm\ket{\lambda,\mu}.
\end{align}
In other words, the states $J_\pm\ket{\lambda,\mu}$ must be proportional to $\ket{\lambda,\mu\pm 1}$:
\be\label{Jpmab}
J_\pm\ket{\lambda,\mu} = A_\pm(\lambda,\mu)\ket{\lambda,\mu\pm 1}.
\ee
This shows that if any one of the eigenstates $\ket{\lambda,\mu}$ is known, then all the eigenstates for that value of $\lambda$ can be derived by repeatedly applying $J_+$ and $J_-$.

On the other hand, the inequality
\be
\bra{\lambda,\mu}J^2-J_z^2\ket{\lambda,\mu} = \bra{\lambda,\mu}J_x^2+J_y^2\ket{\lambda,\mu} \geq 0
\ee
implies that the eigenvalue $\mu$ is restricted by
\be
\mu^2\leq\lambda.
\ee
This condition can be satisfied only if there exists a maximal eigenvalue $\mu_{\rm max}$, such that the action of the raising operator on the state $\ket{\lambda,\mu_{\rm max}}$ does not give a new eigenstate, but instead
\be\label{bmax}
J_+\ket{\lambda,\mu_{\rm max}} = 0.
\ee
Similarly, there must exist a minimal eigenvalue $\mu_{\rm min}$, such that
\be\label{bmin}
J_-\ket{\lambda,\mu_{\rm min}} = 0.
\ee
Using the identities
\begin{align}
J_-J_+ &= J^2 - J_z^2 - J_z, \\
J_+J_- &= J^2 - J_z^2 + J_z
\end{align}
in \Eqs{bmax} and \eqref{bmin}, one finds the relation
\be
\lambda = \mu_{\rm max}(\mu_{\rm max}+1) = -\mu_{\rm min}(-\mu_{\rm min}+1),
\ee
showing that $\mu_{\rm max}$ and $\mu_{\rm min}$ are related to each other by
\be
\mu_{\rm min} = -\mu_{\rm max}.
\ee
Consider now acting repeatedly with $J_-$ on the state $\ket{\lambda,\mu_{\rm max}}$. After a certain number of actions, say $N$, one must arrive at the state $\ket{\lambda,\mu_{\rm min}}$. Since the eigenvalue $\mu$ is lowered by one at each step, the difference $\mu_{\rm max}-\mu_{\rm min}$ must be equal to $N$, giving
\be
\mu_{\rm max} = \frac{N}{2}.
\ee
The eigenvalue $\lambda$ then is
\be
\lambda = \mu_{\rm max}(\mu_{\rm max}+1) = \frac{N}{2}\biggl(\frac{N}{2}+1\biggr).
\ee
The possible eigenvalues of the operators $J^2$ and $J_z$ have therefore been determined. The eigenvalue equations \eqref{J2ab} and \eqref{Jzab} can be written as
\begin{align}
J^2\ket{jm} &= j(j+1)\ket{jm}, \label{J2jm} \\
J_z\ket{jm} &= m\ket{jm}, \label{Jzjm}
\end{align}
where $j$ may be any integer or half-integer, and $m$ ranges from $-j$ to $j$ in steps of $1$.

To complete the solution of the eigenvalue problem, it remains to find the coefficients $A_\pm(\lambda,\mu) \equiv A_\pm(j,m)$ in \Eq{Jpmab}, since this will show how the angular momentum operator acts on the eigenstates $\ket{jm}$. By multiplying the equation with its adjoint, we find
\be
|A_\pm(j,m)|^2 = \bra{jm}J_\mp J_\pm\ket{jm} = \bra{jm}J^2-J_z^2\mp J_z\ket{jm} = j(j+1)-m(m\pm 1),
\ee
which determines $A_\pm(j,m)$ up to a phase. We will follow the nearly universally adopted Condon--Shortley phase convention, according to which $A_\pm(j,m)$ are taken to be real and positive. This leads to
\be\label{Jpm}
J_\pm\ket{jm} = \sqrt{j(j+1)-m(m\pm 1)}\ket{j,m\pm 1}.
\ee

\subsection{Spin-$j$ representation}\label{sec:spin-j}

The states $\ket{jm}$ with a fixed value of $j$ span the $(2j+1)$-dimensional vector space ${\cal H}_j$. A general element of ${\cal H}_j$ has the form
\be
\ket v = \sum_m v^m\ket{jm},
\ee
the index $m$ taking the values $-j$, $-j+1$, $\dots$, $j$. The natural definition
\be
\braket{u}{v} = \sum_m \bar u^m v^m
\ee
for a scalar product between elements of ${\cal H}_j$ promotes ${\cal H}_j$ into a Hilbert space. 

The relevance of the space ${\cal H}_j$ to $SU(2)$ representation theory is due to the fact that the matrices representing the operators $g = e^{-i\alpha\vec n\cdot\vec J}$ on ${\cal H}_j$ define an irreducible representation of $SU(2)$. These matrices, whose elements are given by
\be\label{WignerD}
\D{j}{m}{n}{g} = \bra{jm}e^{-i\alpha\vec n\cdot\vec J}\ket{jn},
\ee
are known as the Wigner matrices. The great orthogonality theorem of group theory implies that the Wigner matrices satisfy
\be\label{int DbarD}
\int dg\,\overline{\displaystyle \D{j}{m}{n}{g}} \D{j'}{m'}{n'}{g} = \frac{1}{d_j}\delta_{jj'}\delta_m^{m'}\delta^n_{n'},
\ee
where $dg$ is the normalized Haar measure\footnote{The Haar measure is determined uniquely by the conditions
\be\tag{1.A}\label{Haar1}
\int dg\,f(g_0g) = \int dg\,f(g), \qquad \int dg\,f(gg_0) = \int dg\,f(g), \qquad \int dg = 1,
\ee
where $g_0$ is an arbitrary, fixed element of $SU(2)$. In the parametrization \eqref{gAB},
\be\tag{1.B}\label{Haar2}
dg = \frac{1}{\pi^2}d({\rm Re}\,\alpha)\,d({\rm Im}\,\alpha)\,d({\rm Re}\,\beta)\,d({\rm Im}\,\beta)\,\delta(1-|\alpha|^2-|\beta|^2),
\ee
from which formulas for other parametrizations can be derived by making the appropriate change of variables and performing one of the integrals to remove the delta function. For more details, see \eg \cite{Sternberg} or \cite{Tung}.
} of $SU(2)$, and
\be
d_j=2j+1
\ee
is a common shorthand for the dimension of ${\cal H}_j$.

To derive the form of the invariant epsilon tensor in the spin-$j$ representation, we may consider the problem of constructing a state of zero total angular momentum on the tensor product space ${\cal H}_j\otimes{\cal H}_j$. Suppose that the state
\be
\ket{\Psi_0} = \sum_{mn} c_{mn}\ket{jm}\ket{jn}
\ee
satisfies
\begin{align}
\bigl(J^{(1)} + J^{(2)}\bigr)^2\ket{\Psi_0} &= 0, \\
\bigl(J^{(1)}_z + J^{(2)}_z\bigr)\ket{\Psi_0} &= 0,
\end{align}
where each angular momentum operator acts on the corresponding factor of the tensor product; for example, $J^{(1)}$ stands for $J^{(1)}\otimes\Id^{(2)}$. This is equivalent to the state being invariant under rotations generated by the total angular momentum $J = J^{(1)} + J^{(2)}$, which in turn implies that the tensor defined by the coefficients $c_{mn}$ is invariant under the action of $SU(2)$ by the matrices $\D{j}{m}{n}{g}$.

The eigenvalue equation for the $z$-component immediately shows that the coefficient $c_{mn} = 0$ unless $n = -m$. Then, requiring that the state
\be
\ket{\Psi_0} = \sum_m c_m\ket{jm}\ket{j,-m}
\ee
is annihilated by either one of $J_+$ or $J_-$ leads to the condition $c_{m+1} = -c_m$. Taken together, the two conditions imply that $c_{mn}$ is proportional to $(-1)^m\delta_{m,-n}$. Defining the tensor $\epsilon^{(j)}_{mn}$ so that $\epsilon^{(j)}_{j,-j} = +1$, we therefore have
\be
\epsilon^{(j)}_{mn} = (-1)^{j-m}\delta_{m,-n}, \qquad \epsilon_{\phantom{m}}^{(j)mn} = (-1)^{j-m}\delta_{m,-n}.
\ee
We see that $\epsilon^{(j)}_{mn}$ satisfies
\be\label{eps_nm}
\epsilon^{(j)}_{nm} = (-1)^{2j}\epsilon^{(j)}_{mn}
\ee
and
\be\label{epseps=delta}
\epsilon^{(j)}_{mm'}\epsilon_{\phantom{m'}}^{(j)m'n} = (-1)^{2j}\delta_m^n.
\ee
The remaining properties of the epsilon tensor are analogous to those in the fundamental representation. For the matrix elements of the inverse matrix, there holds the relation
\be\label{Dj-inv}
\D{j}{m}{n}{g^{-1}} = \epsilon_{\phantom{m'}}^{(j)mm'}\epsilon^{(j)}_{nn'}\D{j}{n'}{m'}{g},
\ee
and indices are raised and lowered according to the convention
\be
v^m = \epsilon_{\phantom{m}}^{(j)mn}v_n, \qquad v_m = v^n\epsilon^{(j)}_{nm}.
\ee
An antisymmetric, $SU(2)$-invariant product on ${\cal H}_j$ can be defined as
\be
(u|v) = \epsilon^{(j)}_{mn}u^mv^n = \sum_m (-1)^{j-m} u^m v^{-m}.
\ee

The way in which we introduced the space ${\cal H}_j$ does not make it particularly clear how the spin-$j$ representation of $SU(2)$ is related to the fundamental representation. We will now clarify this relation by showing how the states $\ket{jm}$ can be constructed from the states $\ket +$ and $\ket -$, which span the space ${\cal H}_{1/2}$, and are eigenstates of the angular momentum operator on ${\cal H}_{1/2}$ with eigenvalues $j=\half$ and $m=\pm\half$. To this end, let us consider the state
\be
\ket{\Psi_j} = \ket +\otimes\ket + \otimes \cdots \otimes\ket +,
\ee
which is an element of the $2j$-fold tensor product space ${\cal H}_{1/2}\otimes\cdots\otimes{\cal H}_{1/2}$. We would like to show that the state $\ket{\Psi_j}$ is an eigenstate of the total angular momentum operator
\be
J^{({\rm tot})} = J^{(1)} + J^{(2)} + \dots + J^{(2j)},
\ee
where each operator $J^{(i)}$ acts on the $i$-th factor of the tensor product $\otimes_{i=1}^{2j} {\cal H}_{1/2}^{(i)}$, \ie
\be
J^{(i)} \equiv \Id^{(1)} \otimes \cdots  \otimes \Id^{(i-1)} \otimes J^{(i)} \otimes \Id^{(i+1)} \otimes \cdots \otimes \Id^{(2j)}.
\ee
For the $z$-component of $J^{({\rm tot})}$, we immediately find
\be\label{JzPsij}
J^{({\rm tot})}_z\ket{\Psi_j} = j\ket{\Psi_j},
\ee
since each state $\ket +$ is an eigenstate of the corresponding operator $J_z$ with the eigenvalue $+\half$. For the square of the total angular momentum, we have
\be\label{Psij-calc}
\bigl(J^{({\rm tot})}\bigr)^2\ket{\Psi_j} = \biggl(\sum_i \bigl(J^{(i)}\bigr)^2 + \sum_{i\neq k} \vec J^{\,(i)}\cdot\vec J^{\,(k)}\biggr)\ket{\Psi_j}.
\ee
The first sum contains $2j$ terms, and in each of them the $J^2$ acts on the corresponding state $\ket +$, producing the eigenvalue $3/4$. To evaluate the cross terms, we note that each of the $2j(2j-1)$ terms in the sum cam be written as
\be
\vec J^{\,(i)}\cdot\vec J^{\,(k)} = J^{(i)}_z J^{(k)}_z + \frac{1}{2}\bigl(J^{(i)}_+ J^{(k)}_- + J^{(i)}_- J^{(k)}_+\bigr).
\ee
Here the action of the first term on the state $\ket{+}_{(i)}\ket{+}_{(k)}$ gives the eigenvalue $m_im_k = 1/4$. The action of the other two terms gives zero, because the state $\ket +$ is annihilated by the raising operator $J_+$. Hence, going back to \Eq{Psij-calc}, we conclude that
\be\label{J2Psij}
\bigl(J^{({\rm tot})}\bigr)^2\ket{\Psi_j} = j(j+1)\ket{\Psi_j}.
\ee
Together, \Eqs{JzPsij} and \eqref{J2Psij} show that $\ket{\Psi_j}$ can be identified with the state $\ket{jj}$. That is,
\be\label{jj}
\ket{jj} = \underbrace{\ket +\otimes\ket + \otimes \cdots \otimes\ket +}_{\text{$2j$ times}}.
\ee
The remaining states $\ket{jm}$ can now be constructed by repeatedly acting on \Eq{jj} with the lowering operator
\be
J_-^{({\rm tot})} = J_-^{(1)} + J_-^{(2)} + \dots + J_-^{(2j)}.
\ee
Recalling \Eq{Jpm}, we obtain
\be\label{jm}
\ket{jm} = \sqrt{\frac{(j+m)!(j-m)!}{(2j)!}}\Bigl(\underbrace{\ket +\otimes\cdots\otimes\ket +}_{\text{$j+m$ times}} \otimes \underbrace{\ket -\otimes \cdots \otimes\ket -}_{\text{$j-m$ times}}\;+\;\text{all permutations}\Bigr).
\ee
In this way we have established a direct relation between the spaces ${\cal H}_j$ and ${\cal H}_{1/2}$.

In particular, an explicit expression for the matrix elements $\D{j}{m}{n}{g}$ can be derived from \Eq{jm} by using the known action of the matrix $g$ on the states $\ket +$ and $\ket -$ in ${\cal H}_{1/2}$. The derivation is valid not only for elements of $SU(2)$, but extends to elements of the group $SL(2,\C)$, which have the form
\be
h = \begin{pmatrix} a&b \\ c&d \end{pmatrix},
\ee
where $\det h = ad-bc=1$, but the matrix elements are otherwise unrestricted. By replacing the states $\ket +$ and $\ket -$ on the right-hand side of \Eq{jm} with
\begin{align}
h\ket + &= a\ket + + c\ket -, \label{h*+} \\
h\ket - &= b\ket + + d\ket -, \label{h*-}
\end{align}
one finds, after a somewhat tedious but in principle straightforward calculation,
\be\label{WignerD-expl}
\D{j}{m}{n}{h} = \sum_k \frac{\sqrt{(j+m)!(j-m)!(j+n)!(j-n)!}}{k!(j-m-k)!(j+n-k)!(m-n+k)!}a^{j+n-k}b^{m-n+k}c^kd^{j-m-k},
\ee
where the sum runs over all the values of $k$ for which the argument of every factorial is non-negative.

Note that the right-hand side of \Eq{jm} is a completely symmetric combination of the states $\ket +$ and $\ket -$. In other words, the states $\ket{jm}$ given by \Eq{jm} belong to the completely symmetric subspace of the tensor product space ${\cal H}_{1/2}\otimes\cdots\otimes{\cal H}_{1/2}$. Indeed, the spin-$j$ representation of $SU(2)$ is often introduced in the literature by defining the space ${\cal H}_j$ as the completely symmetrized part of ${\cal H}_{1/2}\otimes\cdots\otimes{\cal H}_{1/2}$, or equivalently as the space spanned by objects of the form
\be\label{v^A...A}
v^{(A_1\cdots A_{2j})},
\ee
where the ''index'' $(A_1\cdots A_{2j})$ is a completely symmetric combination of the spin-1/2 indices $A_i$. In other words, only the total number of $+$'s and $-$'s among the indices $A_1,\dots,A_{2j}$ is relevant to the value of $(A_1\cdots A_{2j})$. The relation between the symmetrized index $(A_1\cdots A_{2j})$ and the magnetic index $m$ used so far in this chapter is given by $m = \half\sum_i A_i$, where each $A_i$ takes the value $+$ or $-$. The representation matrix of an $SU(2)$ element $g$ in the space of the vectors \eqref{v^A...A} is defined by
\be\label{g...g}
\D{j}{(A_1\cdots A_{2j})}{(B_1\cdots B_{2j})}{g} = g\updown{A_1}{(B_1} \cdots g\updown{A_{2j}}{B_{2j})},
\ee
with $g\updown{A}{B}$ the matrix in the fundamental representation.

In practical $SU(2)$ calculations it is almost always more convenient to use the realization of the space ${\cal H}_j$ in terms of the magnetic indices $m$, rather than the symmetric tensor product indices $(A_1\cdots A_{2j})$. This is especially true when it comes to the graphical techniques of Chapter \ref{ch:graphical}, which are adapted to the magnetic index representation, the corresponding graphical calculus of the symmetric tensor product representation being much more primitive and cumbersome in comparison.

\newpage

\subsection*{Exercises}
\addcontentsline{toc}{subsection}{\hspace{21.4pt} Exercises}

\begin{enumerate}[leftmargin=*]

\item A general rotation in $\R^3$ can be decomposed into the following sequence of rotations: 
\begin{itemize}
\item By an angle $\alpha$ around the $z$-axis;
\item By an angle $\beta$ around the $y$-axis of the rotated coordinate system;
\item By an angle $\gamma$ around the $z$-axis of the rotated coordinate system.
\end{itemize}
The angles $(\alpha,\beta,\gamma)$ are known as the Euler angles.
\begin{itemize}
\item[(a)] Show that the corresponding parametrization of a general $SU(2)$ element is given by
\[
g(\alpha,\beta,\gamma) = e^{-i\alpha\sigma_z/2}e^{-i\beta\sigma_y/2}e^{-i\gamma\sigma_z/2},
\]
where all the rotations refer to the original, unrotated coordinate axes. What are the ranges of the angles $\alpha$, $\beta$ and $\gamma$?
\item[(b)] The sequence of rotations described by the Euler angles is equivalent to a single rotation by some angle around some axis. Derive an expression for the angle and the axis in terms of the Euler angles.
\end{itemize}

\item The spherical components of a vector $\vec v$ are defined as
\[
v^+ = -\frac{1}{\sqrt 2}(v^x - iv^y), \qquad v^0 = v^z, \qquad v^- = \frac{1}{\sqrt 2}(v^x+iv^y).
\]
Under a rotation described by a matrix $R\in SO(3)$, the Cartesian components of $\vec v$ transform as
\[
v^i \to R\updown{i}{j}v^j.
\]
Show that under the same rotation, the spherical components transform according to the matrix $D^{(1)}(g_R)$, \ie
\[
v^m \to \D{1}{m}{n}{g_R}v^n,
\]
where $g_R$ is an $SU(2)$ element corresponding to the rotation. 

\item Derive the expression \eqref{WignerD-expl} for the matrix elements of the Wigner matrices by considering how the state \eqref{jm} transforms under the $SL(2,\C)$ transformation \eqref{h*+}--\eqref{h*-}.

\item
\begin{itemize}
\item[(a)] Check that the measure defined by \Eq{Haar2} satisfies the conditions \eqref{Haar1}, and therefore is a correct expression for the Haar measure of $SU(2)$.
\item[(b)] Show that when the parametrization of $SU(2)$ in terms of the Euler angles is used, the Haar measure is given by
\[
dg = \frac{1}{8\pi^2}\,d\alpha\,d\beta\,d\gamma\,\sin\beta.
\]
\end{itemize}

\item When the space ${\cal H}_j$ is viewed as the completely symmetric part of the $2j$-fold tensor product ${\cal H}_{1/2}\otimes\cdots\otimes{\cal H}_{1/2}$, the tensor product space ${\cal H}_{j_1}\otimes{\cal H}_{j_2}$ consists of objects of the form
\[
u^{(A_1\cdots A_{2j_1})}v^{(B_1\cdots B_{2j_2})},
\]
completely symmetric separately in the two groups of indices. 
\begin{itemize}
\item[(a)] Argue that the space ${\cal H}_{j_1}\otimes{\cal H}_{j_2}$ can be decomposed into invariant subspaces by antisymmetrizing $n$ indices from the group $(A_1\cdots A_{2j_1})$ with $n$ indices from the group $(B_1\cdots B_{2j_2})$, and then completely symmetrizing the remaining set of indices.
\item[(b)] Let $w^{A_1\cdots A_{2j_1}B_1\cdots B_{2j_2}}$ be a tensor in the subspace where $n$ pairs of indices have been antisymmetrized. Show that the state
\[
\sum_{\{A_i\},\{B_k\}}w^{A_1\cdots A_{2j_1}B_1\cdots B_{2j_2}}\ket{A_1}\cdots\ket{A_{2j_1}}\ket{B_1}\cdots\ket{B_{2j_2}}
\]
is an eigenstate of the squared total angular momentum
\[
J^2 = \bigl(J^{(1)} + \dots + J^{(2j_1)} + J^{(2j_1+1)} + \dots + J^{(2j_1+2j_2)}\bigr)^2 
\]
with eigenvalue $j(j+1)$, where $j=j_1+j_2-n$. Here the operator $J^{(i)}$ acts on the $i$-th factor of the tensor product ${\cal H}_{j_1}\otimes{\cal H}_{j_2} = \bigl(\otimes^{2j_1}{\cal H}_{1/2}\bigr)\otimes\bigl(\otimes^{2j_2}{\cal H}_{1/2}\bigr)$.
\item[(c)] Conclude that the decomposition derived in part (a) is the decomposition
\[
{\cal H}_{j_1}\otimes{\cal H}_{j_2} = \bigoplus_{j=|j_1-j_2|}^{j_1+j_2} {\cal H}_j,
\]
familiar from the theory of addition of angular momenta.

\end{itemize}

\item Let ${\cal P}_j$ denote the space of polynomials of degree $2j$ depending on a single complex variable. Thus, a general element of ${\cal P}_j$ has the form $f^{(j)}(z) = \sum_{k=0}^{2j} c_kz^k$ with $z\in\C$.
\begin{itemize}
\item[(a)] Check that the assignment
\[
D^{(j)}(h)f^{(j)}(z) = (bz+d)^{2j}f^{(j)}\biggl(\frac{az+c}{bz+d}\biggr), \qquad h = \begin{pmatrix} a&b \\ c&d \end{pmatrix} \in SL(2,\C)
\]
defines a representation of $SL(2,\C)$, and consequently of the subgroup $SU(2)$, on the space ${\cal P}_j$.
\item[(b)] Verify that
\[
\braket{f^{(j)}}{g^{(j)}} = \frac{d_j}{\pi}\int\frac{d^2z}{(1+|z|^2)^{2j+2}}\,\overline{\displaystyle f^{(j)}(z)}g^{(j)}(z)
\]
defines a scalar product on ${\cal P}_j$. Find an orthonormal basis of ${\cal P}_j$ under the scalar product given above.
\item[(c)] By studying the action of the $SL(2,\C)$ matrices
\[
e^{\epsilon J_+} = \begin{pmatrix} 1&\epsilon \\ 0&1 \end{pmatrix}, \qquad e^{\epsilon J_0} = \begin{pmatrix} e^{\epsilon/2}&0 \\ 0&e^{-\epsilon/2} \end{pmatrix}, \qquad e^{\epsilon J_-} = \begin{pmatrix} 1&0 \\ \epsilon&1 \end{pmatrix}
\]
on a function $f^{(j)}(z)\in{\cal P}_j$, show that the components of the angular momentum operator are represented on ${\cal P}_j$ by the differential operators
\[
J_+ = -z^2\frac{d}{dz} + 2jz, \qquad J_0 = z\frac{d}{dz}-j, \qquad J_- = \frac{d}{dz}.
\]
Show that the Cartesian components of $\vec J$ are self-adjoint under the scalar product defined in part (b).

\end{itemize}

\end{enumerate}

\newpage

\section{Theory of intertwiners}

Intertwiners, or invariant tensors of $SU(2)$, play a crucial role in loop quantum gravity; they lie at the core of the so-called spin network states, which are loop quantum gravity's basic kinematical states of quantized spatial geometry. In this chapter we introduce the Clebsch--Gordan coefficient and the closely related Wigner 3$j$-symbol, show how the latter serves as the elementary building block from which intertwiners can be constructed, and derive the basic properties of the intertwiners obtained in this way. The material presented in this chapter provides the essential foundations underlying the powerful graphical formalism for calculations in $SU(2)$ recoupling theory, which will be introduced in the next chapter.

\subsection{Clebsch--Gordan coefficients}

Consider the tensor product space ${\cal H}_{j_1}\otimes{\cal H}_{j_2}$. An obvious basis on this space is provided by the tensor product states $\ket{j_1m_1}\ket{j_2m_2}$, which are eigenstates of the mutually commuting operators
\be
\bigl(J^{(1)}\bigr)^2, \qquad \bigl(J^{(2)}\bigr)^2, \qquad J^{(1)}_z, \qquad J^{(2)}_z.
\ee
Another complete set of commuting operators on ${\cal H}_{j_1}\otimes{\cal H}_{j_2}$ is formed by the operators
\be
\bigl(J^{(1)}\bigr)^2, \qquad \bigl(J^{(2)}\bigr)^2, \qquad \bigl(J^{(1)}+J^{(2)}\bigr)^2, \qquad J^{(1)}_z+J^{(2)}_z.
\ee
Let us denote their eigenstates by $\ket{j_1j_2;jm}$. Since both sets of states span the space ${\cal H}_{j_1}\otimes{\cal H}_{j_2}$, they must be related to each other by a unitary transformation of the form
\be\label{uncoupled}
\ket{j_1m_1}\ket{j_2m_2} = \sum_{jm} \CG{j_1j_2}{j}{m_1m_2}{m}\ket{j_1j_2;jm}
\ee
and
\be\label{coupled}
\ket{j_1j_2;jm} = \sum_{m_1m_2} \CGi{j_1j_2}{j}{m_1m_2}{m}\ket{j_1m_1}\ket{j_2m_2}.
\ee
The coefficients in these expansions are known as the Clebsch--Gordan coefficients. In the physics literature, a notation such as $\braket{j_1m_1,j_2m_2}{jm}$ is normally used for them. The notation adopted here is designed to display the tensorial structure of the Clebsch--Gordan coefficient when interpreted as an $SU(2)$ tensor; see \Eq{DDC=CD} below.

A number of properties satisfied by the Clebsch--Gordan coefficients follow immediately from their definition:
\begin{itemize}
\item The coefficient $\CG{j_1j_2}{j}{m_1m_2}{m}$ is trivially zero unless the conditions
\be\label{CGtriangle}
|j_1-j_2| \leq j \leq j_1+j_2 \qquad \text{and} \qquad j_1+j_2+j = {\rm integer}
\ee
are met. These conditions are referred to as the Clebsch--Gordan conditions, or the triangular conditions, for the spins $j_1$, $j_2$ and $j$.
\item Moreover, $\CG{j_1j_2}{j}{m_1m_2}{m} = 0$ whenever $m\neq m_1+m_2$.
\item The orthogonality relations of the Clebsch--Gordan coefficients read
\be\label{CG-orth-jm}
\sum_{jm} \CG{j_1j_2}{j}{m_1m_2}{m}\CGi{j_1j_2}{j}{m_1'm_2'}{m} = \delta_{m_1}^{m_1'}\delta_{m_2}^{m_2'}
\ee
\noindent and
\be\label{CG-orth-mm}
\sum_{m_1m_2} \CG{j_1j_2}{j}{m_1m_2}{m}\CGi{j_1j_2}{j'}{m_1m_2}{m'} = \delta_{jj'}\delta^m_{m'}.
\ee
\end{itemize}
The Condon--Shortley phase convention fixes the phases of the Clebsch--Gordan coefficients by requiring that all $\CG{j_1j_2}{j}{m_1m_2}{m}$ are real, and $\CG{j_1j_2}{j}{j_1,j-j_1}{j} > 0$; the relative phases between the coefficients for a fixed value of $j$ are determined by the choice already made in \Eq{Jpm}. Under this convention, the numerical value of the inverse coefficient $\CGi{j_1j_2}{j}{m_1m_2}{m}$ is equal to that of $\CG{j_1j_2}{j}{m_1m_2}{m}$, and for this reason, $\CG{j_1j_2}{j}{m_1m_2}{m}$ and $\CGi{j_1j_2}{j}{m_1m_2}{m}$ are usually not distinguished from each other in the physics literature.

Numerical values of the Clebsch--Gordan coefficients can be derived algebraically, using the properties of the raising and lowering operators $J_\pm$ (though in practice one would of course look up the coefficients using a tool such as Mathematica). For a given value of the total angular momentum $j$, one starts with the state of ''highest weight'' $\ket{j_1j_2;jj}$, in which the magnetic number is equal to its highest possible value. On grounds of the condition $m_1+m_2=m$ for the magnetic numbers in $\CG{j_1j_2}{j}{m_1m_2}{m}$, this state must have the form
\be\label{j1j2jj}
\ket{j_1j_2;jj} = \sum_m c_m \ket{j_1m}\ket{j_2,j-m}.
\ee
Applying the raising operator $J_+ = J^{(1)}_+ + J^{(2)}_+$ now gives
\be\label{sum c_m}
\sum_m c_m\Bigl(A_+(j_1,m)\ket{j_1,m+1}\ket{j_2,j-m} + A_+(j_2,j-m)\ket{j_1m}\ket{j_2,j-m+1}\Bigr) = 0,
\ee
where $A_+(j,m)$ is defined by \Eq{Jpm}. The information contained in \Eq{sum c_m} is sufficient to fully determine the state $\ket{j_1j_2;jj}$, since \Eq{sum c_m} gives $N-1$ conditions for the $N$ coefficients $c_m$, and one more condition is obtained by requiring that the state \eqref{j1j2jj} is normalized. The state $\ket{j_1j_2;jj}$ having been found, the remaining states $\ket{j_1j_2;jm}$ can then be derived by repeatedly applying the lowering operator.

Let us consider the effect of an $SU(2)$ rotation on the equation \eqref{uncoupled}. On the left-hand side, the rotation acts as $D^{(j_1)}(g)\otimes D^{(j_2)}(g)$. On the right-hand side, the terms having a given value of $j$ transform among themselves according to the matrix $D^{(j)}(g)$. Thus,
\be
D^{(j_1)}(g)\ket{j_1m_1}D^{(j_2)}(g)\ket{j_2m_2} = \sum_{jm} \CG{j_1j_2}{j}{m_1m_2}{m}D^{(j)}(g)\ket{j_1j_2;jm}.
\ee
Taking the product of this equation with the state $\bra{j_1n_1}\bra{j_2n_2}$ and using \Eq{uncoupled} on the right-hand side, we obtain the so-called Clebsch--Gordan series
\be\label{CG-ser}
\D{j_1}{m_1}{n_1}{g}\D{j_2}{m_2}{n_2}{g} = \sum_{jmn} \CGi{j_1j_2}{j}{m_1m_2}{m}\CG{j_1j_2}{j}{n_1n_2}{n}\D{j}{m}{n}{g}.
\ee
\newpage Contracting \Eq{CG-ser} with a Clebsch--Gordan coefficient and using the orthogonality relation \eqref{CG-orth-mm}, we further find
\be\label{DDC=CD}
\D{j_1}{n_1}{m_1}{g}\D{j_2}{n_2}{m_2}{g}\CG{j_1j_2}{j}{n_1n_2}{m} = \CG{j_1j_2}{j}{m_1m_2}{n}\D{j}{m}{n}{g},
\ee
showing how the Clebsch--Gordan coefficient itself behaves under $SU(2)$ transformations, and justifying the index structure used in the notation $\CG{j_1j_2}{j}{m_1m_2}{m}$.

\subsection{The 3$j$-symbol}

The Wigner 3$j$-symbol is defined by lowering the upper index of the Clebsch--Gordan coefficient using the epsilon tensor, and multiplying with a numerical factor (which is inserted in order to optimize the symmetry properties of the resulting object):
\begin{align}
\threej{j_1}{j_2}{j_3}{m_1}{m_2}{m_3} &= \frac{1}{\sqrt{d_{j_3}}}(-1)^{j_1-j_2+j_3}\CG{j_1j_2}{j_3}{m_1m_2}{n}\epsilon^{(j_3)}_{nm_3} \notag \\
&= \frac{1}{\sqrt{d_{j_3}}}(-1)^{j_1-j_2-m_3}\CG{j_1j_2}{j_3}{m_1m_2}{-m_3}. \label{3j}
\end{align}
While the 3$j$-symbol is often introduced as a more symmetric version of the Clebsch--Gordan coefficient, its relevance to loop quantum gravity follows from its behaviour under $SU(2)$ transformations. Starting from \Eq{DDC=CD} and recalling \Eq{Dj-inv} for the elements of an inverse Wigner matrix, one can show that the 3$j$-symbol is invariant under the action of $SU(2)$:
\be\label{DDD3j}
\D{j_1}{m_1}{n_1}{g}\D{j_2}{m_2}{n_2}{g}\D{j_3}{m_3}{n_3}{g}\threej{j_1}{j_2}{j_3}{m_1}{m_2}{m_3} = \threej{j_1}{j_2}{j_3}{n_1}{n_2}{n_3}.
\ee
In the language of angular momentum, the Clebsch--Gordan coefficient couples two angular momenta $j_1$ and $j_2$ to a total angular momentum $j$, whereas the 3$j$-symbol couples the three angular momenta $j_1$, $j_2$ and $j_3$ to total angular momentum zero. The $SU(2)$ invariance of the 3$j$-symbol implies that the state
\be
\ket{\Psi_0} = \sum_{m_1m_2m_3} \threej{j_1}{j_2}{j_3}{m_1}{m_2}{m_3} \ket{j_1m_1}\ket{j_2m_2}\ket{j_3m_3}
\ee
is rotationally invariant, and hence is an eigenstate of the total angular momentum operator $J^{(1)} + J^{(2)} + J^{(3)}$ with eigenvalue zero.

The basic properties of the 3$j$-symbol follow from the corresponding properties of the Clebsch--Gordan coefficients:
\begin{itemize}
\item The value of the 3$j$-symbol can be non-zero only if the triangular conditions
\be
|j_1-j_2| \leq j \leq j_1+j_2 \qquad \text{and} \qquad j_1+j_2+j = {\rm integer}
\ee
as well as the condition
\be
m_1+m_2+m_3=0
\ee
are satisfied.
\item The 3$j$-symbol satisfies the orthogonality relations
\be\label{3j-orth-mm}
\sum_{m_1m_2} \threej{j_1}{j_2}{j}{m_1}{m_2}{m}\threej{j_1}{j_2}{j'}{m_1}{m_2}{m'} = \frac{1}{d_j}\delta_{jj'}\delta_m^{m'}
\ee
and
\be\label{3j-orth-jm}
\sum_{jm} d_j\threej{j_1}{j_2}{j}{m_1}{m_2}{m}\threej{j_1}{j_2}{j}{m_1'}{m_2'}{m} = \delta_{m_1}^{m_1'}\delta_{m_2}^{m_2'}.
\ee
\end{itemize}
When the Condon--Shortley phase convention is followed, the 3$j$-symbol is real-valued, and possesses several convenient symmetry properties. Interchanging any two columns in the symbol produces the factor $(-1)^{j_1+j_2+j_3}$; for example,
\be\label{3jperm}
\threej{j_2}{j_1}{j_3}{m_2}{m_1}{m_3} = (-1)^{j_1+j_2+j_3}\threej{j_1}{j_2}{j_3}{m_1}{m_2}{m_3}.
\ee
This implies in particular that the symbol is invariant under cyclic permutations of its columns. The same phase factor results from reversing the sign of all the magnetic numbers:
\be\label{3j-m}
\threej{j_1}{j_2}{j_3}{-m_1}{-m_2}{-m_3} = (-1)^{j_1+j_2+j_3}\threej{j_1}{j_2}{j_3}{m_1}{m_2}{m_3}.
\ee
The definition of the Clebsch--Gordan coefficient immediately implies that the coefficient $\CG{j0}{j}{m0}{n}$ is equal to $\delta_m^n$. Using this in \Eq{3j}, we find that when one of the angular momenta in the 3$j$-symbol is zero, the 3$j$-symbol reduces to the epsilon tensor:
\be\label{3jzero}
\threej{j}{j'}{0}{m}{n}{0} = \delta_{jj'}\frac{1}{\sqrt{d_j}}\epsilon^{(j)}_{mn}.
\ee
Further properties of the 3$j$-symbol, relations satisfied by it, and explicit expressions for its values in particular cases can be found in any of the standard references on angular momentum theory. The most comprehensive source of such information is the encyclopedic collection of formulas by Varshalovich, Moskalev and Khersonskii \cite{Varshalovich}.

\subsection{Three-valent intertwiners}\label{sec:intertwiners-3}

Intertwiners, or invariant tensors of $SU(2)$, play a crucial role in loop quantum gravity as a central ingredient of the so-called spin network states. \Eq{DDD3j} shows that the 3$j$-symbol can be interpreted as such an invariant tensor. To emphasize the tensorial character of the 3$j$-symbol, we introduce the notation
\be\label{iota-3}
\iota_{m_1m_2m_3} = \threej{j_1}{j_2}{j_3}{m_1}{m_2}{m_3}.
\ee
Whenever we want to indicate explicitly the spins entering the 3$j$-symbol, the notation $\iota^{(j_1j_2j_3)}_{m_1m_2m_3}$ will be used for the tensor \eqref{iota-3}. The 3$j$-symbol is in fact (up to normalization) the only three-valent invariant tensor with indices in three given representations $j_1$, $j_2$ and $j_3$. Therefore the 3$j$-symbol alone spans the one-dimensional space of three-valent intertwiners, denoted by ${\rm Inv}\,\bigl({\cal H}_{j_1}\otimes{\cal H}_{j_2}\otimes{\cal H}_{j_3}\bigr)$.

By using epsilon to raise an index of the tensor \eqref{iota-3}, we obtain the tensor
\be
\iota\updown{m_1}{m_2m_3} = \epsilon_{\phantom{m}}^{(j_1)m_1m}\iota_{mm_2m_3},
\ee
which spans the intertwiner space ${\rm Inv}\,\bigl({\cal H}^*_{j_1}\otimes{\cal H}_{j_2}\otimes{\cal H}_{j_3}\bigr)$. Up to a numerical factor, the tensor $\iota\updown{m_1}{m_2m_3}$ is equal to the Clebcsh--Gordan coefficient $\CG{j_2j_3}{j_1}{m_2m_3}{m_1}$. Elements of a space such as ${\rm Inv}\,\bigl({\cal H}^*_{j_1}\otimes{\cal H}_{j_2}\otimes{\cal H}_{j_3}\bigr)$ are invariant under the action of $SU(2)$ in the sense of \Eq{DDC=CD}, \ie when a matrix $D^{(j)}(g)$ acts on each lower index of the tensor, while an inverse matrix $D^{(j)}(g^{-1})$ acts on each upper index.

The symmetry relation \Eq{3j-m} and the condition $m_1+m_2+m_3=0$ imply that the tensor
\be\label{iota3-upper}
\iota^{m_1m_2m_3} = \epsilon_{\phantom{m}}^{(j_1)m_1n_1}\epsilon_{\phantom{m}}^{(j_2)m_2n_2}\epsilon_{\phantom{m}}^{(j_3)m_3n_3}\iota_{n_1n_2n_3},
\ee
obtained by raising all the indices of $\iota_{m_1m_2m_3}$, is numerically equal to $\iota_{m_1m_2m_3}$. The orthogonality relation \eqref{3j-orth-mm} then shows that the three-valent intertwiner defined by \Eq{iota-3} is normalized:
\be
\iota^{m_1m_2m_3}\iota_{m_1m_2m_3} = 1.
\ee

The $SU(2)$ generator $\tau^{(j)}_i$, defined in the spin-$j$ representation as
\be\label{tau-j}
\Tau{j}{i}{m}{n} = -i\bra{jm}J_i\ket{jn},
\ee
can be interpreted as a three-valent intertwiner between the representations $j$, $1$ and $j$. This follows from the Wigner--Eckart theorem, which states that the matrix elements of a spherical tensor operator\footnote{A spherical tensor operator of rank $j$ is an operator whose $2j+1$ components $T^{(j)}_m$ ($m=-j,\dots,j$) transform under $SU(2)$ rotations in the same way as the states $\ket{jm}$. That is,
\be\tag{2.A}\label{sphtensor}
U(g)T^{(j)}_mU^\dagger(g) = \sum_n \D{j}{n}{m}{g}T^{(j)}_n,
\ee
where $U(g)$ is the unitary operator representing the rotation on the Hilbert space in which $T^{(j)}$ acts.} $T^{(j)}_m$ satisfy
\be
\bra{j_1m_1}T^{(j)}_m\ket{j_2m_2} = (j_1||T^{(j)}||j_2)\CG{j_2j}{j_1}{m_2m}{m_1},
\ee
where the so-called reduced matrix element $(j_1||T^{(j)}||j_2)$ is independent of the magnetic numbers; the dependence on $m_1$, $m_2$ and $m$ is given by the Clebsch--Gordan coefficient independently of the operator $T^{(j)}_m$. In \Eq{tau-j}, the vector operator $J_i$ is a spherical tensor operator of rank 1; therefore the matrix elements of the generators are proportional to the Clebsch--Gordan coefficient $\CG{j1}{j}{n1}{m}$. The coefficient of proportionality can be determined by contracting the equation with itself, using the orthogonality of the Clebsch--Gordan coefficient on one side, and
\be
\Tr\bigl(\tau_i^{(j)}\tau_{\phantom{i}}^{(j)i}\bigr) = -\sum_m \bra{jm}J_x^2+J_y^2+J_z^2\ket{jm} = -j(j+1)(2j+1)
\ee
on the other side. In this way one finds
\be\label{tau=C}
\Tau{j}{i}{m}{n} = -i\sqrt{j(j+1)}\CG{j1}{j}{ni}{m}.
\ee

Another familiar object which can be expressed in terms of a three-valent intertwiner is the antisymmetric symbol $\epsilon_{ijk}$. Since the 3$j$-symbol with $j_1=j_2=j_3=1$ is completely antisymmetric in its indices due to the symmetry relation \eqref{3jperm}, one might think that $\epsilon_{ijk}$ is directly proportional to the 3$j$-symbol:
\be\label{eps=3j}
\epsilon_{ijk} = -\sqrt 6\threej{1}{1}{1}{i}{j}{k}.
\ee
While this is a valid numerical relation, some care should be taken when using it, since the indices of $\epsilon_{ijk}$ usually refer to the Cartesian basis, in which an index takes the values $i=x,y,z$, whereas the indices of the 3$j$-symbol take the values $m=+1,0,-1$, and hence refer to the so-called spherical basis. (Strictly speaking, the index $i$ in \Eq{tau=C} should also be interpreted as a spherical index, not a Cartesian index.) The components of a vector with respect to the spherical basis are defined in terms of the Cartesian components by\footnote{Under a rotation descibed by a matrix $R\in SO(3)$, the Cartesian components of a vector $\vec v$ transform as $v^i \to R\updown{i}{j}v^j$. Under the same rotation, the spherical components defined by \Eq{v-spherical} transform according to $v^m \to \D{1}{m}{n}{g_R}v^n$, where $g_R$ is an $SU(2)$ element corresponding to the rotation $R$. Thus the spherical components of a vector are a special case of the definition of a spherical tensor given in the previous footnote.}
\be\label{v-spherical}
v^+ = -\frac{1}{\sqrt 2}(v^x - iv^y), \qquad v^0 = v^z, \qquad v^- = \frac{1}{\sqrt 2}(v^x+iv^y).
\ee
Now one can verify by direct calculation that a triple product of the form $\epsilon_{ijk}u^iv^jw^k$, where the indices $i$, $j$, $k$ refer to the Cartesian basis, is equal to $-i\epsilon_{\lambda\mu\nu}u^\lambda v^\mu w^\nu$, where $\lambda$, $\mu$, $\nu$ refer to the spherical basis, and the antisymmetric symbol in the spherical basis is defined so that $\epsilon_{10{}-1}=+1$. Therefore the correct way to express the triple product in terms of the 3$j$-symbol is given by
\be\label{epsuvw}
\epsilon_{ijk}u^iv^jw^k = i\sqrt 6\threej{1}{1}{1}{\lambda}{\mu}{\nu}u^\lambda v^\mu w^\nu.
\ee

\subsection{Intertwiners of higher valence}\label{sec:intertwiners-N}

The invariant tensors $\iota_{m_1m_2m_3}$ and $\epsilon^{(j)}_{mn}$ are the basic building blocks out of which intertwiners of higher valence can be constructed. For example, by using epsilon to contract two three-valent intertwiners on one index, we obtain the four-valent intertwiner
\begin{align}
\bigl(\iota_{12}^{(k)}\bigr)_{m_1m_2m_3m_4} &= \threej{j_1}{j_2}{k}{m_1}{m_2}{m}\epsilon_{\phantom{m}}^{(k)mn}\threej{k}{j_3}{j_4}{n}{m_3}{m_4} \notag \\
&= \sum_m (-1)^{k-m}\threej{j_1}{j_2}{k}{m_1}{m_2}{m}\threej{k}{j_3}{j_4}{-m}{m_3}{m_4}. \label{iota412}
\end{align}
The invariance of $\iota_{m_1m_2m_3}$ and $\epsilon^{(j)}_{mn}$ implies that the intertwiner \eqref{iota412} is invariant under the action of $SU(2)$ on its indices:
\be
\D{j_1}{m_1}{n_1}{g}\D{j_2}{m_2}{n_2}{g}\D{j_3}{m_3}{n_3}{g}\D{j_4}{m_4}{n_4}{g}\bigl(\iota_{12}^{(k)}\bigr)_{m_1m_2m_3m_4} = \bigl(\iota_{12}^{(k)}\bigr)_{n_1n_2n_3n_4}.
\ee
When the internal spin $k$ ranges over all the values allowed by the Clebsch--Gordan conditions, the tensors \eqref{iota412} span the intertwiner space ${\rm Inv}\,\bigl({\cal H}_{j_1}\otimes{\cal H}_{j_2}\otimes{\cal H}_{j_3}\otimes{\cal H}_{j_4}\bigr)$. Using the orthogonality relation of the 3$j$-symbols, one finds
\be\label{iota4-prod}
\bbraket{\iota_{12}^{(k)}}{\iota_{12}^{(k')}} = \bigl(\iota_{12}^{(k)}\bigr)^{m_1m_2m_3m_4}\bigl(\iota_{12}^{(k')}\bigr)_{m_1m_2m_3m_4} = \frac{1}{d_k}\delta_{kk'},
\ee
showing that the basis given by the intertwiners \eqref{iota412} is orthogonal but not normalized. To obtain normalized intertwiners, \Eq{iota412} should be multiplied by $\sqrt{d_k}$. 

Another basis on the four-valent intertwiner space is provided by the intertwiners
\be\label{iota413}
\bigl(\iota_{13}^{(l)}\bigr)_{m_1m_2m_3m_4} = \threej{j_1}{j_3}{l}{m_1}{m_3}{m}\epsilon_{\phantom{m}}^{(l)mn}\threej{l}{j_2}{j_4}{n}{m_2}{m_4},
\ee
in which the spins $j_1$ and $j_3$ have been coupled to the internal spin. The change of basis between the bases \eqref{iota412} and \eqref{iota413} will be discussed in section \ref{sec:6j9j}.

Intertwiners of arbitrarily high valence can evidently be derived by continuing to attach three-valent intertwiners to each other by contraction with epsilon. An $N$-valent intertwiner constructed according to this scheme has the form
\begin{align}
\iota^{(k_1\cdots k_{N-2})}_{m_1\cdots m_N} &= \threej{j_1}{j_2}{k_1}{m_1}{m_2}{\mu_1}\epsilon_{\phantom{m}}^{(k_1)\mu_1\nu_1} \threej{k_1}{j_3}{k_2}{\nu_1}{m_3}{\mu_2}\epsilon_{\phantom{m}}^{(k_2)\mu_2\nu_2} \cdots \notag \\
&\quad\cdots \threej{k_{N-3}}{j_{N-2}}{k_{N-2}}{\nu_{N-3}}{m_{N-2}}{\mu_{N-2}}\epsilon_{\phantom{m}}^{(k_{N-2})\mu_{N-2}\nu_{N-2}} \threej{k_{N-2}}{j_{N-1}}{j_N}{\nu_{N-2}}{m_{N-1}}{m_N}. \label{iotaN}
\end{align}
It is labeled by $N-2$ internal spins, which determine the eigenvalues of the operators $\bigl(J^{(1)} + J^{(2)}\bigr)^2$, $\bigl(J^{(1)} + J^{(2)} + J^{(3)})^2$, $\dots$, $\bigl(J^{(1)} + \dots + J^{(N-2)}\bigr)^2$. The intertwiner \eqref{iotaN} is not normalized; to normalize it, it should be multiplied by $\sqrt{d_{k_1}}\cdots\sqrt{d_{k_{N-2}}}$. Just as in the three-valent case, the intertwiner $\iota^{(k_1\cdots k_{N-2})m_1\cdots m_N}$, obtained by raising all the indices of the intertwiner \eqref{iotaN} by means of the epsilon tensor, is numerically equal to $\iota^{(k_1\cdots k_{N-2})}_{m_1\cdots m_N}$.

A useful fact of the $N$-valent intertwiner space concerns the integral
\be
I^{(j_1\cdots j_N)} = \int dg\,D^{(j_1)}(g)\cdots D^{(j_N)}(g).
\ee
The invariance and normalization of the Haar measure imply that the integral is invariant under the action of $SU(2)$, and satisfies $(I^{(j_1\cdots j_N)})^2 = I^{(j_1\cdots j_N)}$. Therefore $I^{(j_1\cdots j_N)}$, viewed as an operator on ${\cal H}_{j_1}\otimes\cdots\otimes{\cal H}_{j_N}$, must be the projection \mbox{operator} onto the $SU(2)$ invariant subspace of ${\cal H}_{j_1}\otimes\cdots\otimes{\cal H}_{j_N}$, \ie onto the intertwiner space \mbox{${\rm Inv}\,\bigl({\cal H}_{j_1}\otimes\cdots\otimes{\cal H}_{j_N}\bigr)$. Hence} the integral can be expressed as $I^{(j_1\cdots j_N)} = \sum_\iota \ket{\iota}\bra{\iota}$, or
\be\label{int D...D}
\int dg\,\D{j_1}{m_1}{n_1}{g}\cdots\D{j_N}{m_N}{n_N}{g} = \sum_\iota \overline{\displaystyle \iota^{m_1\cdots m_N}}\iota_{n_1\cdots n_N},
\ee
where the sum runs over any orthonormal basis of ${\rm Inv}\,\bigl({\cal H}_{j_1}\otimes\cdots\otimes{\cal H}_{j_N}\bigr)$; in general the basis may be complex-valued, although the basis given by the intertwiners \eqref{iotaN} is always real.

\subsection{6$j$- and 9$j$-symbols}\label{sec:6j9j}

The intertwiners \eqref{iota412} and \eqref{iota413} provide two inequivalent bases of the four-valent intertwiner space. One basis is expressed in terms of the other by the relation
\be\label{6j-def}
\bket{\iota_{13}^{(l)}} = \sum_k d_k(-1)^{j_2+j_3+k+l}\sixj{j_1}{j_2}{k}{j_4}{j_3}{l}\bket{\iota_{12}^{(k)}},
\ee
where the object with curly brackets is the Wigner 6$j$-symbol. From \Eq{6j-def} one can derive the expression
\be\label{6j=iiii}
\sixj{j_1}{j_2}{j_3}{k_1}{k_2}{k_3} = \bigl(\iota^{(j_1j_2j_3)}\bigr)^{m_1m_2m_3}\bigl(\iota^{(j_1k_2k_3)}\bigr)\downup{m_1}{n_2}{\vphantom{\bigl(\iota^{(j_1k_2k_3)}\bigr)}}_{n_3}\bigl(\iota^{(k_1j_2k_3)}\bigr)\downup{n_1m_2}{n_3}\bigl(\iota^{(k_1k_2j_3)}\bigr)\updown{n_1}{n_2m_3}
\ee
\noindent for the 6$j$-symbol as a contraction of four three-valent intertwiners. This shows that the 6$j$-symbol vanishes unless the triples of spins indicated by 
\be
\sixj{\circ}{\circ}{\circ}{}{}{} \qquad \sixj{\circ}{}{}{}{\circ}{\circ} \qquad \sixj{}{\circ}{}{\circ}{}{\circ} \qquad \sixj{}{}{\circ}{\circ}{\circ}{}{}
\ee
satisfy the Clebsch--Gordan conditions. Furthermore, orthogonality of the states \eqref{6j-def} for different values of $l$ implies that the 6$j$-symbol satisfies the orthogonality relation
\be\label{6j-orth}
\sum_x d_x\sixj{j_1}{j_2}{x}{j_3}{j_4}{k}\sixj{j_1}{j_2}{x}{j_3}{j_4}{l} = \frac{1}{d_k}\delta_{kl}.
\ee
Symmetry properties of the 6$j$-symbol can be derived from \Eq{6j=iiii}, though they are more easily seen using the graphical representation of the symbol, which will be introduced in the next chapter. The value of the 6$j$-symbol is unchanged by any permutation of its columns:
\be
\sixj{j_1}{j_2}{j_3}{k_1}{k_2}{k_3} = \sixj{j_1}{j_3}{j_2}{k_1}{k_3}{k_2} = \sixj{j_2}{j_3}{j_1}{k_2}{k_3}{k_1}, \quad \text{etc.}
\ee
and by interchanging the upper and lower spins simultaneously in any two columns:
\be
\sixj{j_1}{j_2}{j_3}{k_1}{k_2}{k_3} = \sixj{j_1}{k_2}{k_3}{k_1}{j_2}{j_3}, \quad \text{etc.}
\ee

The Wigner 9$j$-symbol arises when changes of basis between five-valent intertwiners are performed. Two different bases in the intertwiner space ${\rm Inv}\,\bigl({\cal H}_{j_1}\otimes\cdots\otimes{\cal H}_{j_5}\bigr)$ are provided by the intertwiners
\be\label{iota512}
\bigl(\iota_{12}^{(k_{12}k_{34})}\bigr)_{m_1\cdots m_5} = \bigl(\iota^{(j_1j_2k_{12})}\bigr)\downup{m_1m_2}{n_{12}}\bigl(\iota^{(j_3j_4k_{34})}\bigr)\downup{m_3m_4}{n_{34}}\bigl(\iota^{(k_{12}k_{34}j_5)}\bigr)_{n_{12}n_{34}m_5}
\ee
and
\be\label{iota513}
\bigl(\iota_{13}^{(l_{13}l_{24})}\bigr)_{m_1\cdots m_5} = \bigl(\iota^{(j_1j_3l_{13})}\bigr)\downup{m_1m_3}{n_{13}}\bigl(\iota^{(j_2j_4l_{24})}\bigr)\downup{m_2m_4}{n_{24}}\bigl(\iota^{(l_{13}l_{24}j_5)}\bigr)_{n_{13}n_{24}m_5}.
\ee
Elements of one basis are expanded in the other basis as
\be\label{9j-def}
\bket{\iota_{13}^{(l_{13}l_{24})}} = \sum_{k_{12}k_{34}} d_{k_{12}}d_{k_{34}}\ninej{j_1}{j_2}{k_{12}}{j_3}{j_4}{k_{34}}{l_{13}}{l_{24}}{j_5}\bket{\iota_{12}^{(k_{12}k_{34})}},
\ee
where the 9$j$-symbol appears on the right-hand side. From this definition it follows that the 6$j$-symbol is given by a contraction of six three-valent intertwiners as
\begin{align}
\ninej{j_1}{j_2}{j_3}{k_1}{k_2}{k_3}{l_1}{l_2}{l_3} &= \bigl(\iota^{(j_1j_2j_3)}\bigr)_{m_1m_2m_3}\bigl(\iota^{(k_1k_2k_3)}\bigr)_{n_1n_2n_3}\bigl(\iota^{(l_1l_2l_3)}\bigr)_{\mu_1\mu_2\mu_3} \notag \\
&\quad\times \bigl(\iota^{(j_1k_1l_1)}\bigr)^{m_1n_1\mu_1}\bigl(\iota^{(j_2k_2l_2)}\bigr)^{m_2n_2\mu_2}\bigl(\iota^{(j_3k_3l_3)}\bigr)^{m_3n_3\mu_3}.\label{9j=i^6}
\end{align}
Hence the 9$j$-symbol can have a non-zero value only if the Clebsch--Gordan conditions are satisfied by the spins in each row and each column.

The 9$j$-symbol possesses a high degree of symmetry. Transposing the array of spins in the symbol preserves its value:
\be
\ninej{j_1}{k_1}{l_1}{j_2}{k_2}{l_2}{j_3}{k_3}{l_3} = \ninej{j_1}{j_2}{j_3}{k_1}{k_2}{k_3}{l_1}{l_2}{l_3}.
\ee
Moreover, interchanging any rows or any two columns produces a sign factor:
\be
\ninej{j_1}{j_2}{j_3}{k_1}{k_2}{k_3}{l_1}{l_2}{l_3} = (-1)^S\ninej{j_2}{j_1}{j_3}{k_2}{k_1}{k_3}{l_2}{l_1}{l_3} = (-1)^S\ninej{k_1}{k_2}{k_3}{j_1}{j_2}{j_3}{l_1}{l_2}{l_3}, \quad \text{etc.} \\
\ee
where
\be
S = j_1+j_2+j_3+k_1+k_2+k_3+l_1+l_2+l_3.
\ee
These symmetry relations again become the most transparent when the 9$j$-symbol is expressed in graphical form. A large number of further properties satisfied by the 6$j$- and 9$j$-symbols can be found in any of the standard references, such as \cite{Varshalovich}.

\subsection{Intertwiners in the symmetric tensor product representation}\label{sec:int-stp}

When the representation of the space ${\cal H}_j$ as a completely symmetrized tensor product of the spaces ${\cal H}_{1/2}$ is used, intertwiners can be expressed directly in terms of the fundamental invariant tensor $\epsilon_{AB}$, which is the only invariant tensor available in the space ${\cal H}_{1/2}$. The basic three-valent intertwiner
\be\label{qwerty}
\iota_{(A_1\cdots A_{2j_1})(B_1\cdots B_{2j_2})(C_1\cdots C_{2j_3})},
\ee
out of which higher intertwiners can be derived, must be constructed from the object
\be\label{eps...eps}
\epsilon_{A_1B_1}\cdots\epsilon_{A_aB_a}\epsilon_{B_{a+1}C_1}\cdots\epsilon_{B_{a+b}C_b}\epsilon_{C_{b+1}A_{a+1}}\cdots\epsilon_{C_{b+c}A_{a+c}},
\ee
which is the only possible combination of epsilons with the correct index structure. The requirement that the total number of $A$, $B$ and $C$ indices is respectively $2j_1$, $2j_2$ and $2j_3$ determines the numbers $a$, $b$ and $c$ as
\be
a = j_1+j_2-j_3, \qquad b = j_2+j_3-j_1, \qquad c = j_3+j_1-j_2.
\ee
The three-valent intertwiner \eqref{qwerty} is then obtained from the expression \eqref{eps...eps} by symmetrizing each group of indices. This gives
\be\label{iota_ABC}
\iota_{(A_1\cdots A_{2j_1})(B_1\cdots B_{2j_2})(C_1\cdots C_{2j_3})} = N_{j_1j_2j_3}I_{(A_1\cdots A_{2j_1})(B_1\cdots B_{2j_2})(C_1\cdots C_{2j_3})},
\ee
where
\be
I_{A_1\cdots A_{2j_1}B_1\cdots B_{2j_2}C_1\cdots C_{2j_3}} = \epsilon_{A_1B_1}\cdots\epsilon_{A_aB_a}\epsilon_{B_{a+1}C_1}\cdots\epsilon_{B_{a+b}C_b}\epsilon_{C_{b+1}A_{a+1}}\cdots\epsilon_{C_{b+c}A_{a+c}},
\ee
and the normalization factor is
\be\label{Njjj}
N_{j_1j_2j_3} = \sqrt{\frac{(2j_1)!(2j_2)!(2j_3)!}{(j_1+j_2+j_3+1)!(j_1+j_2-j_3)!(j_2+j_3-j_1)!(j_3+j_1-j_2)!}}.
\ee
The three-valent intertwiner having been found, intertwiners of higher valence can be constructed by contracting several three-valent intertwiners, in the same way as in \Eqs{iota412} and \eqref{iotaN}. For example, a four-valent intertwiner is given by
\be\label{iota_ABCD}
\iota^{(k)}_{(A_1\cdots A_{2j_1})\cdots(D_1\cdots D_{2j_4})} = \sqrt{d_k}\iota_{(A_1\cdots A_{2j_1})(B_1\cdots B_{2j_2})(E_1\cdots E_{2k})}\iota\updown{(E_1\cdots E_{2k})}{(C_1\cdots C_{2j_3})(D_1\cdots D_{2j_4})},
\ee
where the spin-1/2 indices are raised with $\epsilon^{AB}$ according to the convention \eqref{raise-1/2}, and the factor $\sqrt{d_k}$ is inserted for normalization.

Intertwiners in the symmetric tensor product representation can be expressed graphically by introducing a primitive graphical notation, in which
\begin{itemize}
\item $\epsilon_{AB}$ or $\epsilon^{AB}$ is represented by a line with an arrow pointing from index $A$ to $B$;
\item $\delta^A_B = \epsilon^{AC}\epsilon_{BC}$ is represented by a line with no arrow;
\item Contraction of an index is represented by connecting two lines at the contracted index.
\end{itemize}
Then we can write the intertwiner \eqref{iota_ABC} in the form 
\be\label{i3graphic}
\iota_{(A_1\cdots A_{2j_1})(B_1\cdots B_{2j_2})(C_1\cdots C_{2j_3})} = N_{j_1j_2j_3}\;\RealSymb{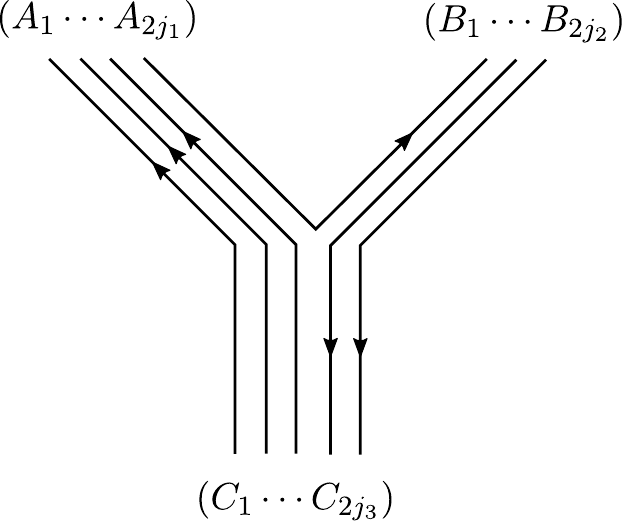}{0.8}
\ee
where each external group of lines is understood to be completely symmetrized. Similarly, the four-valent intertwiner \eqref{iota_ABCD} is given by
\be\label{i4graphic}
\iota^{(k)}_{(A_1\cdots A_{2j_1})\cdots(D_1\cdots D_{2j_4})} = \sqrt{d_k}N_{j_1j_2k}N_{kj_3j_4}\;\RealSymb{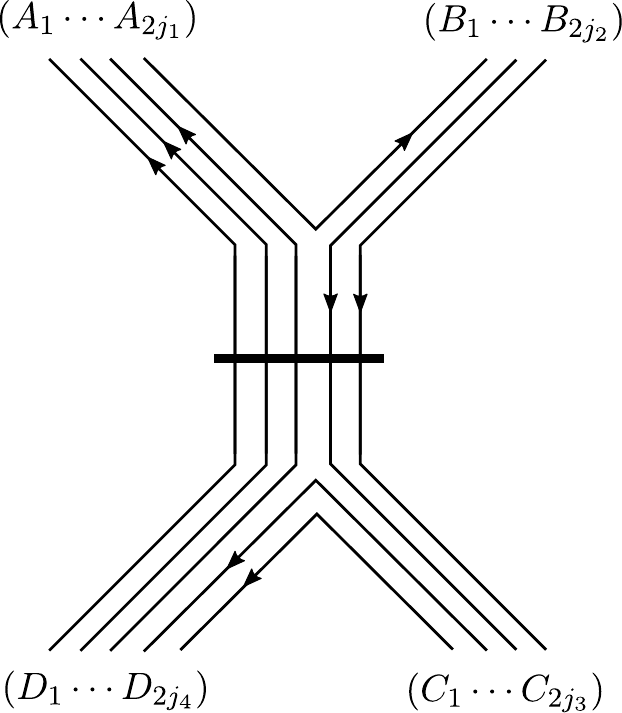}{0.8}
\ee
where we draw the horizontal bar to indicate a complete symmetrization of the internal group of lines.

\subsection*{Exercises}
\addcontentsline{toc}{subsection}{{\hspace{21.4pt} Exercises}}

\begin{enumerate}[leftmargin=*]

\item Verify that the $N$-valent intertwiner \eqref{iotaN} is invariant under the action of $SU(2)$ on its indices.

\item Calculate the scalar product between two intertwiners of the form \eqref{iotaN}.

\item Show that
\begin{align*}
\D{j_1}{m_1}{n_1}{g^{-1}}&\cdots\D{j_M}{m_M}{n_M}{g^{-1}}\D{j_{M+1}}{n_{M+1}}{m_{M+1}}{g}\cdots \\
&\qquad\qquad\cdots\D{j_N}{n_N}{m_N}{g}\iota\updown{n_1\cdots n_M}{n_{M+1}\cdots n_N} = \iota\updown{m_1\cdots m_M}{m_{M+1}\cdots m_N}
\end{align*}
where $\iota\updown{n_1\cdots n_M}{n_{M+1}\cdots n_N}$ is obtained by raising some indices of an $N$-valent intertwiner $\iota_{n_1\cdots n_N}$ with the epsilon tensor.

\item Prove the Wigner--Eckart theorem for the matrix elements of a spherical tensor operator by studying how the object
\[
T\updown{m_1}{mm_2} \equiv \bra{j_1m_1}T^{(j)}_m\ket{j_2m_2}
\]
transforms under the action of $SU(2)$ on its indices, and concluding that $T\updown{m_1}{mm_2}$ must be an element of the intertwiner space ${\rm Inv}\,\bigl({\cal H}_{j_1}^*\otimes{\cal H}_j\otimes{\cal H}_{j_2}\bigr)$.

\item Show that the Clebsch--Gordan coefficients satisfy the recursion relation
\begin{align*}
A_\pm(j,m)&\CG{j_1j_2}{j}{m_1m_2}{m\pm 1} \\
&= A_\mp(j_1,m_1)\CG{j_1j_2}{j}{m_1\mp 1,m_2}{m} + A_\mp(j_2,m_2)\CG{j_1j_2}{j}{m_1,m_2\mp 1}{m},
\end{align*}
where $A_\pm(j,m) = \sqrt{j(j+1)-m(m\pm 1)}$.

\item Show that the definition of a spherical tensor operator given by \Eq{sphtensor} is equivalent to the components of the operator satisfying the commutation relations
\begin{align*}
[J_\pm,T^{(j)}_m] &= \sqrt{j(j+1)-m(m\pm 1)}T^{(j)}_{m\pm 1}, \\
[J_z,T^{(j)}_m] &= mT^{(j)}_m.
\end{align*}

\item Give an alternative proof of the Wigner--Eckart theorem by using the results of the previous two problems to show that the reduced matrix element defined by the equation
\[
\bra{j_1m_1}T^{(j)}_m\ket{j_2m_2} = (j_1||T^{(j)}||j_2)\CG{j_2j}{j_1}{m_2m}{m_1}
\]
is independent of $m_1$, $m_2$ and $m$.

\item Consider the coupling of three angular momenta $j_1$, $j_2$ and $j_3$ into a total angular momentum characterized by the quantum numbers $j$ and $m$. This can be done in the following two inequivalent ways:
\begin{itemize}
\item First $j_1$ and $j_2$ are coupled to an intermediate angular momentum $k_{12}$; then $k_{12}$ and $j_3$ are coupled to produce the total angular momentum.
\item First $j_2$ and $j_3$ are coupled to an intermediate angular momentum $k_{23}$; then $k_{23}$ and $j_1$ are coupled to produce the total angular momentum.
\end{itemize}
Show that the states obtained from the two coupling schemes are related to each other by
\[
\ket{j_1j_2j_3(k_{23});jm} = \sum_{k_{12}}\sqrt{d_{k_{12}}d_{k_{23}}}\sixj{j_1}{j_2}{k_{12}}{j_3}{j}{k_{23}}\ket{j_1j_2(k_{12})j_3;jm},
\]
taking \Eq{6j-def} or \Eq{6j=iiii} as the definition of the 6$j$-symbol.

\item Consider the coupling of four angular momenta $j_1$, $j_2$, $j_3$ and $j_4$ into a total angular momentum characterized by the quantum numbers $j$ and $m$. This can be done in the following two inequivalent ways:
\begin{itemize}
\item First $j_1$ and $j_2$ are coupled to an intermediate angular momentum $k_{12}$, while $j_3$ and $j_4$ are coupled into an intermediate angular momentum $k_{34}$. Then $k_{12}$ and $k_{34}$ are coupled to produce the total angular momentum.
\item First $j_1$ and $j_3$ are coupled to an intermediate angular momentum $l_{13}$, while $j_2$ and $j_4$ are coupled into an intermediate angular momentum $l_{24}$. Then $l_{13}$ and $l_{24}$ are coupled to produce the total angular momentum.
\end{itemize}
Show that the states obtained from the two coupling schemes are related to each other by
\begin{align*}
&\ket{j_1j_3(l_{13})j_2j_4(l_{24});jm} \\
&= \sum_{k_{12}k_{34}}\sqrt{d_{k_{12}}d_{k_{34}}d_{l_{13}}d_{l_{24}}}\ninej{j_1}{j_2}{k_{12}}{j_3}{j_4}{k_{34}}{l_{13}}{l_{24}}{j}\ket{j_1j_2(k_{12})j_3j_4(k_{34});jm},
\end{align*}
taking \Eq{9j-def} or \Eq{9j=i^6} as the definition of the 9$j$-symbol.

\item Let $T^{(k_1)}$ and $U^{(k_2)}$ be spherical tensor operators of rank $k_1$ and $k_2$.
\begin{itemize}
\item[(a)] Show that the operator defined by
\[
\bigl[T^{(k_1)}\otimes U^{(k_2)}\bigr]^{(k)}_m \equiv \sum_{m_1m_2} \CGi{k_1k_2}{k}{m_1m_2}{m}T^{(k_1)}_{m_1}U^{(k_2)}_{m_2}
\]
(and referred to as the tensor product of $T^{(k_1)}$ and $U^{(k_2)}$) is a spherical tensor operator of rank $k$.
\item[(b)] Suppose now that $T^{(k_1)}$ and $U^{(k_2)}$ act on two different Hilbert spaces ${\cal H}_1$ and ${\cal H}_2$. Show that the reduced matrix elements of the tensor product operator $\bigl[T^{(k_1)}\otimes U^{(k_2)}\bigr]^{(k)}$ with respect to the coupled basis $\ket{j_1j_2;jm}$ on ${\cal H}_1\otimes{\cal H}_2$ are given in terms of the reduced matrix elements of $T^{(k_1)}$ and $U^{(k_2)}$ by
\begin{align*}
(j_1j_2;j||&\bigl[T^{(k_1)}\otimes U^{(k_2)}\bigr]^{(k)}||j_1'j_2';j') \\
&= \sqrt{d_{j_1}d_{j_2}d_{k_1}d_{k_2}d_k}\ninej{j_1}{j_1'}{k_1}{j_2}{j_2'}{k_2}{j}{j'}{k}(j_1||T^{(k_1)}||j_1')(j_2||U^{(k_2)}||j_2').
\end{align*}
\end{itemize}

\item Use the graphical representation \eqref{i3graphic} to derive the symmetry relations satisfied by the 3$j$-symbol.

\item Determine the value of the normalization factor $N_{j_1j_2j_3}$ in \Eq{iota_ABC}.

\end{enumerate}

\newpage

\section{The graphical method}\label{ch:graphical}

The last two chapters of these notes deal with the powerful graphical techniques for calculations in $SU(2)$ recoupling theory. In this chapter we introduce the graphical framework, which on the surface may simply seem to consist of adopting a diagrammatic notation for $SU(2)$ calculations. However, in reality the graphical approach serves a highly efficient computational device, since the visual graphical diagrams are invariably easier to comprehend and simpler to work with than the corresponding non-graphical expressions. In loop quantum gravity, the graphical techniques can be generally used in any calculation involving intertwiners, and we will give several detailed examples of such calculations in the next chapter.

Several closely related but slightly different versions of the graphical formalism can be found in the literature of quantum angular momentum \cite{BrinkSatchler, Varshalovich, YLV}, and in articles in which graphical methods are used for calculations in loop quantum gravity \cite{AlesciLiegenerZipfel, AlesciRovelli, AlesciThiemannZipfel, YangMa1, YangMa2, paper1}. In these circumstances it seems preferable (and it certainly requires less effort) to develop a consistent set of conventions for the graphical formalism on one's own, using the available references as a guide, rather than trying to ensure that one is in full agreement with the conventions chosen in some previous treatment of the subject. When compared to conventions available in the literature, the conventions adopted in this work match the most closely with those of Brink and Satchler \cite{BrinkSatchler}, which was the main reference used by the author to learn the graphical techniques in the early stages of his PhD studies.

\subsection{Elements of the graphical formalism}

The basic elements of the graphical formalism are provided by the graphical representations of the elementary objects of $SU(2)$ representation theory: the delta and epsilon tensors, the 3$j$-symbol, and $SU(2)$ group elements. The properties satisfied by these objects lead to graphical rules according to which diagrams appearing in a graphical calculation can be manipulated.

\subsubsection*{Delta and epsilon tensors}

The unit tensor $\delta_m^n$ is represented graphically by a line which carries the indices $m$ and $n$ at its two ends:
\be\label{delta g}
\delta_m^n\; = \;\;\RealSymb{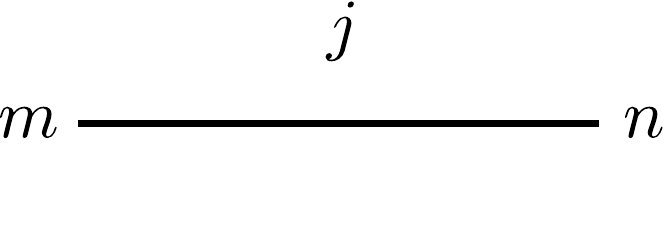}{0.6}\vspace{-12pt}
\ee
If necessary, the line can be labeled with a spin $j$ to indicate the representation which the indices of $\delta_m^n$ refer to.

The invariant tensor $\epsilon^{(j)}_{mn}$ is represented by a line with an arrow pointing from index $m$ to index $n$:
\be\label{epsilon g}
\epsilon^{(j)}_{mn}\; = \;\;\RealSymb{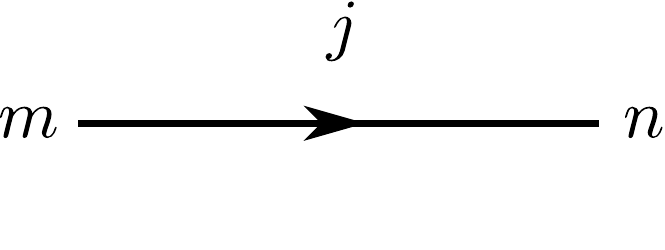}{0.6}\vspace{-12pt}
\ee
Since tensor $\epsilon^{(j)mn}_{\phantom{m}}$ is numerically equal to $\epsilon^{(j)}_{mn}$, it is represented by the same diagram:
\be\label{epsilon2 g}
\epsilon^{(j)mn}_{\phantom{m}}\; = \;\;\RealSymb{figA-epsilon-indices.pdf}{0.6}\vspace{-12pt}
\ee
The magnetic indices in graphical diagrams such as \eqref{delta g}--\eqref{epsilon2 g} are usually not shown explicitly, if leaving them out is not likely to cause any confusion. For example, the indices of the epsilon tensor can be safely omitted, since the direction of the arrow shows which end of the line corresponds to which index of $\epsilon^{(j)}_{mn}$. From now on we will follow this practice and not write out the magnetic indices unless there is a particular reason for doing so.

The relations \eqref{eps_nm} and \eqref{epseps=delta} satisfied by the epsilon tensor imply that the arrow behaves according to the rules
\be\label{invarrow}
\RealSymb{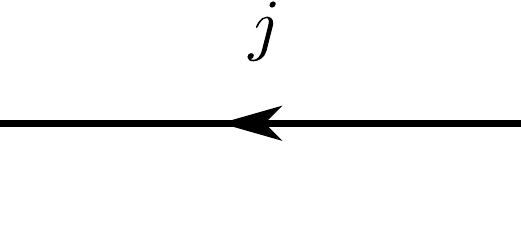}{0.6}\quad = \quad(-1)^{2j}\;\RealSymb{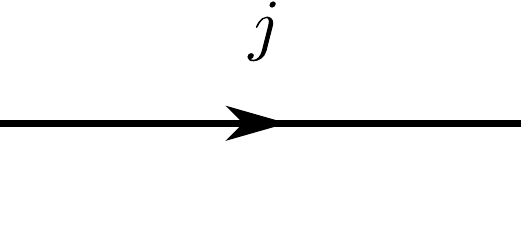}{0.6}\vspace{-16pt}
\ee
and
\be\label{2arrows-same}
\RealSymb{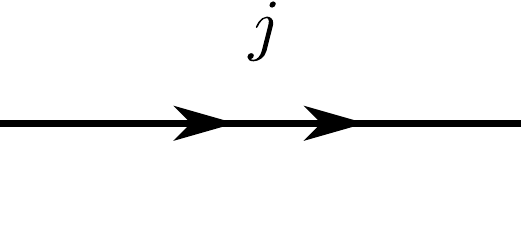}{0.6}\quad = \quad(-1)^{2j}\;\RealSymb{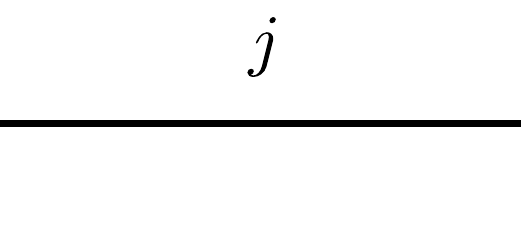}{0.6}\vspace{-8pt}
\ee
In the second relation we have introduced an important element of the graphical formalism: Contraction of indices is carried out by connecting the ends of the two lines corresponding to the contracted index. Using the first relation, the second relation can be equivalently written as
\be\label{2arrows-opp}
\RealSymb{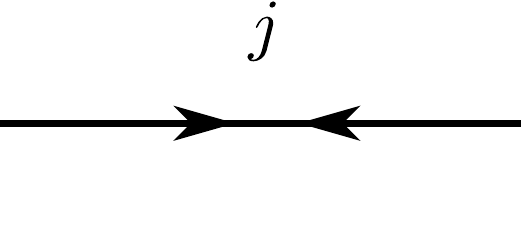}{0.6}\quad = \quad\RealSymb{figA-delta.pdf}{0.6}\vspace{-12pt}
\ee

\subsubsection*{The 3$j$-symbol}

The graphical representation of the 3$j$-symbol is given by three lines connected at a node:
\be\label{3j g}
\threej{j_1}{j_2}{j_3}{m_1}{m_2}{m_3}\; = \;\RealSymb{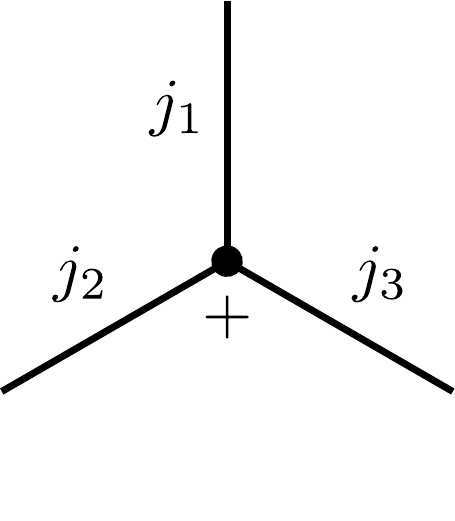}{0.6}\vspace{-20pt}
\ee
The sign at the node encodes the cyclic order of the spins in the symbol, a $+$ sign corresponding to the anticlockwise order of \Eq{3j g}, while a $-$ sign corresponds to a clockwise order. Thus, \vspace{-8pt}
\be\label{3j- g}
\threej{j_1}{j_3}{j_2}{m_1}{m_3}{m_2}\; = \quad\RealSymb{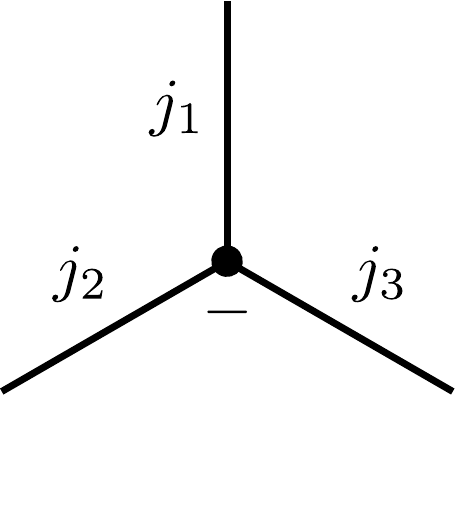}{0.6}\quad = \quad \RealSymb{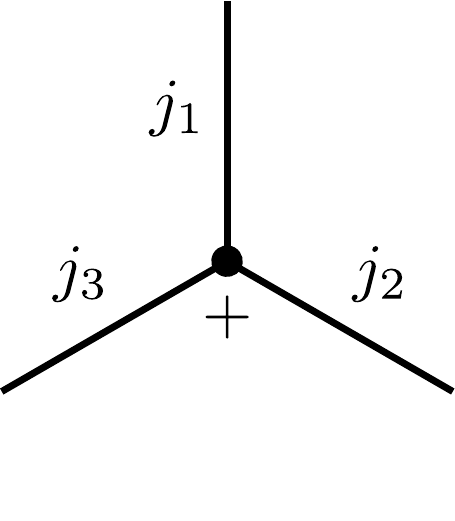}{0.6}\vspace{-20pt}
\ee
\Eq{3jperm} then implies that reversing the sign produces the factor $(-1)^{j_1+j_2+j_3}$:
\be\label{3jminus g}
\RealSymb{figA-3j-minus.pdf}{0.6} \quad = \quad (-1)^{j_1+j_2+j_3}\RealSymb{figA-3j.pdf}{0.6}\vspace{-18pt}
\ee
From the other symmetry relation \eqref{3j-m} it follows that
\be\label{3jarrows g}
\RealSymb{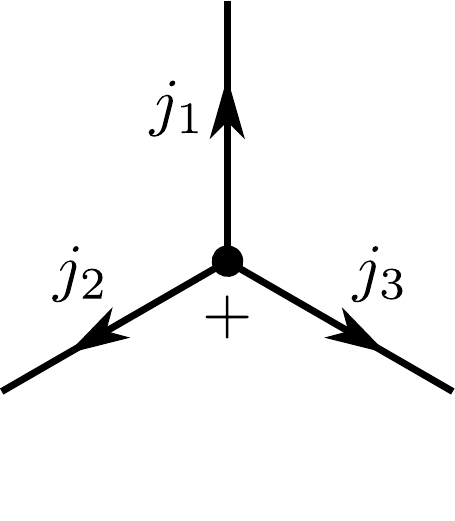}{0.6} \quad = \quad \RealSymb{figA-3j.pdf}{0.6}\vspace{-16pt}
\ee
which is a graphical representation of the fact that $\iota^{m_1m_2m_3}$ is numerically equal to $\iota_{m_1m_2m_3}$. Furthermore, the orthogonality relations \eqref{3j-orth-mm} and \eqref{3j-orth-jm} read
\be\label{3j-orth-mm g}
\RealSymb{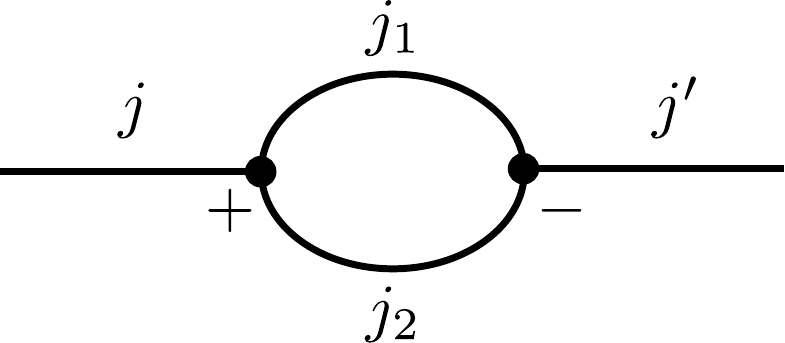}{0.6} \quad = \quad\delta_{jj'}\frac{1}{d_j}\;\RealSymb{figA-delta.pdf}{0.6}\vspace{6pt}
\ee
and
\be\label{3j-orth-jm g}
\sum_j d_j\RealSymb{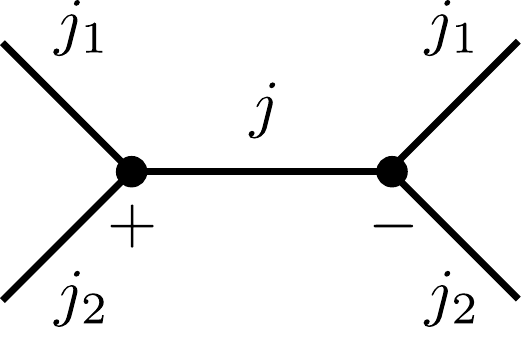}{0.6} \quad = \quad \RealSymb{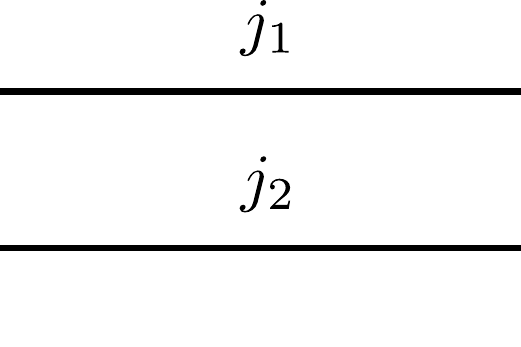}{0.6}\vspace{6pt}
\ee
The normalization of the 3$j$-symbol, implied by the first orthogonality relation, gives
\be\label{3j-theta g}
\RealSymb{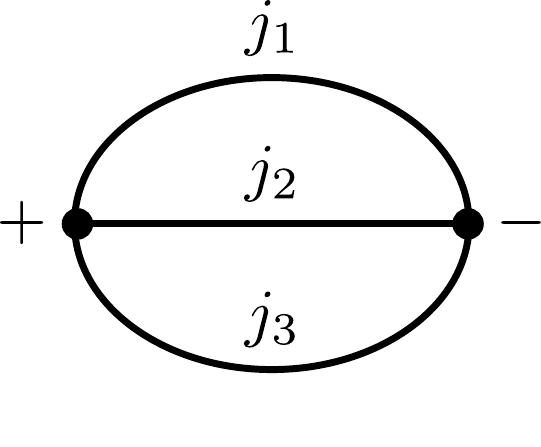}{0.6}\quad =\;\;1.\vspace{-8pt}
\ee
\Eq{3jzero}, which shows how the 3$j$-symbol reduces to the epsilon tensor when one of the spins is zero, is written graphically as
\be\label{3j-zero g}
\RealSymb{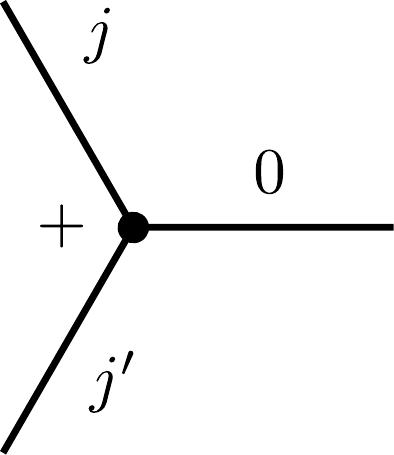}{0.6} \quad = \quad\delta_{jj'}\frac{1}{\sqrt{d_j}}\quad\RealSymb{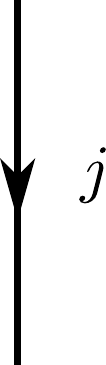}{0.6}\vspace{6pt}
\ee
In equations such as \eqref{3j-orth-jm g} and \eqref{3j-zero g}, where the relation between the indices on the two sides of the equation is in principle ambiguous unless one writes them out explicitly, we follow the convention that the relative position of the indices is the same on both sides. For example, if we consider the index at the top left corner on the left-hand side of \Eq{3j-orth-jm g} as the index $m_1$ of the corresponding non-graphical equation \eqref{3j-orth-jm}, then $m_1$ is at the top left corner also on the right-hand side of \Eq{3j-orth-jm g}.

\newpage

From \Eq{3j}, which defines the 3$j$-symbol in terms of the Clebsch--Gordan coefficient, we find that the graphical representation of the Clebsch--Gordan coefficient is
\be\label{clebsch g}
\CG{j_1j_2}{j}{m_1m_2}{m} \; = \; (-1)^{j_1-j_2-j}\sqrt{d_j}\;\RealSymb{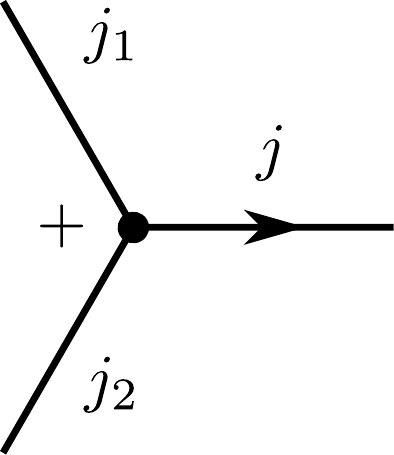}{0.6}
\ee
\Eq{tau=C} then shows that the matrix elements of the $SU(2)$ generators are given by \vspace{-20pt}
\be\label{tau=C g}
\Tau{j}{i}{m}{n} \; = \; iW_j\,\RealSymb{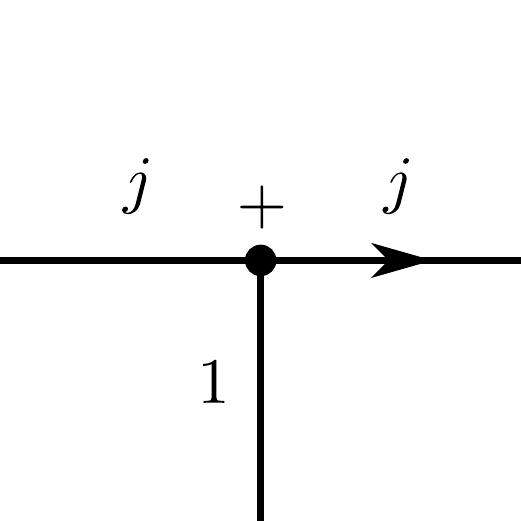}{0.6}
\ee
where we used the shorthand notation
\be
W_j = \sqrt{j(j+1)(2j+1)}.
\ee
To write the triple product $\epsilon_{ijk}u^iv^jw^j$ in graphical form, let us introduce the diagram
\be
v^m \; = \; \RealSymb{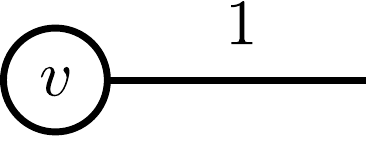}{0.6}
\ee
to represent a vector with an index in the $j=1$ representation. Then, according to \Eq{epsuvw}, the triple product can be expressed graphically as
\be\label{epsuvw g}
\epsilon_{ijk}u^iv^jw^j \; = \;i\sqrt{6}\;\RealSymb{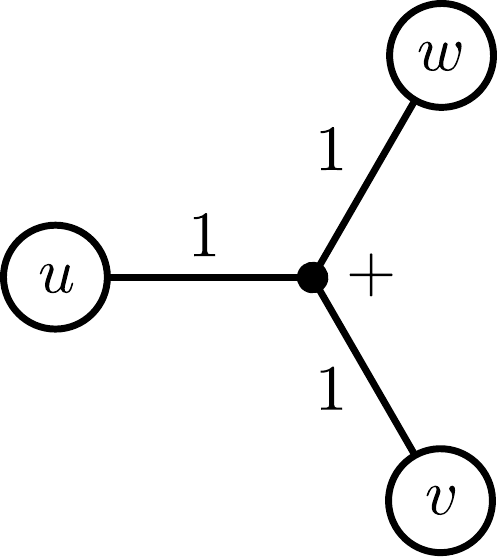}{0.6}
\ee

\subsubsection*{Group elements}

For the matrix elements of the Wigner matrices $D^{(j)}(g)$, we adopt the graphical representation
\be\label{WignerD g}
\D{j}{m}{n}{g} \; = \; \RealSymb{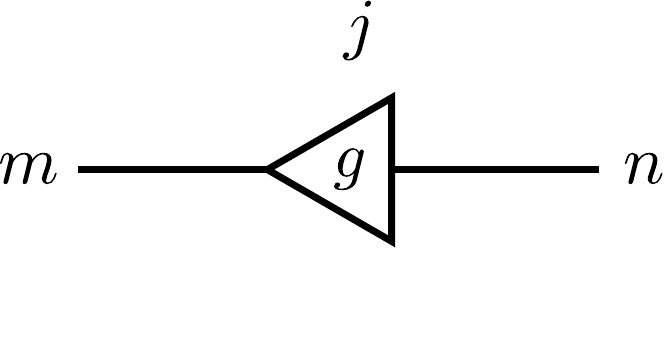}{0.6}\vspace{-12pt}
\ee
The group multiplication law is expressed graphically as
\be
\RealSymb{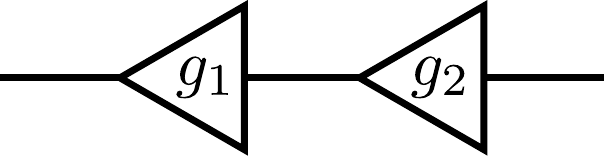}{0.6} \quad = \quad \RealSymb{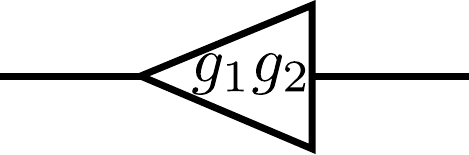}{0.6}
\ee
From \Eq{Dj-inv}, we see that that the inverse matrix is given by
\be\label{D^-1 g}
\RealSymb{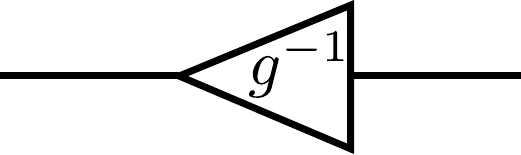}{0.6} \quad = \quad \RealSymb{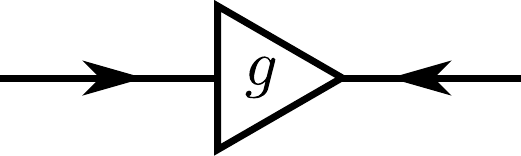}{0.6}
\ee
The orthogonality theorem for the Wigner matrices can be translated into graphical form, if we denote the complex conjugate of the matrix elements by a bar over the group element $g$ inside the triangle in \Eq{WignerD g}. We then have
\be
\int dg\quad \RealSymb{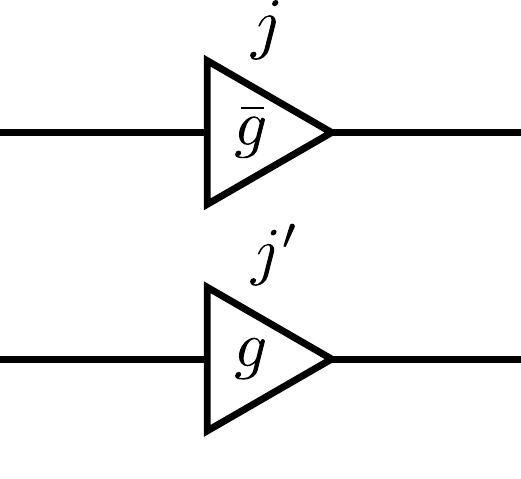}{0.6} \quad\; = \quad \delta_{jj'}\frac{1}{d_j}\quad\RealSymb{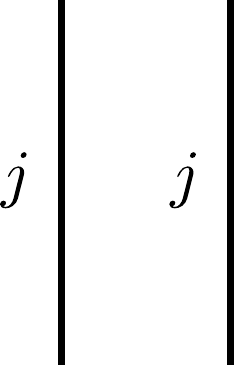}{0.6}
\ee
Alternatively, the relation $D^{(j)}(g^\dagger) = D^{(j)}(g^{-1})$ can be used to eliminate the complex conjugate. This leads to
\begin{align}
\int dg\,&\D{j}{m}{n}{g}\D{j'}{m'}{n'}{g} \notag \\
&= \epsilon^{(j)m\mu}_{\phantom{m}}\epsilon^{(j)}_{n\nu}\int dg\,\D{j}{\nu}{\mu}{g^{-1}}\D{j'}{m'}{n'}{g} = \delta_{jj'}\frac{1}{d_j}\epsilon^{(j)mm'}_{\phantom{m}}\epsilon^{(j)}_{nn'} \label{int DD}
\end{align}
The corresponding graphical equation is
\be\label{int DD g}
\int dg\quad \RealSymb{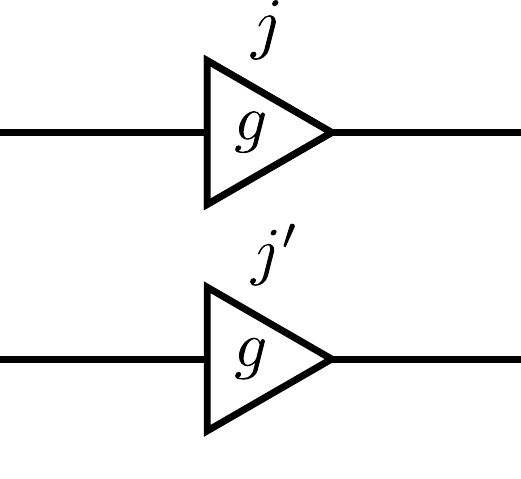}{0.6} \quad\; = \quad \delta_{jj'}\frac{1}{d_j}\quad\RealSymb{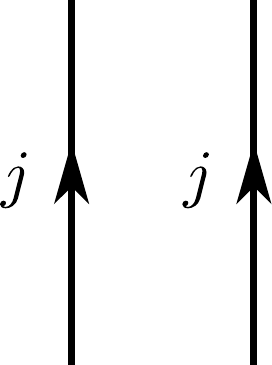}{0.6}
\ee
The Clebsch--Gordan series \eqref{CG-ser} is expressed graphically as
\begin{align}
\RealSymb{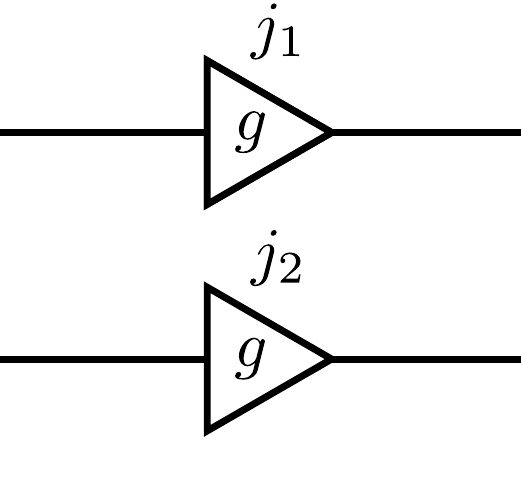}{0.6} \quad = \quad \sum_j d_j\;\RealSymb{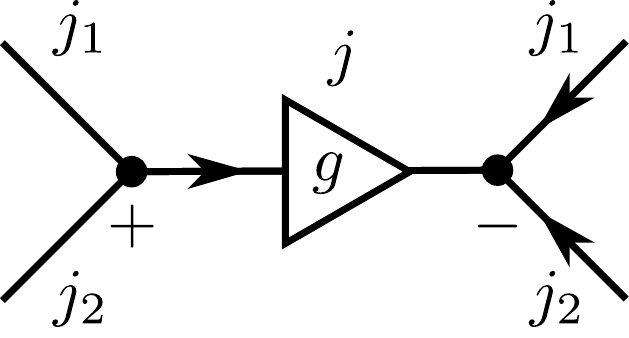}{0.6} \label{DD g} \\
\intertext{Using \Eq{D^-1 g}, we can also derive the coupling of a group element with its inverse:} 
\RealSymb{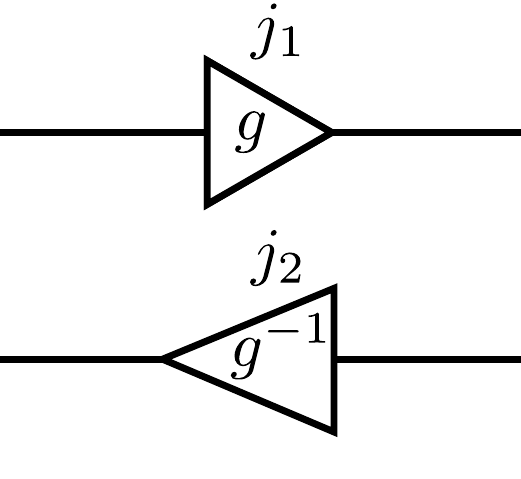}{0.6} \quad = \quad \sum_j d_j\;\RealSymb{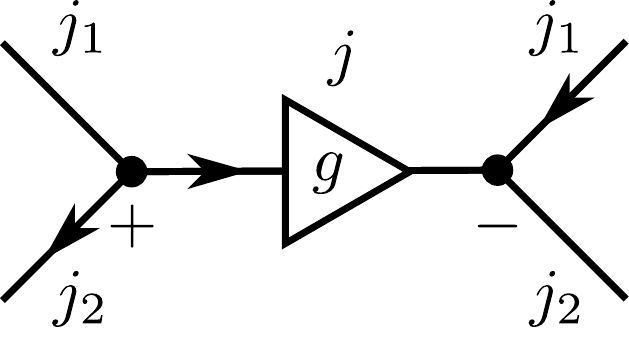}{0.6}\label{D^-1D g}
\end{align}

\newpage

\subsection{Intertwiners in graphical form}

The graphical representation of the basic three-valent intertwiner $\iota_{m_1m_2m_3}$ is given by \Eq{3j g}. Intertwiners of higher valence can be constructed by using the epsilon tensor to attach several three-valent intertwiners to each other, as discussed in section \ref{sec:intertwiners-N}. Hence the four-valent intertwiner of \Eq{iota412} is represented graphically as
\be\label{iota412 g}
\bigl(\iota_{12}^{(k)}\bigr)_{m_1m_2m_3m_4} \; = \;\RealSymb{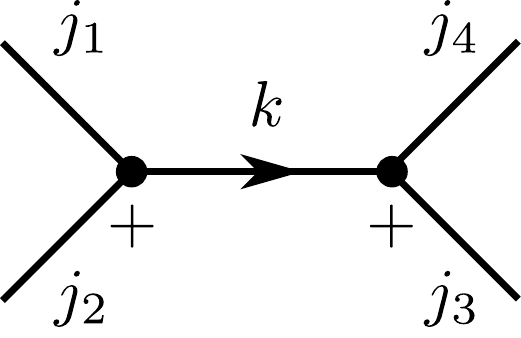}{0.6}
\ee
where the epsilon tensor appears as an arrow on the internal line. By graphical means it is easy to show that the scalar product between two intertwiners of this form is indeed given by \Eq{iota4-prod}. Using \Eqs{3j-orth-mm g} and \eqref{2arrows-opp}, we find
\be
\bbraket{\iota_{12}^{(k)}}{\iota_{12}^{(k')}} \; = \quad \RealSymb{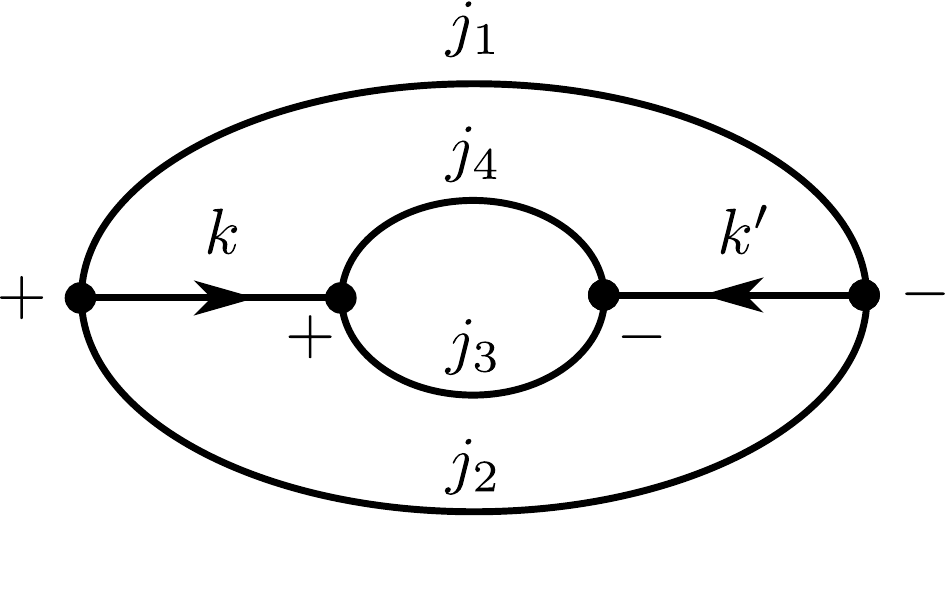}{0.6} \quad = \; \frac{1}{d_k}\delta_{kk'}.\vspace{-12pt}
\ee
The intertwiner \eqref{iota413}, in which the spins $j_1$ and $j_3$ have been coupled to the internal spin, has the graphical form
\be\label{iota413 g}
\bigl(\iota_{13}^{(l)}\bigr)_{m_1m_2m_3m_4} \; = \;\RealSymb{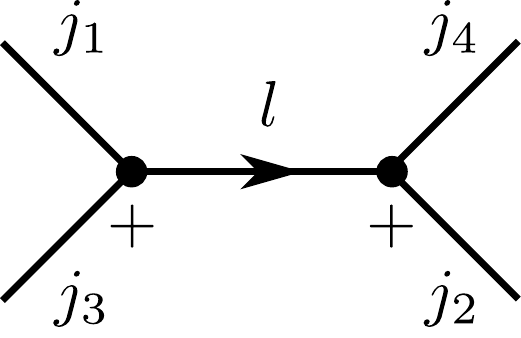}{0.6}
\ee
The change of basis between the bases \eqref{iota412 g} and \eqref{iota413 g} is given by
\be\label{6j-def g}
\RealSymb{figA-iota413.pdf}{0.6} \quad = \quad \sum_k d_k(-1)^{j_2+j_3+k+l}\sixj{j_1}{j_2}{k}{j_4}{j_3}{l}\;\RealSymb{figA-iota412.pdf}{0.6}
\ee
By contracting both sides of this equation with the intertwiner \eqref{iota412 g}, one can derive the graphical expression \vspace{-12pt}
\be\label{6j g}
\sixj{j_1}{j_2}{j_3}{k_1}{k_2}{k_3} \quad = \; \RealSymb{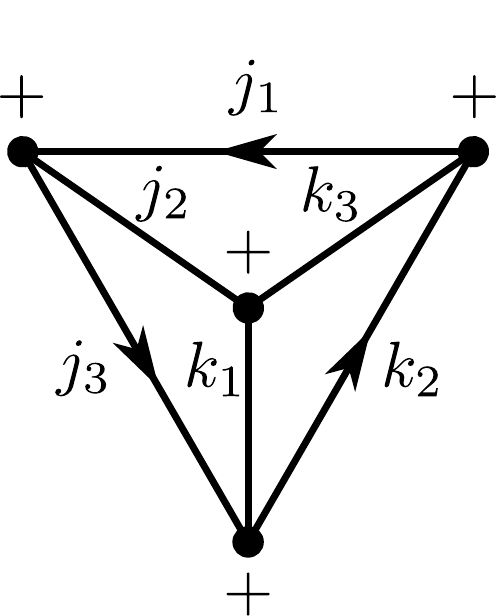}{0.6}
\ee
for the 6$j$-symbol. \Eq{6j g} is of course the graphical equivalent of \Eq{6j=iiii}. If one is given the diagram on the right-hand side of the above equation, the 6$j$-symbol can be read off from the diagram as follows: Pick any node of the diagram and write the spins connected to the node (in any order) in the top row of the 6$j$-symbol. Then the spin which is ''opposite'' a given spin of the top row in the diagram (for example, in the diagram \eqref{6j g} the spin opposite $j_1$ is $k_1$, and so on) goes under that spin in the bottom row of the 6$j$-symbol. The symmetries of the 6$j$-symbol guarantee that it makes no difference which node or which order of spins in the top row is chosen.

The five-valent intertwiners \eqref{iota512} and \eqref{iota513} are written graphically as
\begin{align}
\bigl(\iota_{12}^{(k_{12}k_{34})}\bigr)_{m_1\cdots m_5} \; &= \; \RealSymb{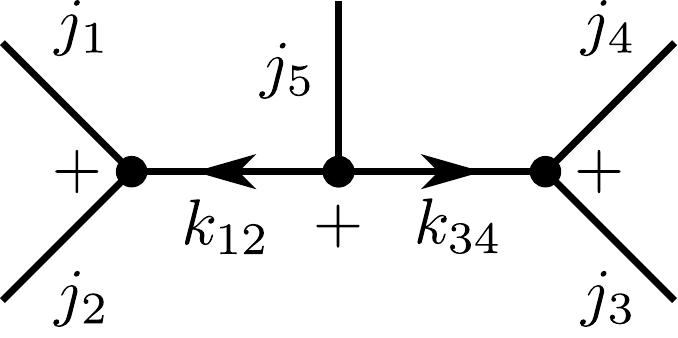}{0.6} \label{iota512 g} \\
\intertext{and}
\bigl(\iota_{13}^{(l_{13}l_{24})}\bigr)_{m_1\cdots m_5} \; &= \; \RealSymb{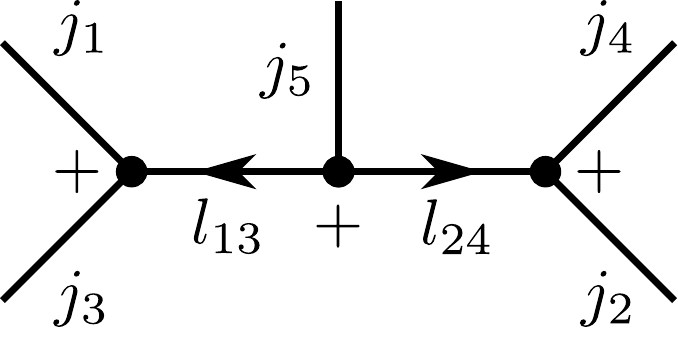}{0.6} \label{iota513 g}
\end{align}
The intertwiner \eqref{iota513 g} is expressed in the basis \eqref{iota512 g} by the relation
\begin{align}
&\RealSymb{figA-iota513-l.pdf}{0.6} \notag \\
&\qquad\qquad=\quad\sum_{k_{12}k_{34}} d_{k_{12}}d_{k_{34}}\ninej{j_1}{j_2}{k_{12}}{j_3}{j_4}{k_{34}}{l_{13}}{l_{24}}{j_5}\;\RealSymb{figA-iota512.pdf}{0.6}\label{9j-def g}
\end{align}
Contracting both sides with the intertwiner \eqref{iota512 g}, one finds that the 9$j$-symbol has the graphical representation \vspace{-12pt}
\be\label{9j g}
\ninej{j_1}{j_2}{j_3}{k_1}{k_2}{k_3}{l_1}{l_2}{l_3} \quad = \quad \RealSymb{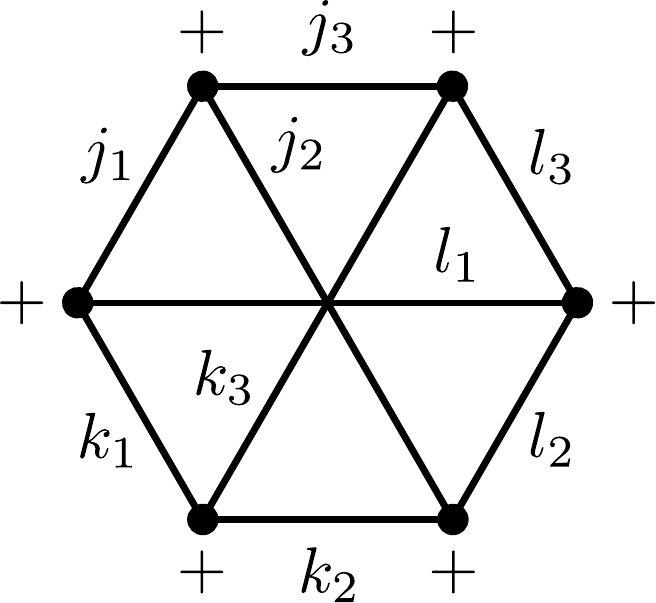}{0.6}
\ee
which is just \Eq{9j=i^6} translated into graphical form. When it comes to remembering how to read off the 9$j$-symbol from the diagram, note that the spins of the 9$j$-symbol appear in the diagram consistently in a counterclockwise order. When going around the different nodes of the diagram, the rows and columns of the 9$j$-symbol are encountered in a counterclockwise order, and likewise the spins of each row and each column appear in the corresponding node in a counterclockwise order.

$N$-valent intertwiners can be derived by continuing to contract three-valent intertwiners in the way indicated by \Eqs{iota412 g} and \eqref{iota512 g}, as shown by the non-graphical equation \eqref{iotaN}. However, it can sometimes be convenient to use a different pattern of contractions to construct a basis of the $N$-valent intertwiner space. An example of such a basis is given by the intertwiners
\be\label{iotaN g}
\RealSymb{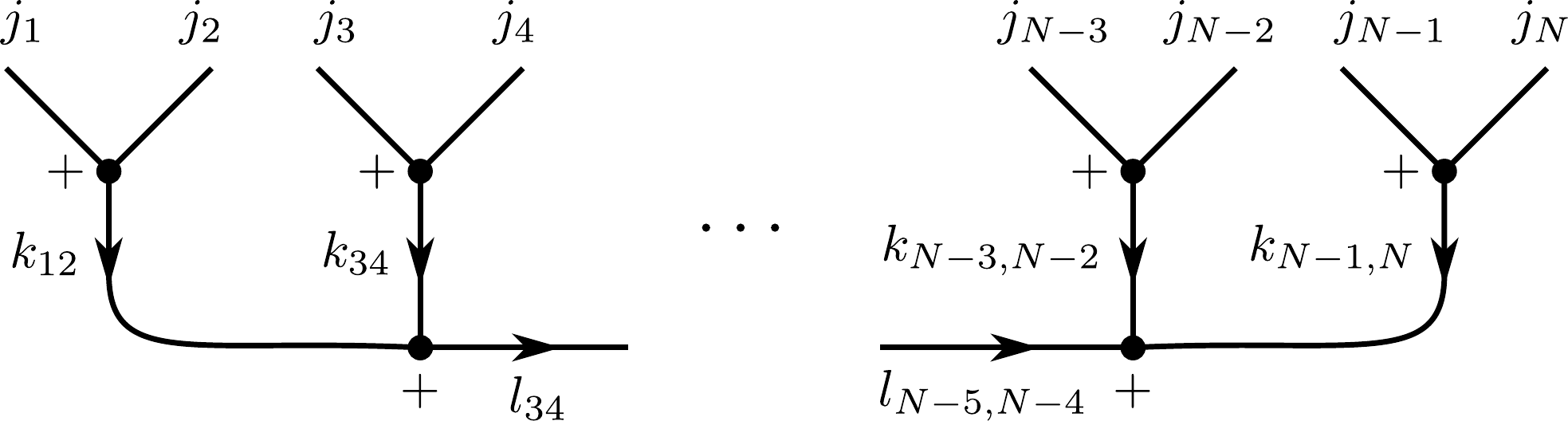}{0.6}
\ee
in which the spins of the intertwiner are coupled pairwise to internal spins. Just as the intertwiners given by \Eq{iotaN}, the intertwiners \eqref{iotaN g} are not normalized, but intertwiners carrying different internal spins are orthogonal to each other. To normalize the intertwiner \eqref{iotaN g}, it should be multiplied by the factor
\be
\sqrt{\displaystyle d_{k_{12}}\cdots d_{k_{N-1,N}}d_{l_{34}}\cdots d_{l_{N-5,N-4}}}.
\ee
Note that there is no loss of generality in using intertwiners of the form \eqref{iotaN g}, even though at a first sight it may seem that one is assuming the valence $N$ to be even. However, to obtain an intertwiner of odd valence from \eqref{iotaN g}, it suffices to set one of the spins, say $j_N$, equal to zero.

\subsection{The fundamental theorem of graphical calculus}\label{sec:ftgc}

The basic rules for working with graphical diagrams are given by \Eqs{invarrow}--\eqref{2arrows-opp} and \eqref{3j- g}--\eqref{3jarrows g}, which show how the arrows and signs in a graphical expression can be manipulated. However, the most essential tool for performing graphical calculations with intertwiners arises from the seemingly simple observation that an invariant tensor having $N$ indices can be expanded in a basis of the corresponding space of $N$-valent intertwiners. To express the implications of this observation in graphical form, we will represent a $N$-valent invariant tensor graphically as a rectangular block with $N$ lines attached to it.

A tensor carrying a single index cannot be invariant, unless the index belongs to the trivial representation. Thus,
\be\label{thm1}
\RealSymb{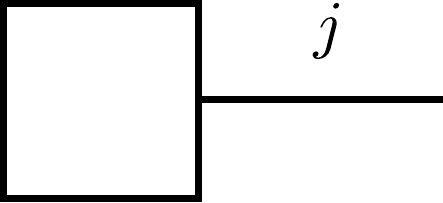}{0.6} \quad = \quad \delta_{j,0}\quad\RealSymb{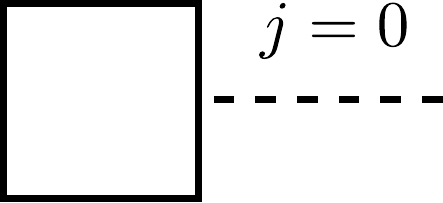}{0.6}
\ee
A two-valent tensor $T_{mn}$ can be invariant only if both of its indices belong to the same representation. In this case, $T_{mn}$ must be proportional to $\epsilon^{(j)}_{mn}$, which is the only invariant tensor having two lower indices. The coefficient of proportionality can be determined by contracting both sides of the equation with the epsilon tensor. Since the contraction of $\epsilon^{(j)}_{mn}$ with itself gives $d_j$, we have the graphical rule
\newpage
\be\label{thm2}
\RealSymb{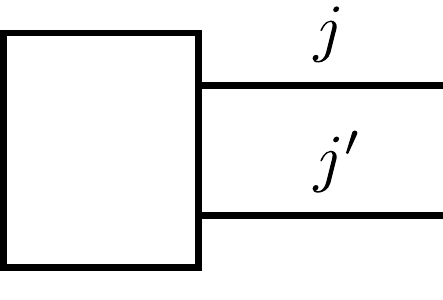}{0.6} \quad = \quad \delta_{jj'}\frac{1}{d_j}\quad\RealSymb{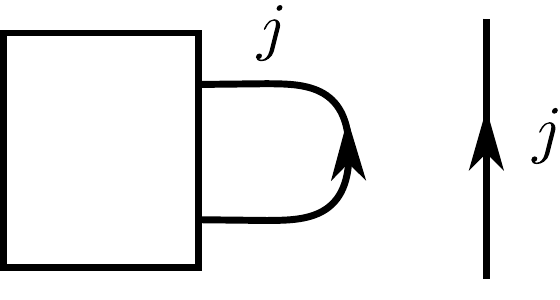}{0.6}\vspace{4pt}
\ee
A three-valent invariant tensor $T_{m_1m_2m_3}$ belongs to the space ${\rm Inv}\,\bigl({\cal H}_{j_1}\otimes{\cal H}_{j_2}\otimes{\cal H}_{j_3}\bigr)$, which is again one-dimensional (provided that the spins $j_1$, $j_2$ and $j_3$ satisfy the Clebsch--Gordan conditions). Therefore $T_{m_1m_2m_3}$ must be proportional to the 3$j$-symbol, the coefficient of proportionality again being determined by contracting the equation with the 3$j$-symbol. That is,
\be\label{thm3}
\RealSymb{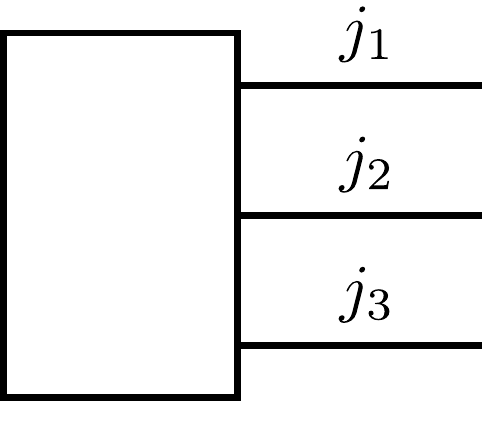}{0.6} \quad = \quad \RealSymb{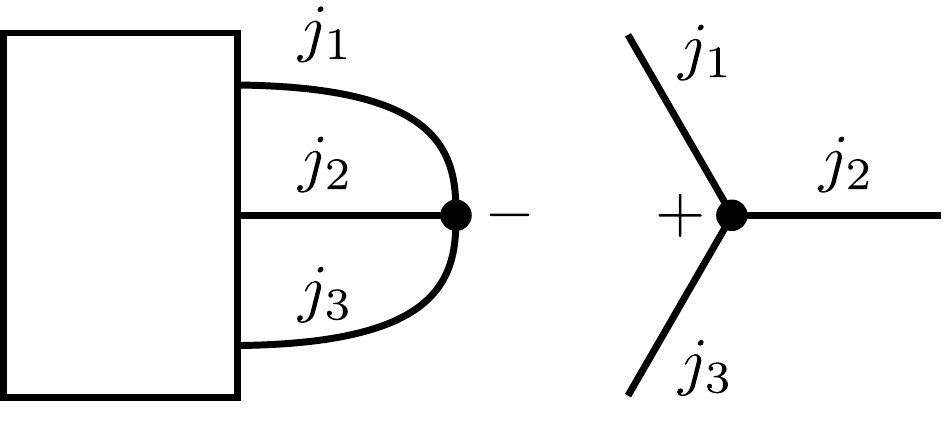}{0.6}
\ee
For invariant tensors of valence four or higher, there generally no longer is a unique intertwiner to which the tensor would have to be proportional. Nevertheless, an invariant tensor with $N$ indices can be expanded in any basis of the corresponding $N$-valent intertwiner space. For example, expanding a four-valent invariant tensor in the basis given by the intertwiners \eqref{iota412 g}, and recalling that the norm of the intertwiner \eqref{iota412 g} is $1/\sqrt{d_k}$, we obtain
\be\label{thm4}
\RealSymb{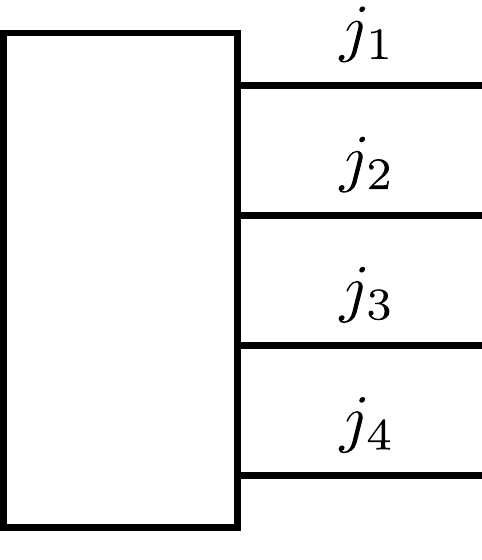}{0.6} \quad = \quad \sum_x d_x\quad\RealSymb{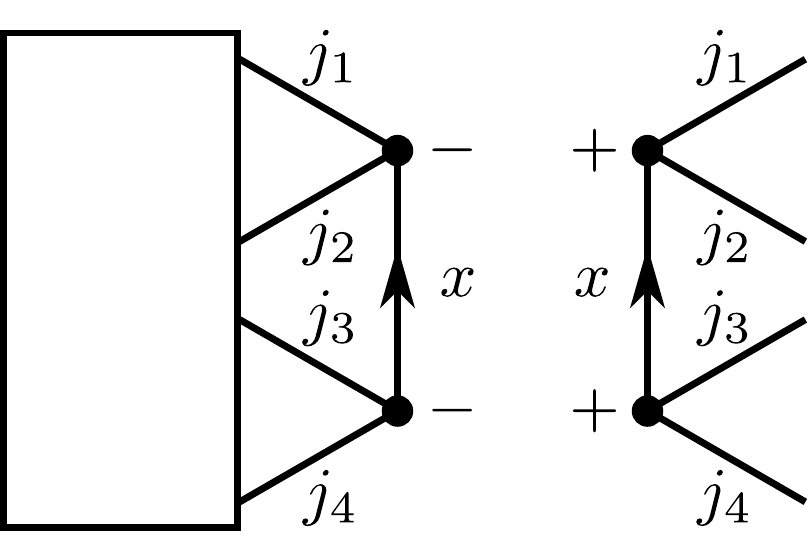}{0.6}
\ee

The relations \eqref{thm1}--\eqref{thm4} play such a central role in graphical calculations that they would fully deserve to be known as the fundamental theorem of graphical calculus. The last part of the theorem, given by \Eq{thm4}, generalizes in a straightforward way to tensors carrying more than four indices. Moreover, even though we wrote \Eqs{thm1}--\eqref{thm4} for tensors having only lower indices, similar relations are naturally valid for tensors having a different index structure. \Eqs{thm1}--\eqref{thm4} can be extended to such tensors simply by using the epsilon tensor to raise indices, corresponding graphically to attaching arrows to some lines on both sides of the equation.

An important way in which the fundamental theorem can be used is to break down a complicated graphical diagram into simpler constituents. Consider a graphical diagram representing an arbitrary invariant contraction of 3$j$-symbols or other invariant tensors. Suppose that the diagram contains a subdiagram with $N$ external lines, such that the subdiagram itself is an $N$-valent invariant tensor, and the graph of the entire diagram can be separated into two disconnected pieces by cutting the $N$ lines of the subdiagram. In this case the diagram can potentially be simplified by using the fundamental theorem to expand the subdiagram in a basis of the $N$-valent intertwiner space.

If a diagram can be divided into two pieces by cutting one, two or three lines, then \Eqs{thm1}--\eqref{thm3} imply that the diagram simply splits into a product of two factors according to the rules
\begin{align}
\RealSymb{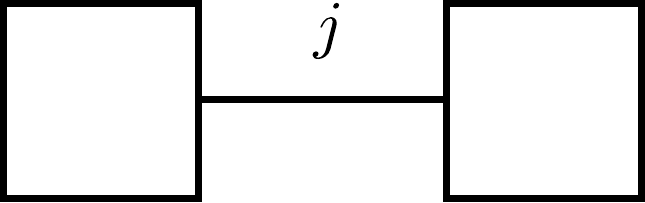}{0.6} \quad &= \quad \delta_{j,0}\quad\RealSymb{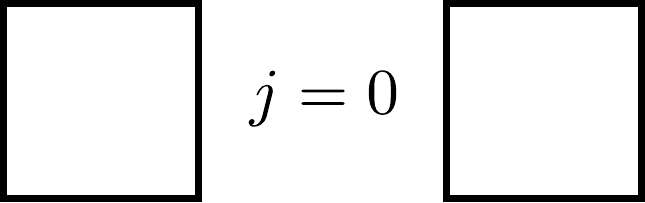}{0.6} \label{thm1'} \\[8pt]
\RealSymb{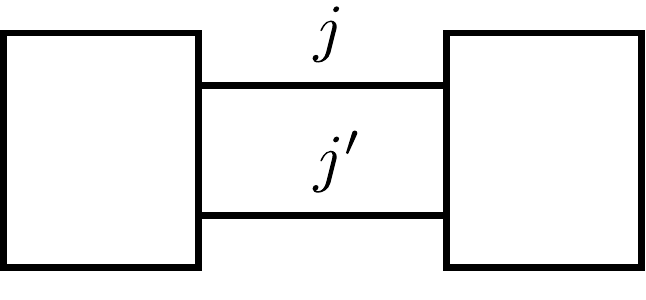}{0.6} \quad &= \quad \delta_{jj'}\frac{1}{d_j}\quad\RealSymb{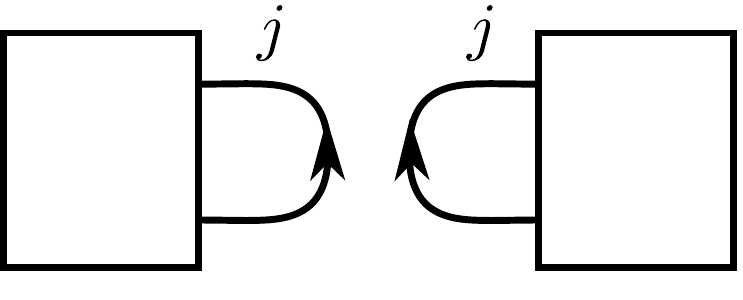}{0.6} \label{thm2'} \\[8pt]
\RealSymb{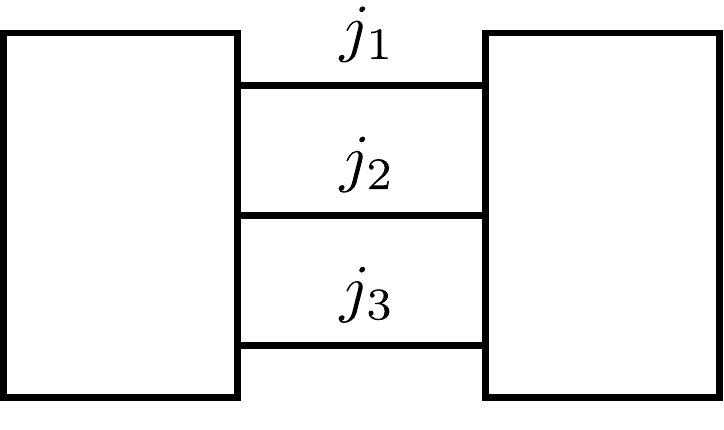}{0.6} \quad &= \quad \RealSymb{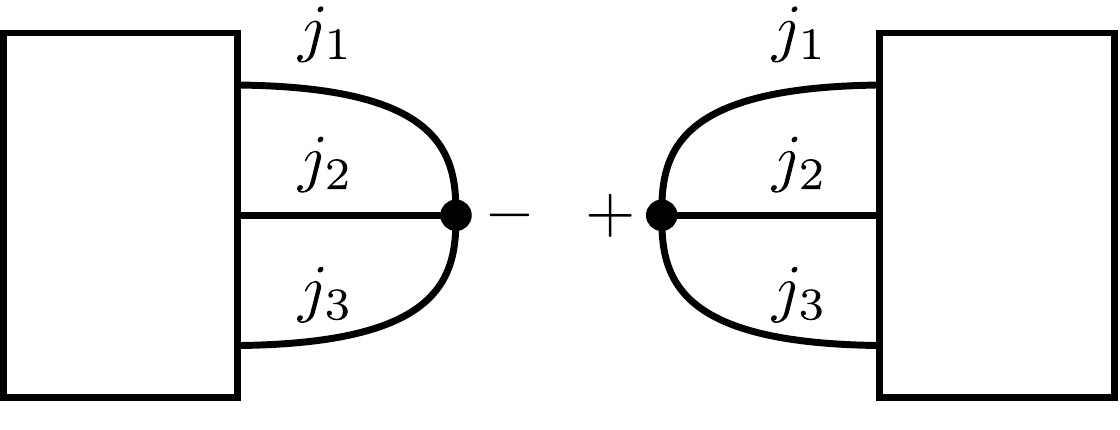}{0.6} \label{thm3'} \\
\intertext{In the case of cutting four lines to break a diagram into two, \Eq{thm4} shows that one does not obtain simply a product of two factors, but rather a sum in which each term is a product of two disconnected diagrams:}
\RealSymb{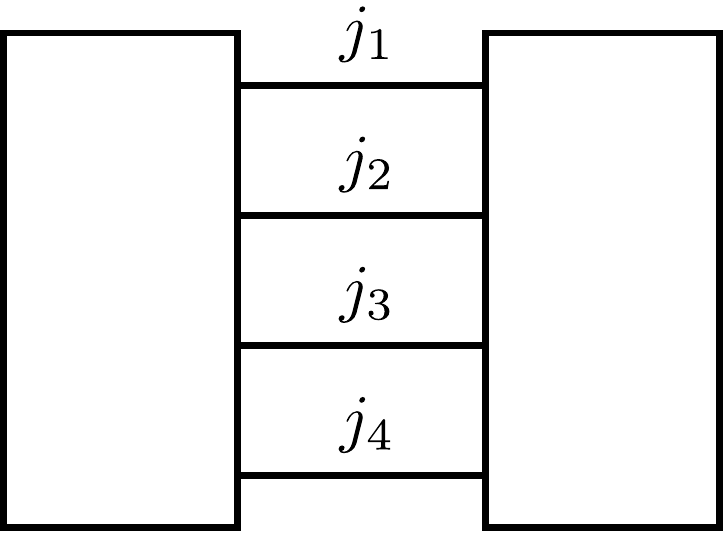}{0.6} \quad &= \quad \sum_x d_x\quad\RealSymb{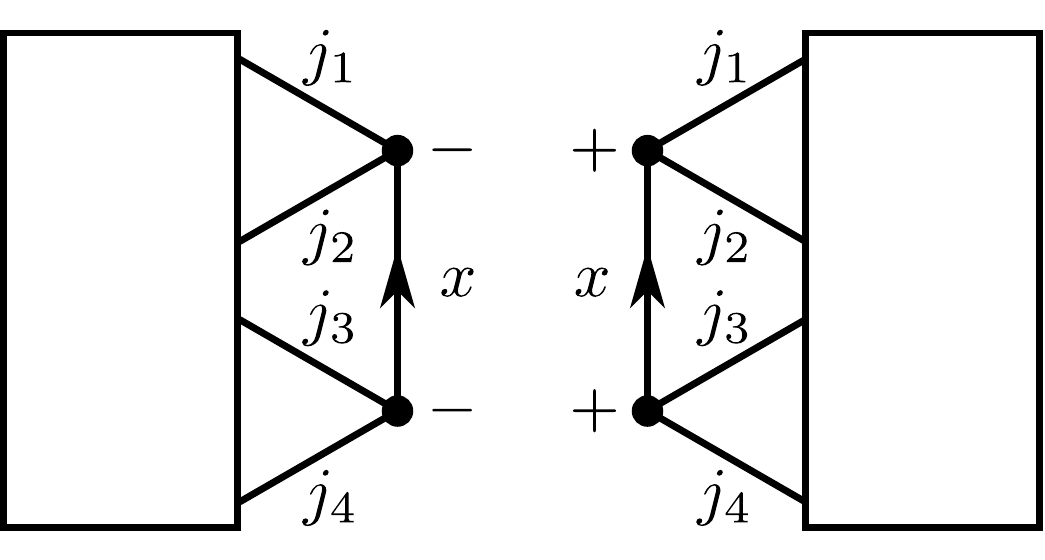}{0.6} \label{thm4'}
\end{align}
It is again straightforward to generalize \Eq{thm4'} to the case where a diagram is cut on more than four lines (though it seems that such generalizations are very rarely needed in practice).

When using the fundamental theorem in the form \eqref{thm1'}--\eqref{thm4'}, it is important to make sure that each of the two blocks actually represents a proper invariant tensor. If there is doubt as to whether this is the case, the invariance of the block can be checked by using the following criterion: A diagram representing a tensor constructed by contracting 3$j$-symbols corresponds to an invariant tensor if and only if it is possible to use \Eqs{2arrows-opp} and \eqref{3jarrows g} to bring the diagram into a form in which each internal line of the diagram carries exactly one arrow. It may sometimes be necessary to use \Eq{2arrows-opp} to introduce additional arrows into a diagram before the fundamental theorem can be used to correctly cut it into two invariant pieces.

\newpage

Let us mention at this point a useful theorem concerning diagrams which represent invariant scalar contractions of 3$j$-symbols, and hence carry no external, uncontracted lines. The theorem states that the value of such a diagram remains unchanged if one simultaneously reverses the direction of every arrow and the sign at every node in the diagram. This theorem implies, for example, that the 6$j$- and 9$j$-symbols of \Eqs{6j g} and \eqref{9j g} are equivalently represented by the diagrams
\be
\sixj{j_1}{j_2}{j_3}{k_1}{k_2}{k_3} \quad = \quad \RealSymb{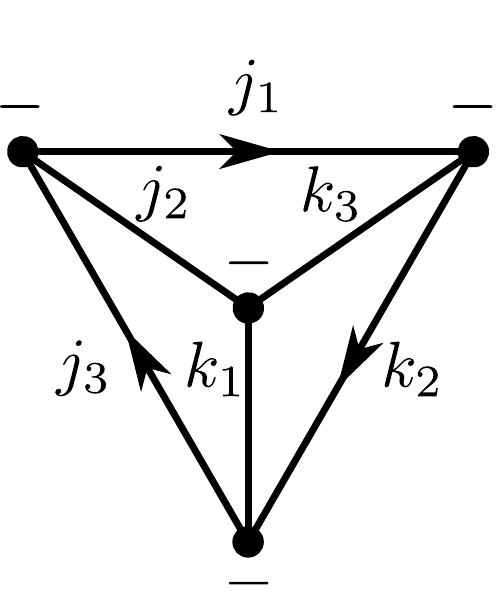}{0.6} \label{6jminus g}
\ee
and
\be
\ninej{j_1}{j_2}{j_3}{k_1}{k_2}{k_3}{l_1}{l_2}{l_3} \quad = \quad \RealSymb{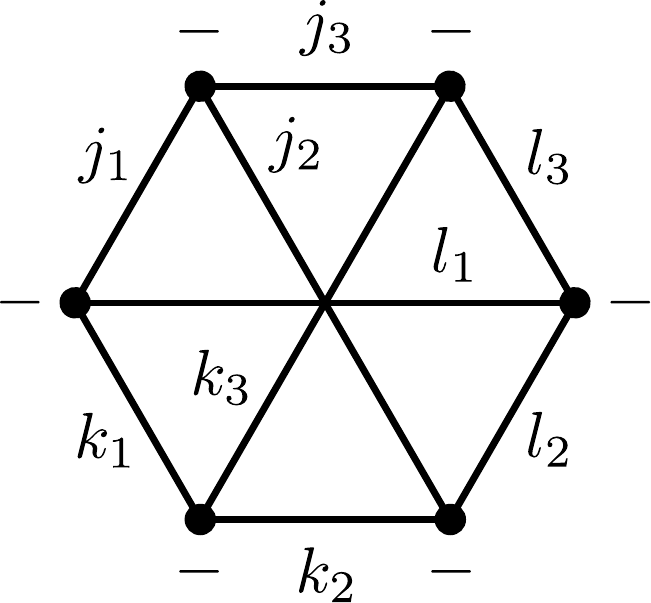}{0.6} \label{9jminus g}
\ee
To prove the theorem, suppose that we have brought a given invariant diagram into a form where each line carries precisely one arrow. Since reversing an arrow on a line carrying spin $j$ produces the factor $(-1)^{2j}$, we see that reversing all the arrows in the diagram multiplies the diagram by the factor $(-1)^{2J}$, where $J$ is the sum of all the spins appearing in the diagram. Similarly, reversing the sign at a node where spins $j_1$, $j_2$ and $j_3$ meet produces the factor $(-1)^{j_1+j_2+j_3}$, so reversing all the signs in the diagram also multiplies the diagram by $(-1)^{2J}$, because every line in the diagram is connected to two nodes. In total, the diagram is therefore multiplied by $(-1)^{4J} = +1$, since $2J$ is an integer, and so $4J$ is even.

While the principal use of \Eqs{thm1'}--\eqref{thm4'} is to break large diagrams into smaller pieces, in some cases \Eq{thm4'} can be used in the reverse direction in order to evaluate a sum over an internal spin in a graphical expression. A very simple example of using the fundamental theorem in this way is provided by a graphical proof of the orthogonality relation \eqref{6j-orth} for the 6$j$-symbol. Using the graphical representation \eqref{6j g}, we can write the sum as
\be
\sum_x d_x\sixj{j_1}{j_2}{x}{j_3}{j_4}{k}\sixj{j_1}{j_2}{x}{j_3}{j_4}{l} \;\; = \;\; \sum_x d_x\;\;\RealSymb{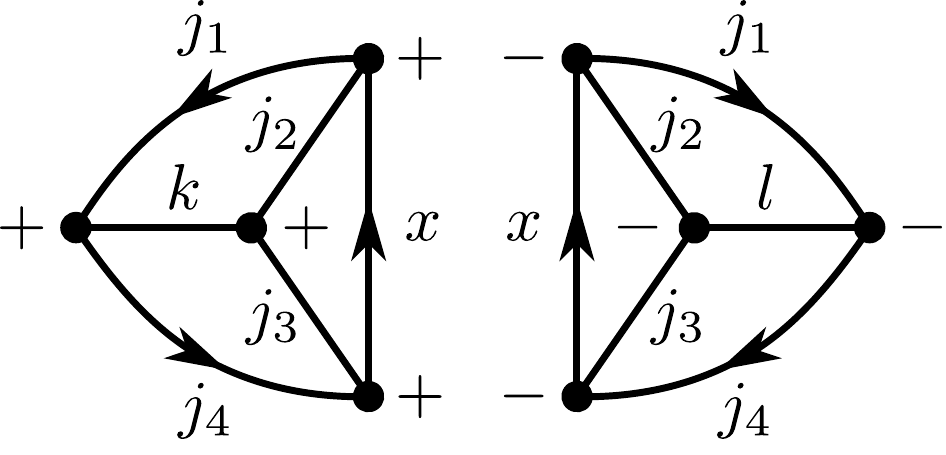}{0.6} \vspace{12pt}
\ee
\newpage
\noindent We see that the form of this expression matches the right-hand side of \Eq{thm4'}, the two blocks with four lines being
\be
\RealSymb{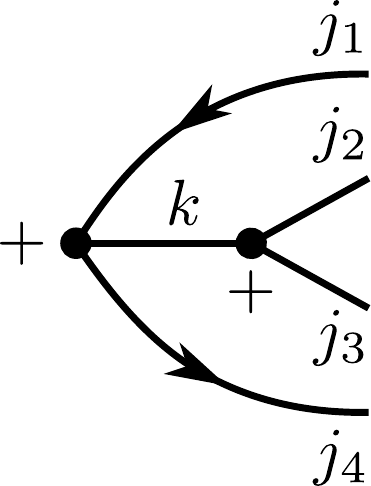}{0.6} \qquad\qquad \text{and} \qquad\qquad \RealSymb{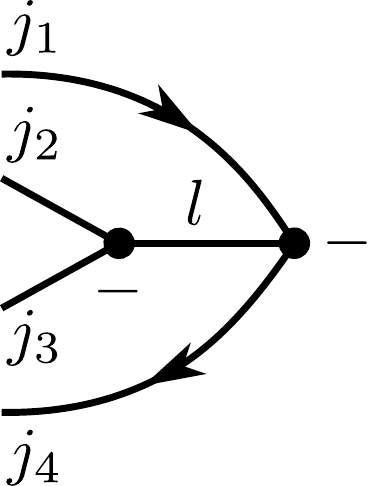}{0.6}
\ee
Therefore \Eq{thm4'} gives
\be
\sum_x d_x\sixj{j_1}{j_2}{x}{j_3}{j_4}{k}\sixj{j_1}{j_2}{x}{j_3}{j_4}{l} \quad = \quad \RealSymb{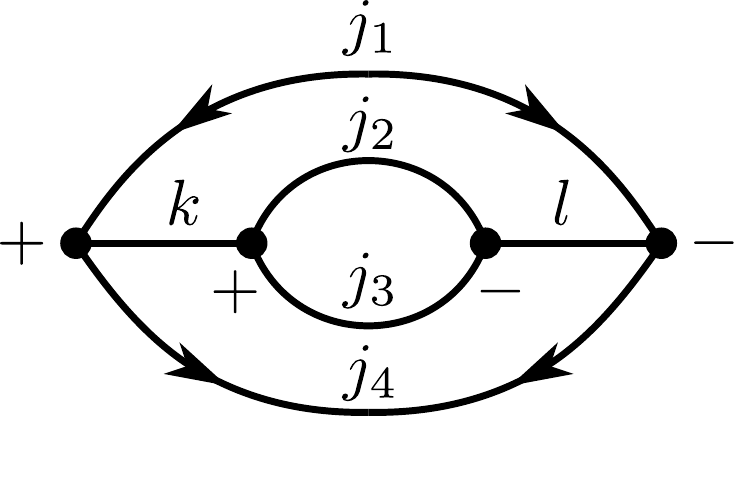}{0.6} \quad = \; \frac{1}{d_k}\delta_{kl},
\ee
where \Eqs{2arrows-opp}, \eqref{3j-orth-mm g} and \eqref{3j-theta g} were used to evaluate the resulting diagram.

Let us conclude this chapter by considering a more involved example of using the graphical machinery introduced in this chapter in a calculation involving intertwiners. (Many further examples are given in the exercises, and in the following chapter, in which we illustrate the applications of the graphical formalism to calculations in loop quantum gravity.) The problem consists of expressing the six-valent intertwiner
\vspace{10pt}
\be\label{iota612 g}
\RealSymb{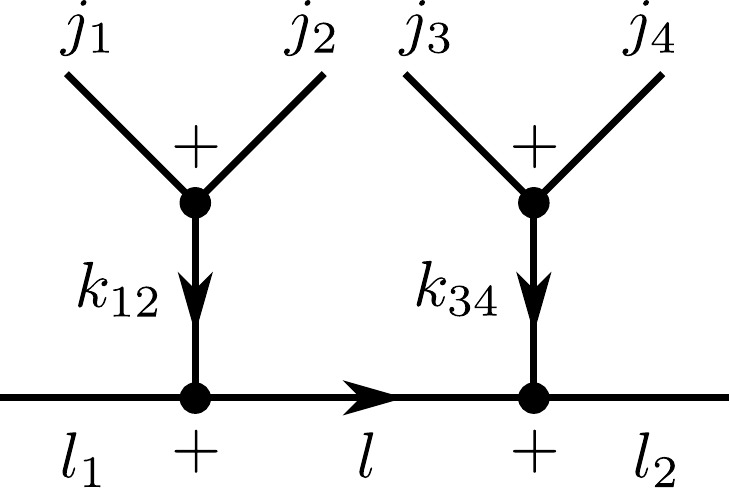}{0.6} \vspace{10pt}
\ee
in the basis formed by the intertwiners
\vspace{10pt}
\be\label{iota613 g}
\RealSymb{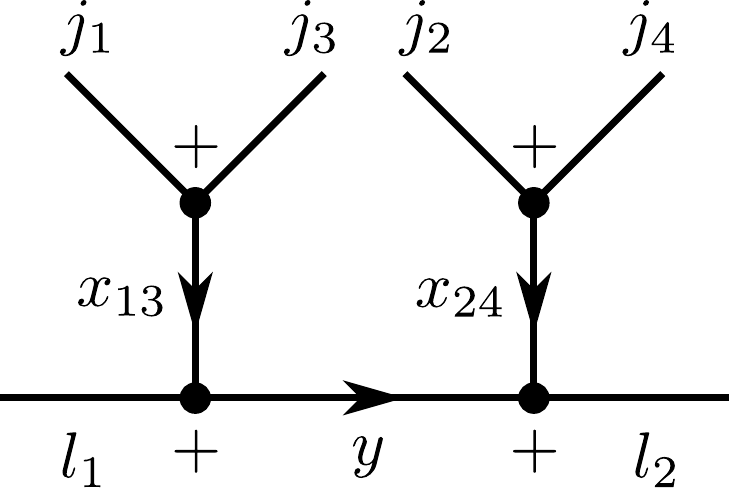}{0.6} \vspace{10pt}
\ee
\newpage According to the generalization of \Eq{thm4} to six-valent invariant tensors, the intertwiner \eqref{iota612 g} can be expressed in terms of the intertwiners \eqref{iota613 g} as
\begin{align}
&\RealSymb{figA-iota612.pdf}{0.6} \notag \\[-12pt]
&\qquad\qquad = \quad \sum_{x_{13}x_{24}y} d_{x_{13}}d_{x_{24}}d_y\,K(x_{13},x_{24},y|k_{12},k_{34},l)\;\RealSymb{figA-iota613.pdf}{0.6}\label{iota612-start}
\end{align}
where the coefficient $K(x_{13},x_{24},y|k_{12},k_{34},l)$ is given by the contraction of the intertwiners \eqref{iota612 g} and \eqref{iota613 g}: \vspace{-12pt}
\be\label{iota6-K}
K(x_{13},x_{24},y|k_{12},k_{34},l) \quad = \quad \RealSymb{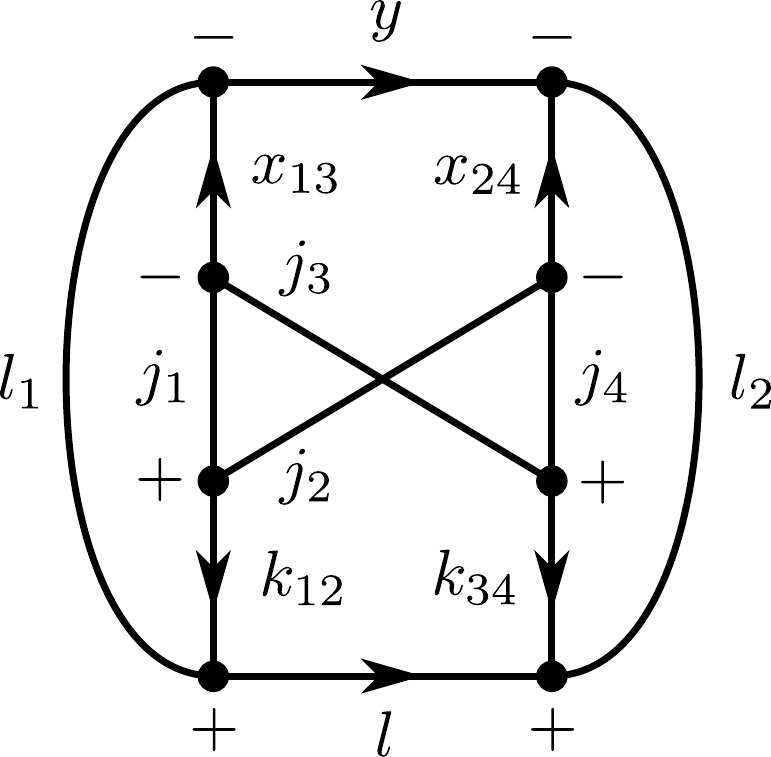}{0.6}
\ee
This diagram can be recognized as the so-called 12$j$-symbol of the first kind (see \eg \cite{Varshalovich} or \cite{YLV}), but we can use the fundamental theorem to break it down to the more familiar 6$j$- and 9$j$-symbols. The diagram clearly cannot be separated into two non-trivial pieces by cutting only two or three lines, so we will use \Eq{thm4'} to cut out the part consisting of the four nodes in the middle of the diagram, as indicated by the red dashed line in \Eq{iota6-cut}. For convenience, we make use of \Eqs{invarrow} and \eqref{3jarrows g} to reverse the direction of the arrows and interchange one pair of signs in \Eq{thm4'}, leading to
\be\label{iota6-cut}
\RealSymb{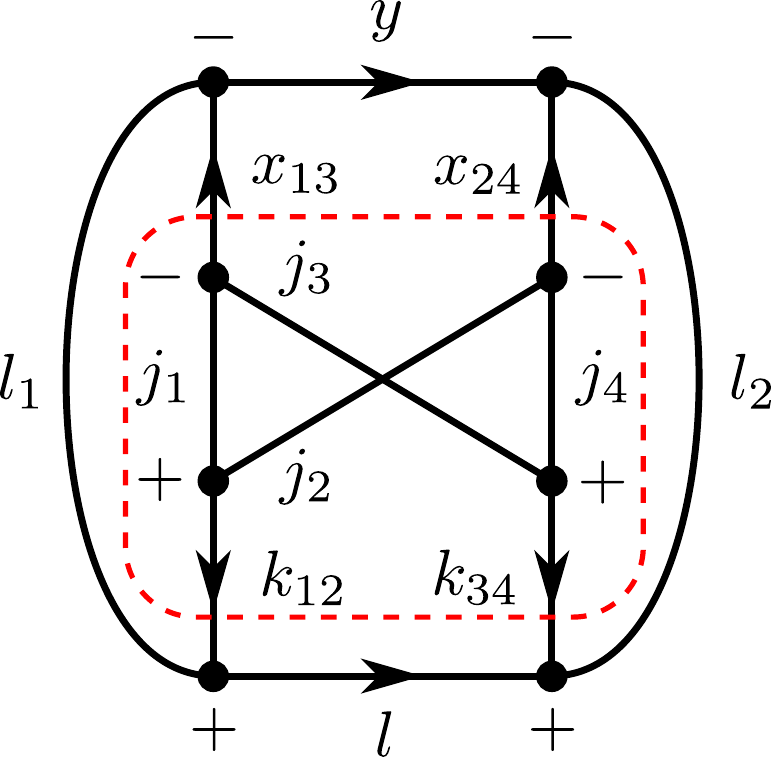}{0.6} \quad = \quad \sum_s d_s\quad\RealSymb{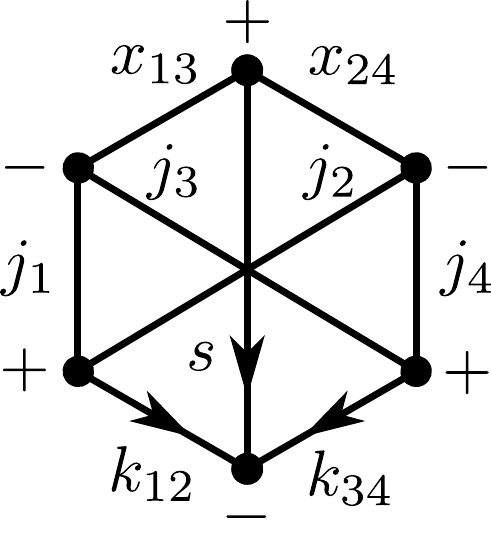}{0.6}\quad\RealSymb{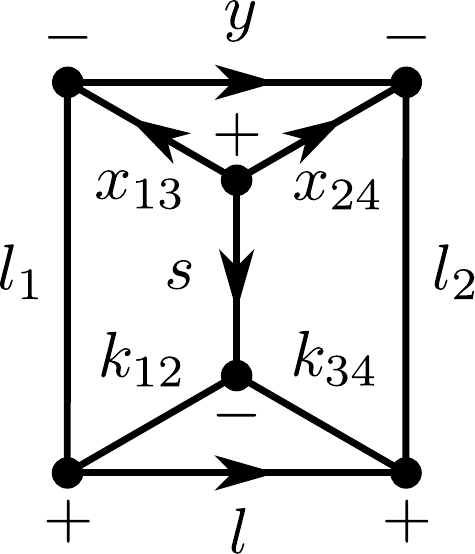}{0.6}
\ee
The first factor in the sum is simply the 9$j$-symbol of \Eq{9j g},
\be\label{iota6-cut1}
\RealSymb{figA-iota6-cut-part1.pdf}{0.6} \quad = \quad \RealSymb{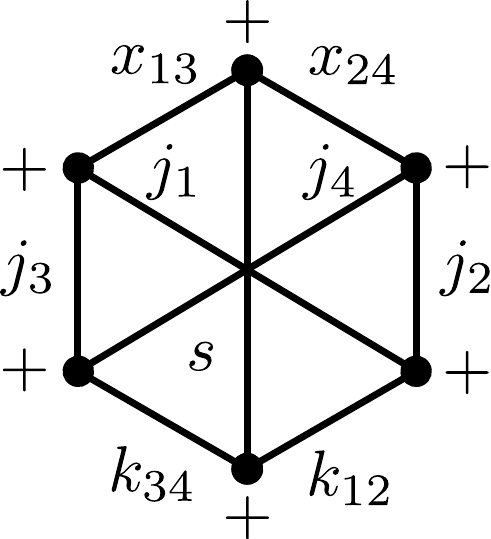}{0.6} \quad = \quad \ninej{j_1}{j_2}{k_{12}}{j_3}{j_4}{k_{34}}{x_{13}}{x_{24}}{s}.
\ee
Here we have brought the diagram into the form \eqref{9j g} by interchanging the two nodes carrying spins $j_1,j_2,k_{12}$ and $j_3,j_4,k_{34}$ -- noting that this changes the three minus signs into plus signs -- and using \Eq{3jarrows g} to remove the three arrows. 

In the second factor in \Eq{iota6-cut}, we cancel the three arrows in the middle, and use \Eq{2arrows-opp} to introduce oppositely directed arrows on the lines carrying spins $l_1$ and $l_2$. After this we cut the diagram into two pieces along the three vertical lines according to \Eq{thm3'}. The piece resulting from the upper part of the diagram is the 6$j$-symbol in the form \eqref{6jminus g}, up to the factor $(-1)^{x_{13}+x_{24}+s}$, which arises when the sign of the node in the middle is changed from $+$ to $-$. Similarly, the lower piece is the 6$j$-symbol in the form \eqref{6j g}, multiplied by the factor $(-1)^{k_{12}+k_{34}+s}$. We therefore find
\be\label{iota6-cut2}
\RealSymb{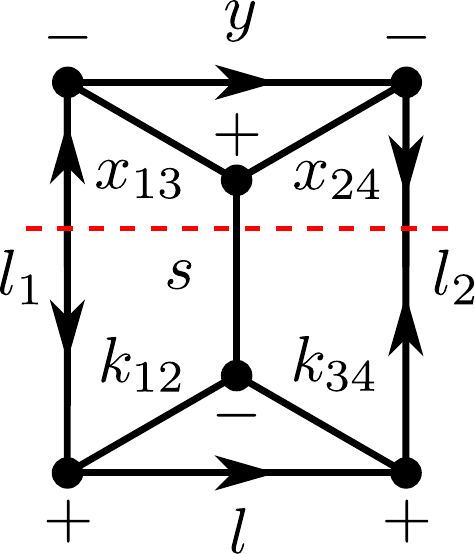}{0.6} \quad = \;\; (-1)^{k_{12}+k_{34}+s}\sixj{k_{12}}{k_{34}}{s}{l_2}{l_1}{l} (-1)^{x_{13}+x_{24}+s}\sixj{x_{13}}{x_{24}}{s}{l_2}{l_1}{y},
\ee
which completes the calculation. Inserting \Eqs{iota6-K}--\eqref{iota6-cut2} back into \Eq{iota612-start}, we conclude that the expansion of the intertwiner \eqref{iota612 g} in the basis of the intertwiners \eqref{iota613 g} is given by
\vspace{10pt}
\begin{align}
&\RealSymb{figA-iota612.pdf}{0.6} \quad = \quad \sum_{x_{13}x_{24}y} d_{x_{13}}d_{x_{24}}d_y(-1)^{k_{12}+k_{34}-x_{13}-x_{24}} \notag \\
&\times\sum_s d_s\sixj{k_{12}}{k_{34}}{s}{l_2}{l_1}{l}\sixj{x_{13}}{x_{24}}{s}{l_2}{l_1}{y}\ninej{j_1}{j_2}{k_{12}}{j_3}{j_4}{k_{34}}{x_{13}}{x_{24}}{s}\quad\RealSymb{figA-iota613.pdf}{0.6}\label{iota6change}
\end{align}

\newpage

\subsection*{Exercises}
\addcontentsline{toc}{subsection}{{\hspace{21.4pt} Exercises}}

\begin{enumerate}[leftmargin=*]

\item 
\begin{itemize}
\item[(a)] Draw the diagram representing the $N$-valent intertwiner \eqref{iotaN}.
\item[(b)] Write the intertwiner \eqref{iotaN g} in non-graphical form.
\end{itemize}

\item Calculate the scalar product between two intertwiners of the form \eqref{iotaN}, and between two intertwiners of the form \eqref{iotaN g}.

\item Show that the components of the intertwiner defined by \Eq{iotaN} satisfy $\iota^{m_1\cdots m_N} = \iota_{m_1\cdots m_N}$ (as a numerical, rather than tensorial, equation).

\item Prove the identity \eqref{int D...D} by repeatedly applying \Eq{DD g}, and using \Eq{int DD g} in the end.

\item Starting from \Eqs{6j-def g} and \eqref{9j-def g}, verify that the 6$j$- and 9$j$-symbols are indeed represented by the graphical expressions \eqref{6j g} and \eqref{9j g}.

\item Derive the symmetry properties of the 6$j$- and 9$j$-symbols from the graphical representation of the symbols. 

\item Show that
\[
\displaystyle \sixj{j_1}{j_2}{j_3}{k_1}{k_2}{0} = \delta_{j_1k_2}\delta_{j_2k_1}\frac{(-1)^{j_1+j_2+j_3}}{\sqrt{d_{j_1}d_{j_2}}}
\]
and
\[
\displaystyle \ninej{j_1}{j_2}{k_{12}}{j_3}{j_4}{k_{34}}{l_{13}}{l_{24}}{0} = \delta_{k_{12}k_{34}}\delta_{l_{13}l_{24}}\frac{(-1)^{j_2+j_3+k_{12}+l_{13}}}{\sqrt{d_{k_{12}}d_{l_{13}}}}\sixj{j_1}{j_2}{k_{12}}{j_4}{j_3}{l_{13}}.
\]

\item Show that
\begin{align*}
\sum_{n_1n_2n_3} (-1)^{k_1+k_2+k_3-n_1-n_2-n_3}\threej{j_1}{k_2}{k_3}{m_1}{-n_2}{n_3}&\threej{k_1}{j_2}{k_3}{n_1}{m_2}{-n_3}\threej{k_1}{k_2}{j_3}{-n_1}{n_2}{m_3} \\
&= \sixj{j_1}{j_2}{j_3}{k_1}{k_2}{k_3}\threej{j_1}{j_2}{j_3}{m_1}{m_2}{m_3}.
\end{align*}

\item Show that
\[
\ninej{j_1}{j_2}{j_3}{k_1}{k_2}{k_3}{l_1}{l_2}{l_3} = \sum_x (-1)^{2x}\sixj{j_1}{j_2}{j_3}{k_3}{l_3}{x}\sixj{k_1}{k_2}{k_3}{j_2}{x}{l_2}\sixj{l_1}{l_2}{l_3}{x}{j_1}{k_1}.
\]

\item Show that
\[
\sum_x d_x(-1)^{k+l+x}\sixj{j_1}{j_2}{x}{j_3}{j_4}{k}\sixj{j_1}{j_2}{x}{j_4}{j_3}{l} = \sixj{j_1}{j_4}{k}{j_2}{j_3}{l}.
\]

\item Express the $N$-valent intertwiner \eqref{iotaN g} in the basis given by the intertwiners \eqref{iotaN}.

\item Prove the graphical identity \cite{BrinkSatchler}
\begin{align*}
&\RealSymb{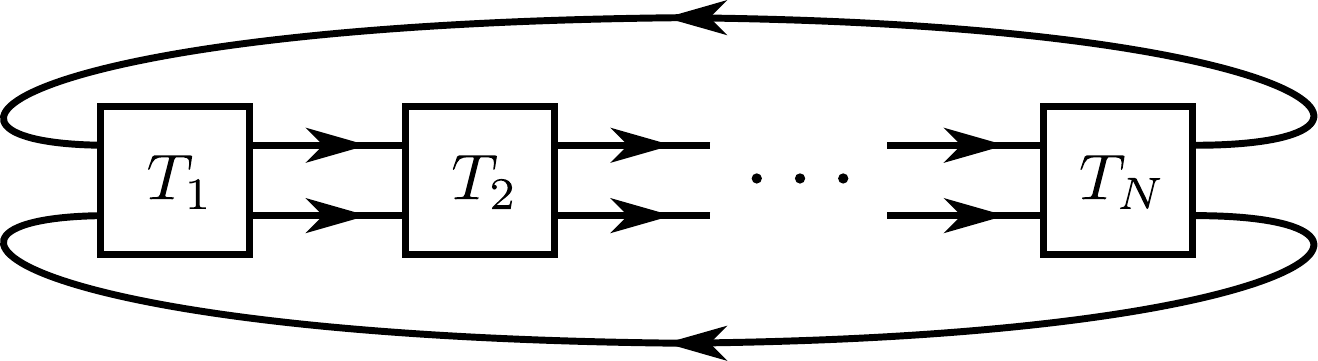}{0.6} \\
&\vspace{3cm} \quad = \quad \sum_x d_x \; \RealSymb{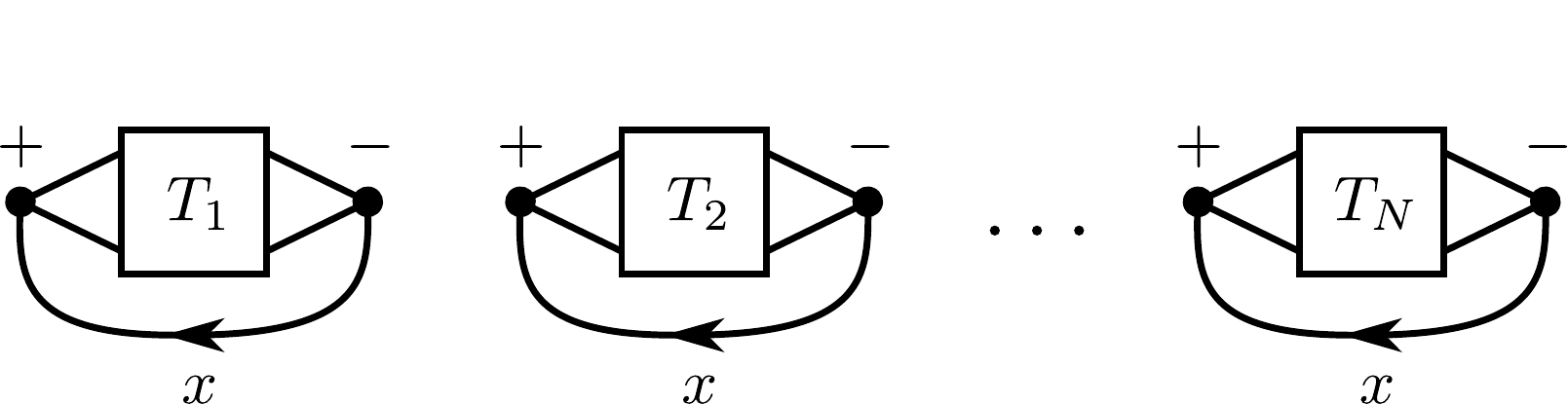}{0.6} 
\end{align*}
where each block represents a four-valent invariant tensor.

\item Derive the Biedenharn--Elliot identity
\begin{align*}
&\quad\sixj{j_1}{j_2}{j_3}{k_1}{k_2}{k_3}\sixj{j_1}{j_2}{j_3}{l_1}{l_2}{l_3} \\
&=\sum_x d_x(-1)^{j_1+j_2+j_3+k_1+k_2+k_3+l_1+l_2+l_3+x}\sixj{j_1}{l_2}{l_3}{x}{k_3}{k_2}\sixj{l_1}{j_2}{l_3}{k_3}{x}{k_1}\sixj{l_1}{l_2}{j_3}{k_2}{k_1}{x}
\end{align*}
\item[(a)] By using \Eq{6j-def g} to carry out the change of basis
\[
\RealSymb{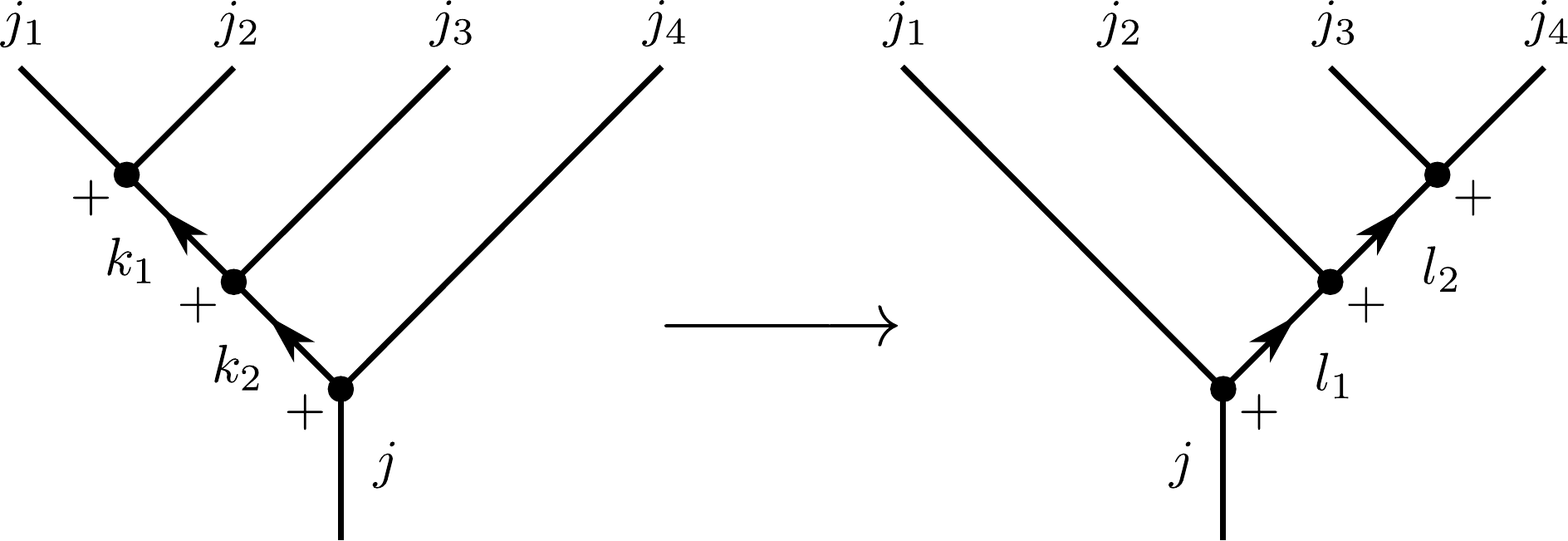}{0.6}
\]
in two different ways;
\item[(b)] By using the fundamental theorem to decompose the diagram
\[
\RealSymb{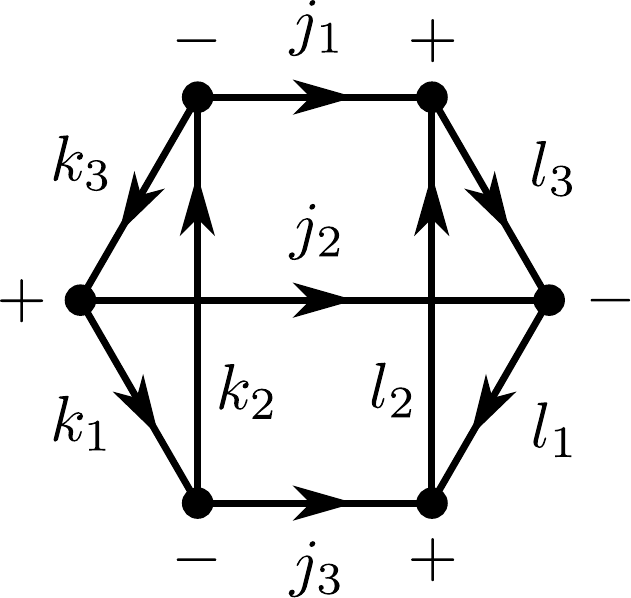}{0.6}
\]
in two different ways \cite{BrinkSatchler}.

\end{enumerate}

\newpage

\section{Examples from loop quantum gravity}

The graphical formalism presented in the previous chapter is an extremely effective tool for practical calculations in loop quantum gravity. In this chapter we illustrate the use of the graphical machinery in loop quantum gravity by going through several detailed example calculations. The chapter begins with a brief outline of the kinematical framework of loop quantum gravity, whose primary purpose is to set down our notation and conventions for the basic kinematical states of the theory and the elementary operators acting on them. We then present three examples where graphical methods are used to compute the action of a given loop quantum gravity operator in the basis of spin network states. We will consider the volume operator \cite{ALvolume, RSvolume}, the Euclidean part of the so-called Warsaw Hamiltonian \cite{paper1, paper2}, and the curvature operator \cite{curvature}. The reader who wishes to see more examples of graphical calculations in loop quantum gravity is encouraged to take a look at the calculations displayed in \cite{thesis}.

\subsection{The kinematical structure of loop quantum gravity}

\subsubsection*{Spin network states}

The basic kinematical states of loop quantum gravity are the so-called spin network states. A spin network state is characterized by the following quantum numbers:
\begin{itemize}
\item A graph $\Gamma$.
\item A spin $j_e$ associated to each edge of the graph.
\item An intertwiner $\iota_v$ associated to each node of the graph. The intertwiner carries a lower index for each edge coming in to the node, and an upper index for each edge going out of the node. If the edges coming in to the node are labeled by the spins $j_1,\dots,j_I$, while the edges going out of the node are labeled by the spins $j_1',\dots,j_O'$, the intertwiner $\iota_v$ is an element of the space
\[
{\rm Inv}\,\bigl({\cal H}_{j_1}\otimes\cdots\otimes{\cal H}_{j_I}\otimes{\cal H}_{j_1'}^*\otimes\cdots\otimes{\cal H}_{j_O'}^*\bigr).
\]
\end{itemize}
The ''wave function'' of a spin network state based on a graph having $N$ edges is a function of $N$ group elements of $SU(2)$, one for each edge of the graph. The wave function has the form
\be\label{spinnetwork}
\Psi_{\Gamma,\{j_e\},\{\iota_v\}}(h_{e_1},\dots,h_{e_N}) = \biggl(\prod_{v\in\Gamma}\iota_v\biggr)^{n_1\cdots n_N}_{m_1\cdots m_N}\biggl(\prod_{e\in\Gamma}\D{j_e}{m_e}{n_e}{h_e}\biggr),
\ee
where the contraction of indices is carried out according to the pattern of the graph. That is, for the upper index of each representation matrix $\D{j_e}{m_e}{n_e}{h_e}$ there is a corresponding lower index in the intertwiner at the endpoint of the edge $e$, and these indices are contracted against each other in \Eq{spinnetwork}. Similarly, the lower index of the representation matrix is contracted against an upper index of the intertwiner at the beginning point of $e$.

The group elements $h_e$ originate from holonomies of the Ashtekar connection in the classical theory, and for this reason they are referred to as holonomies also in the quantum theory. The holonomy satisfies certain algebraic properties, reflecting its classical interpretation as a parallel propagator. Letting $e^{-1}$ denote the edge $e$ taken with the opposite orientation, we have
\be
h_{e^{-1}} = h_e^{-1}.
\ee
Furthermore, if $e_1$ and $e_2$ are two edges such that the endpoint of $e_1$ coincides with the beginning point of $e_2$, we have
\be\label{h2h1}
h_{e_2}h_{e_1} = h_{e_2\circ e_1},
\ee
where $e_2\circ e_1$ stands for the edge composed of $e_1$ followed by $e_2$. 

To give a concrete example of a spin network state, let us consider a graph consisting of two nodes connected to each other by four edges (all of which assumed to have the same orientation). An example of a spin network state defined on this graph is given by
\be\label{spinnetwork-ex}
\bigl(\iota_{12}^{(k)}\bigr)_{m_1m_2m_3m_4} \D{j_1}{m_1}{n_1}{h_{e_1}}\D{j_2}{m_2}{n_2}{h_{e_3}}\D{j_3}{m_3}{n_3}{h_{e_3}}\D{j_4}{m_4}{n_4}{h_{e_4}} \bigl(\iota_{13}^{(l)}\bigr)^{n_1n_2n_3n_4}.
\ee
Recalling the graphical representation of the four-valent intertwiner and the $SU(2)$ representation matrix\footnote{Under a local $SU(2)$ gauge transformation described by the gauge function $g(x)\in SU(2)$, the holonomy transforms as $h_e \to g_t h_e g_s^{-1}$, or
\[
\D{j}{m}{n}{h_e} \to \D{j}{m}{\mu}{g_t}\D{j}{\mu}{\nu}{h_e}\D{j}{\nu}{n}{g_s^{-1}},
\]
where $g_s$ and $g_t$ denote the values of $g(x)$ at the beginning point and endpoint (''source'' and ''target'') of the edge $e$. This suggests that the upper index of $\D{j}{m}{n}{h_e}$ \vadjust{\vskip\maxdimen} is associated with the endpoint of the edge, while the lower index is associated with its beginning point. In this sense, the orientation of the triangle in the graphical representation of the holonomy
\[
\D{j}{m}{n}{h_e} \; = \; \RealSymb{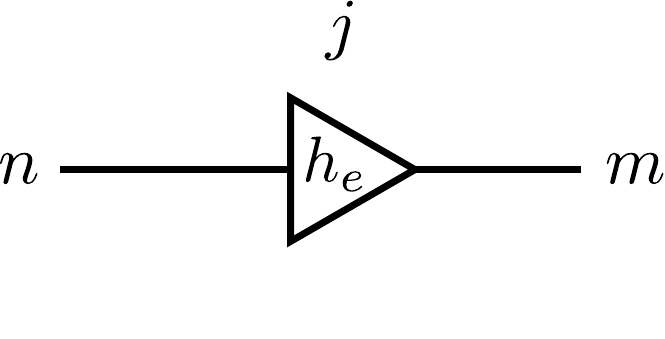}{0.5}
\]
is consistent with the orientation of the edge $e$.}
from the previous chapter, we may write the spin network \eqref{spinnetwork-ex} in graphical form as
\vspace{12pt}
\be\label{spinnetwork-gr}
\RealSymb{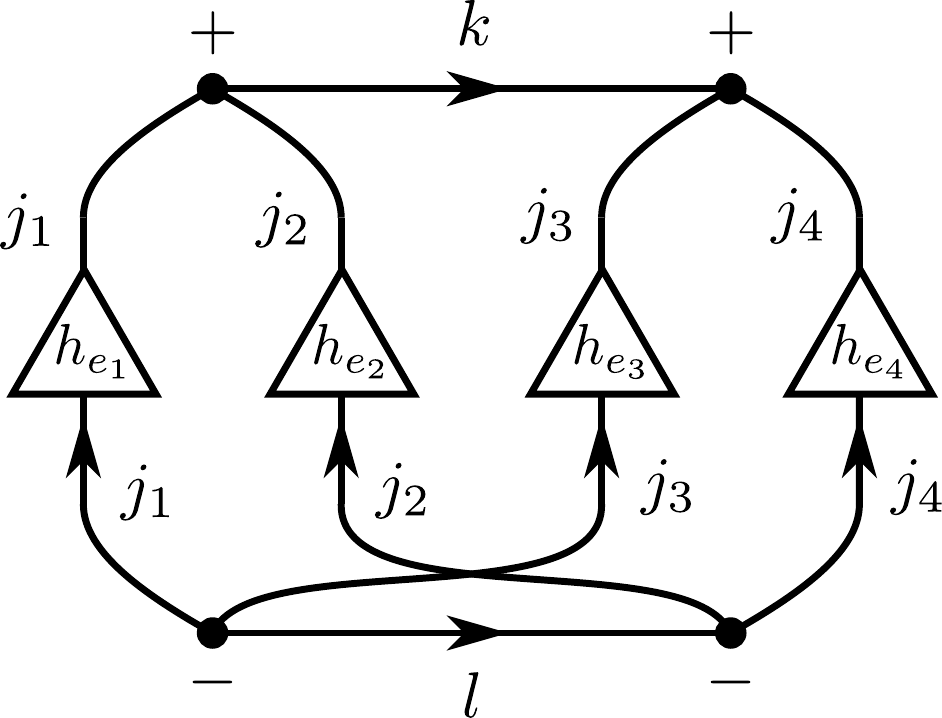}{0.6}\vspace{12pt}
\ee
However, this kind of detailed diagrams of spin networks are rarely invoked in the loop quantum gravity literature. Instead, the usual way to represent a spin network state pictorially is just to draw the graph of the state and label its edges and nodes with their corresponding quantum numbers. For instance, the spin network \eqref{spinnetwork} would be described by the drawing
\vspace{12pt}
\be\label{spinnetwork-drawing}
\RealSymb{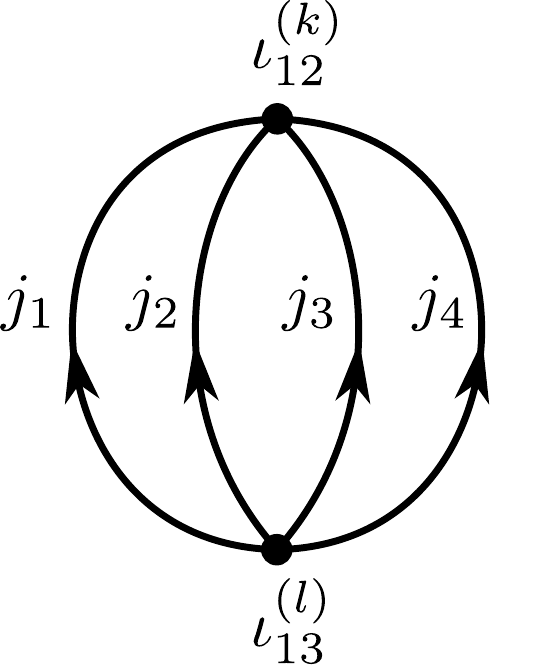}{0.6}
\ee
\newpage
It is important not to confuse such generic pictures of spin network states with expressions written down using the graphical notation introduced in the previous chapter, even though the two types of diagrams may often look quite similar. In particular, the arrows in the drawing \eqref{spinnetwork-drawing} indicate the orientation of the edges of the graph, and have no relation at all to the identical-looking arrows which are seen for example in the diagram \eqref{spinnetwork-gr}, and which represent the epsilon tensor in the graphical calculus of $SU(2)$.

\subsubsection*{Elementary operators}

The holonomy operator $\D{j}{m}{n}{h_e}$ acts on the state $\Psi_\Gamma(h_{e_1},\dots,h_{e_N})$ by multiplication:
\be\label{Dh*Psi}
\D{j}{m}{n}{h_e}\Psi_\Gamma(h_{e_1},\dots,h_{e_N}).
\ee
The character of the result depends on whether the edge $e$ is contained among the edges of the graph $\Gamma$. If $e$ is not an edge of $\Gamma$, the function \eqref{Dh*Psi} defines a state based on the graph $\Gamma\cup e$; in effect, the action of the holonomy operator has added a new edge to the graph of the state on which it acted. On the other hand, if $e$ coincides with one of the edges of $\Gamma$, the state \eqref{Dh*Psi} is still based on the graph $\Gamma$. In this case, the basic tool for computing the action of the holonomy operator is the Clebsch--Gordan series
\be
\RealSymb{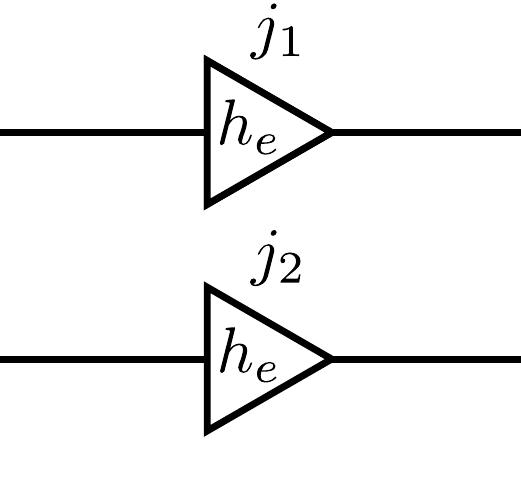}{0.6} \quad = \quad \sum_j d_j\;\RealSymb{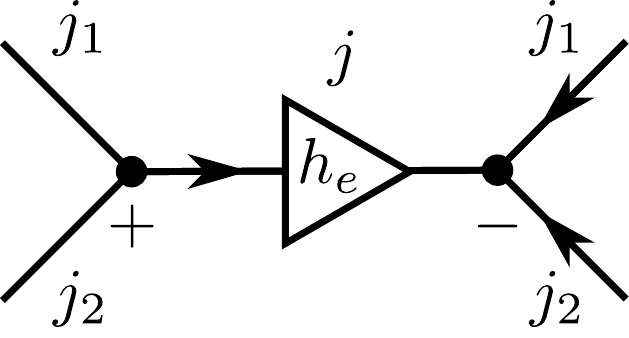}{0.6}
\ee
which indicates how the holonomy $\D{j}{m}{n}{h_e}$ is coupled with the group element $h_e$ appearing in the function $\Psi_\Gamma(h_{e_1},\dots,h_{e_N})$ in \eqref{Dh*Psi}. If the orientation of the holonomy operator is opposite to the orientation of the edge $e$, the action of the operator is computed using the Clebsch--Gordan series for the coupling of a group element with its inverse:
\be
\RealSymb{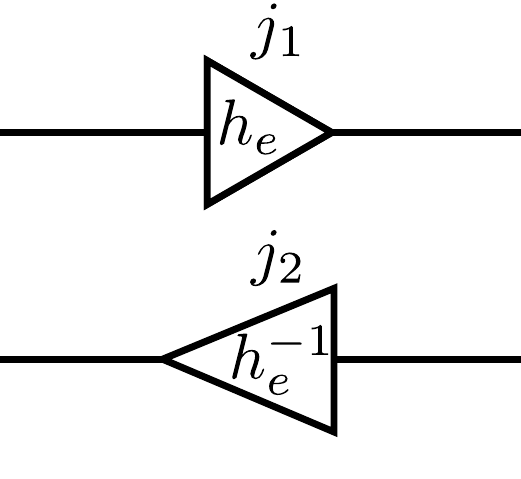}{0.6} \quad = \quad \sum_j d_j\;\RealSymb{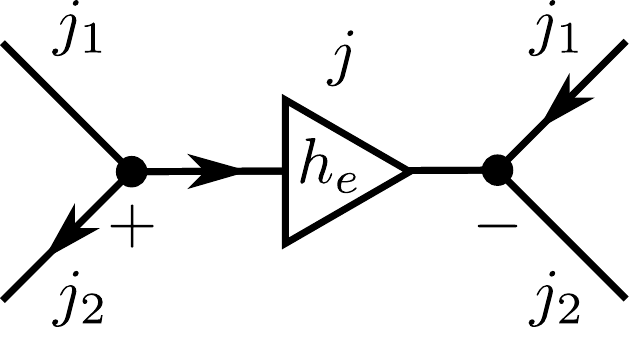}{0.6}
\ee
The case where the edge $e$ partially overlaps with an edge (or multiple edges) of the graph $\Gamma$ can be reduced to the two cases discussed above by using the multiplicative property \eqref{h2h1} of the holonomy.

The conjugate operator of the holonomy is (somewhat loosely speaking) the derivative operator $J_i^{(v,e)}$, which carries an $SU(2)$ vector index $i$, and is labeled by a node $v$ and an edge $e$ such that $v$ is either the beginning point or endpoint of $e$. The action of $J_i^{(v,e)}$ on the state $\Psi_\Gamma(h_{e_1},\dots,h_{e_N})$ is defined to be
\begin{align}
J_i^{(v,e)}&\Psi_\Gamma(h_{e_1},\dots,h_{e_N}) \notag \\
&= \begin{cases} i\dfrac{d}{d\epsilon}\bigg|_{\epsilon=0}\Psi_\Gamma(h_{e_1},\dots,h_{e_k}e^{\epsilon\tau_i},\dots,h_{e_N}) & \text{if $e=e_k$ and $e$ begins at $v$} \\ \vspace{-8pt}  \\
-i\dfrac{d}{d\epsilon}\bigg|_{\epsilon=0}\Psi_\Gamma(h_{e_1},\dots,e^{\epsilon\tau_i}h_{e_k},\dots,h_{e_N}) & \text{if $e=e_k$ and $e$ ends at $v$} \end{cases}
\end{align}
In the case that $v$ is not a node of $\Gamma$, or $e$ is not an edge of $\Gamma$, we set
\be
J_i^{(v,e)}\Psi_\Gamma(h_{e_1},\dots,h_{e_N}) = 0.
\ee
It is immediate to see that the action of $J_i^{(v,e)}$ on the holonomy $\D{j}{m}{n}{h_e}$ is given by
\be\label{J*h_e}
J_i^{(v,e)}\D{j}{m}{n}{h_e} = \begin{cases} i\D{j}{m}{\mu}{h_e}\Tau{j}{i}{\mu}{n} & \text{($e$ begins at $v$)} \\
-i\Tau{j}{i}{m}{\mu}\D{j}{\mu}{n}{h_e} & \text{($e$ ends at $v$)} \end{cases}
\ee
It also follows from the definition that the operator $J_i^{(v,e)}$ satisfies the $SU(2)$ algebra
\be
\bigl[J_i^{(v,e)},J_j^{(v',e')}\bigr] = \delta_{vv'}\delta_{ee'}i\epsilon\downup{ij}{k}J_k^{(v,e)}.
\ee
Making use of the graphical representation of the $SU(2)$ generators from \Eq{tau=C g}, we may cast the action of $J_i^{(v,e)}$ in graphical form as follows: \vspace{-24pt}
\begin{align}
J_i^{(v,e)}\,\RealSymb{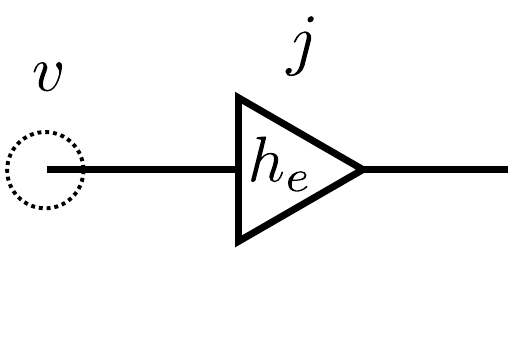}{0.6} \quad &= \quad -W_j\;\RealSymb{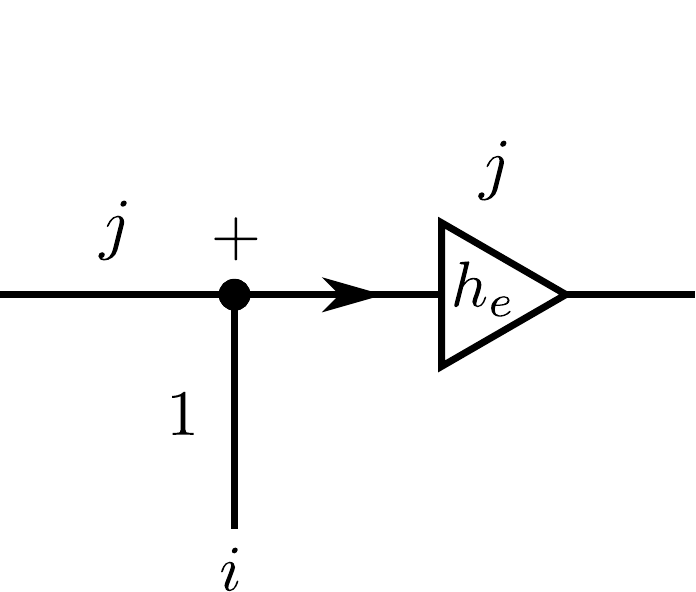}{0.6} \label{J*h source}\\
\intertext{when the operator acts at the beginning point of the edge $e$, and \vspace{-24pt}}
J_i^{(v,e)}\;\RealSymb{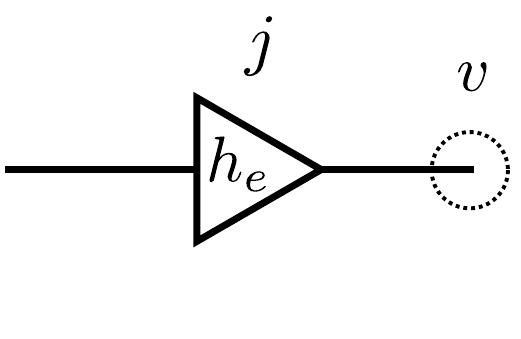}{0.6} \quad &= \quad W_j\;\RealSymb{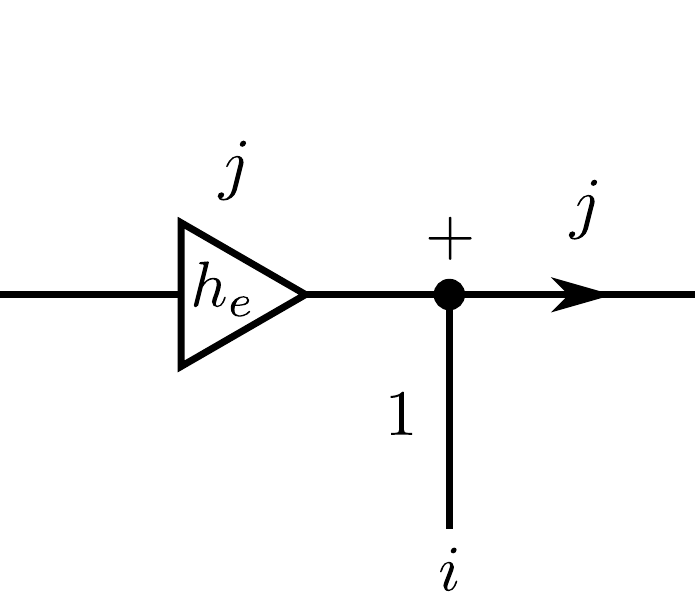}{0.6} \label{J*h target}
\end{align}
when the operator acts at the endpoint of $e$.

\newpage

\subsection{Volume operator}

The volume operator in loop quantum gravity \cite{ALvolume, RSvolume} is constructed out of operators of the form
\be
q_v^{(e_1,e_2,e_3)} = \epsilon^{ijk}J_i^{(v,e_1)}J_j^{(v,e_2)}J_k^{(v,e_3)}.
\ee
The operator $q_v^{(e_1,e_2,e_3)}$ acts on a triple of edges $(e_1,e_2,e_3)$ all belonging to the node $v$. We will also refer to the operator $q_v^{(e_1,e_2,e_3)}$ itself as the ''volume operator'', even though this is strictly speaking not completely correct. The same remark also applies to most of the other operators discussed in this chapter, such as the ''curvature operator'' $R_v^{(e_1,e_2)}$ considered in section \ref{sec:curvature}, and the ''angle operator'' $\Delta_v^{(e_1,e_2)}$ encountered in the exercises. 

As the simplest possible example of computing the action of the volume operator, let us consider the three-valent spin network node 
\vspace{12pt}
\be\label{vol-node}
\RealSymb{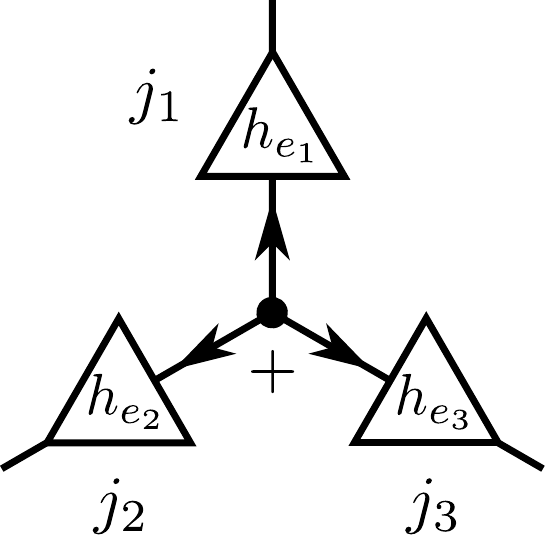}{0.6} \quad = \quad \iota^{n_1n_2n_3}\D{j_1}{m_1}{n_1}{h_{e_1}}\D{j_2}{m_2}{n_2}{h_{e_2}}\D{j_3}{m_3}{n_3}{h_{e_3}}.\vspace{12pt}
\ee
Since some orientation of the edges must be chosen so that the calculation can be conveniently carried out, we have assumed that all the edges are oriented outwards from the node. At this point it is not obvious whether the result of the calculation will be affected by this choice. However, at the end of the calculation we will argue that the result is in fact independent of the orientation of the edges, not only in the case of the volume operator, but for any operator constructed out of the angular momentum operators $J_i^{(v,e)}$.

When the operator $q_v^{(e_1,e_2,e_3)}$ acts on the state \eqref{vol-node}, the action of each angular momentum operator is given by \Eq{J*h source} as \vspace{-24pt}
\be
J_i^{(v,e)}\,\RealSymb{figA-h_e-source.pdf}{0.6} \quad = \quad -W_j\;\RealSymb{figA-tauh_e.pdf}{0.6}
\ee
Recalling also \Eq{epsuvw g}, which instructs us to translate the epsilon symbol into graphical notation as
\vspace{12pt}
\be
\epsilon^{ijk} \quad \longrightarrow \quad i\sqrt 6\RealSymb{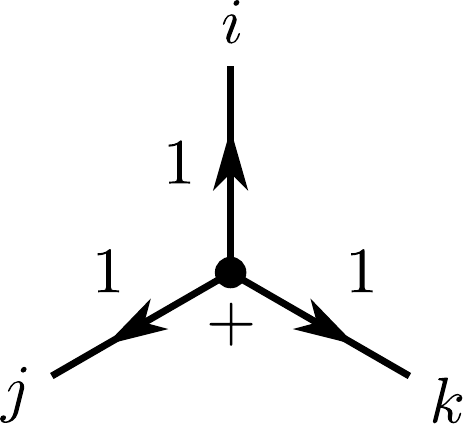}{0.6}
\ee
\newpage 
\noindent we find that the action of the volume operator on the three-valent node gives
\vspace{12pt}
\begin{align}
q_v^{(e_1,e_2,e_3)}&\;\RealSymb{state123_vol.pdf}{0.6} \notag \\
&=-i\sqrt 6W_{j_1}W_{j_2}W_{j_3}\;\RealSymb{eps_ijk.pdf}{0.6}\;\RealSymb{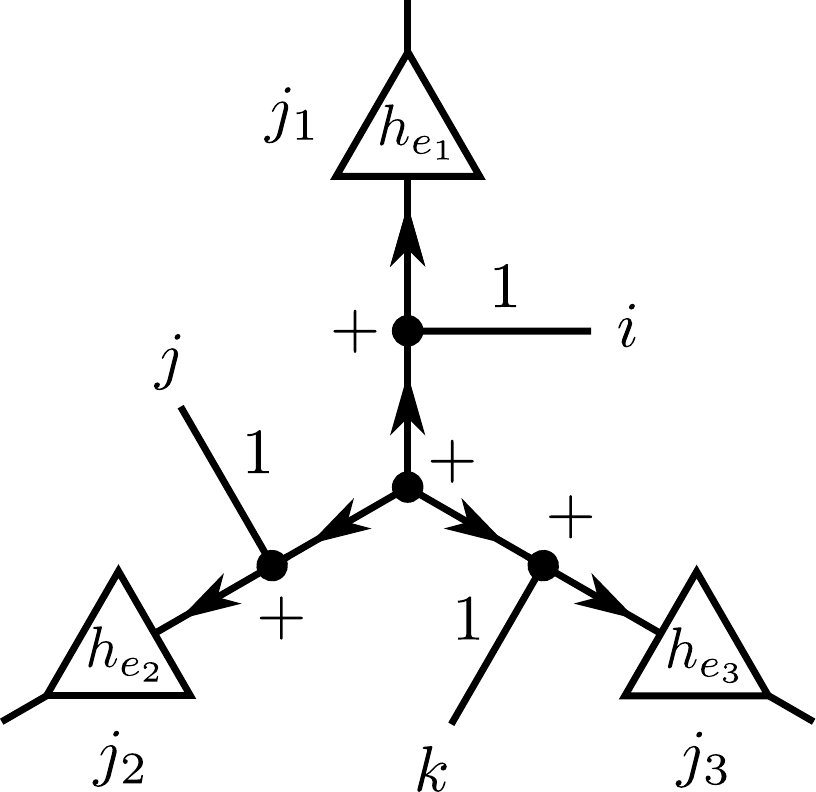}{0.6} \vspace{12pt}
\end{align}
When the vector indices are contracted by joining the free ends of the corresponding lines, this becomes \vspace{12pt}
\be\label{q*vol-node}
q_v^{(e_1,e_2,e_3)}\;\RealSymb{state123_vol.pdf}{0.6} \quad = \quad -i\sqrt 6W_{j_1}W_{j_2}W_{j_3}\;\RealSymb{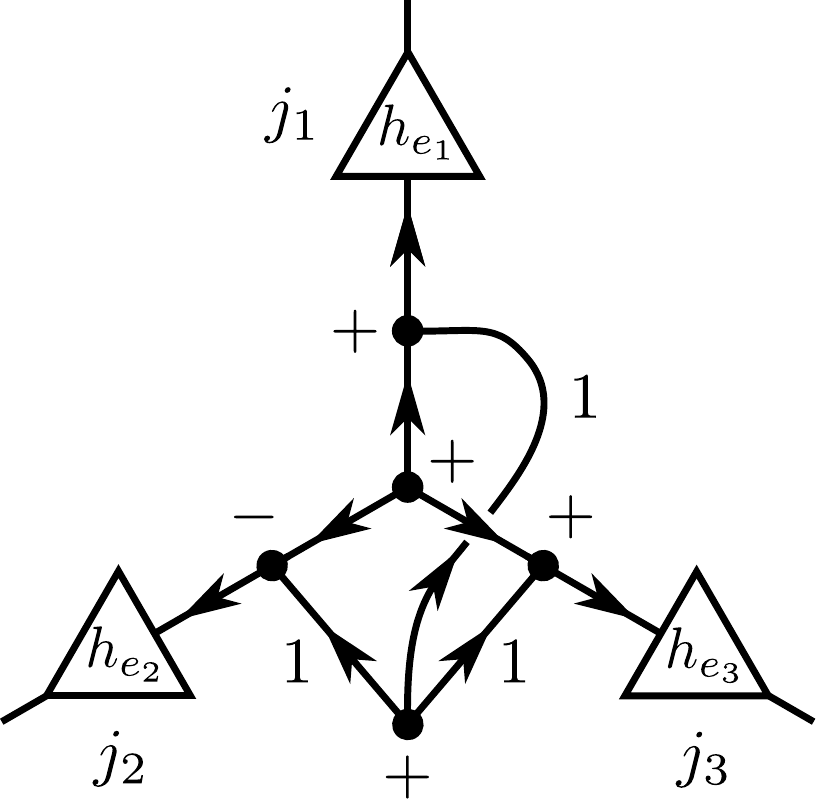}{0.6}\vspace{12pt}
\ee
Now the way to analyze an object such as the one appearing here on the right-hand side is to momentarily forget about the holonomies, and focus on the intertwiner to which they are attached. We see that the action of the volume operator has effectively replaced the intertwiner of the state \eqref{vol-node},
\vspace{12pt}
\be\label{vol_intw}
\RealSymb{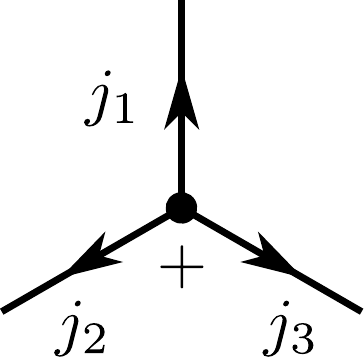}{0.6}
\ee
with the intertwiner
\be
-i\sqrt 6W_{j_1}W_{j_2}W_{j_3}\;\RealSymb{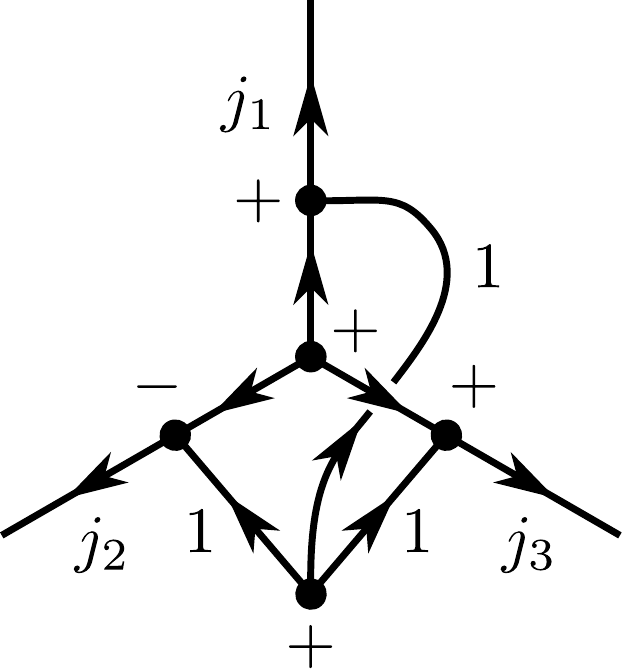}{0.6}
\ee
This is again a three-valent intertwiner having three upper indices, so we know that it must be proportional to the original intertwiner \eqref{vol_intw}. By applying the fundamental theorem of graphical calculus in the form \eqref{thm3}, we deduce
\be\label{q*vol_intw}
\RealSymb{epsJJJ_intw.pdf}{0.6} \quad = \quad \RealSymb{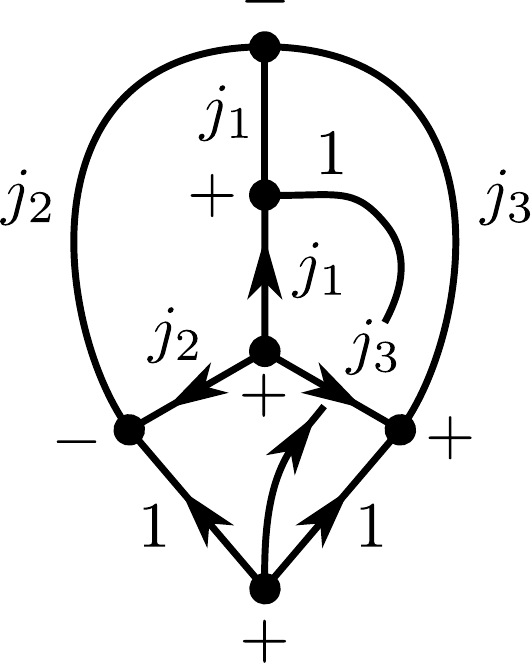}{0.6} \quad \RealSymb{vol_intw.pdf}{0.6}
\ee
where the coefficient of proportionality is
\be\label{q*vol_coeff}
\RealSymb{epsJJJ_intw_contr.pdf}{0.6} \quad = \quad \RealSymb{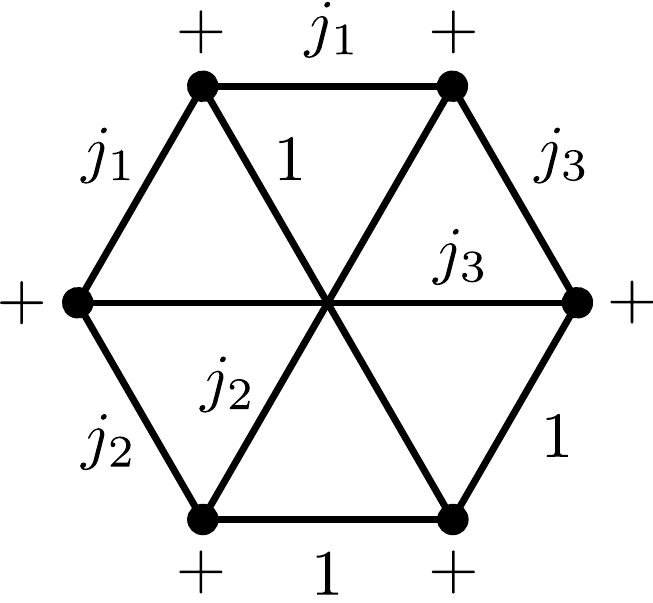}{0.6} \quad = \quad \ninej{j_1}{1}{j_1}{j_2}{1}{j_2}{j_3}{1}{j_3}.
\ee
In the first step, we have simply rearranged the nodes of the diagram into a hexagonal formation, and then we have determined the value of the resulting diagram by comparing it with \Eq{9j g}.

Going back to \Eq{q*vol-node} with \Eqs{q*vol_intw} and \eqref{q*vol_coeff}, we find that the action of the volume operator on the three-valent node reads
\be
q_v^{(e_1,e_2,e_3)}\RealSymb{state123_vol.pdf}{0.6} \;\; = \;\; -i\sqrt 6W_{j_1}W_{j_2}W_{j_3}\ninej{j_1}{1}{j_1}{j_2}{1}{j_2}{j_3}{1}{j_3}\RealSymb{state123_vol.pdf}{0.6}
\ee
Now there is still a significant observation to be made. The 9$j$-symbol appearing in the above equation contains two identical columns, so the value of the symbol is obviously unchanged if we interchange these two columns with each other. On the other hand, the symmetries of the 9$j$-symbol imply that the interchange is equivalent to multiplying the symbol by $(-1)^{2(j_1+j_2+j_3)+1} = -1$. From this it follows that the value of the symbol must be zero. Therefore we conclude that the three-valent spin network node is annihilated by the volume operator:
\be\label{q=0}
q_v^{(e_1,e_2,e_3)}\;\RealSymb{state123_vol.pdf}{0.6} \quad = \quad 0.
\ee
A four-valent node is the simplest node on which the volume operator will act in a non-trivial way.

Our calculation having been completed, let us then come back to the question of how the calculation is affected by the orientation of the edges in the state \eqref{vol-node}. Suppose that one of the edges, say $e_1$, is oriented into the node. In this case we would have to consider the action of the operator $q_v^{(e_1,e_2,e_3)}$ on the state
\be\label{vol-node_inv}
\RealSymb{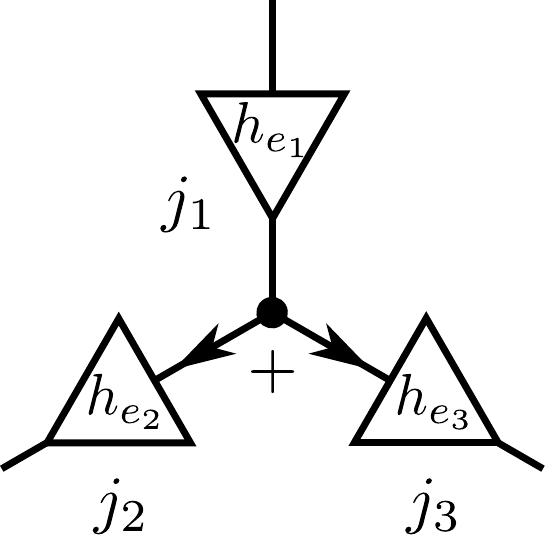}{0.6}
\ee
The action of the angular momentum operator $J_i^{(v,e_1)}$ on the first holonomy is now given by \eqref{J*h target}. We then obtain, instead of \Eq{q*vol-node},
\be\label{q*vol-node_inv}
q_v^{(e_1,e_2,e_3)}\;\RealSymb{state123_vol_inv.pdf}{0.6} \quad = \quad i\sqrt 6W_{j_1}W_{j_2}W_{j_3}\;\RealSymb{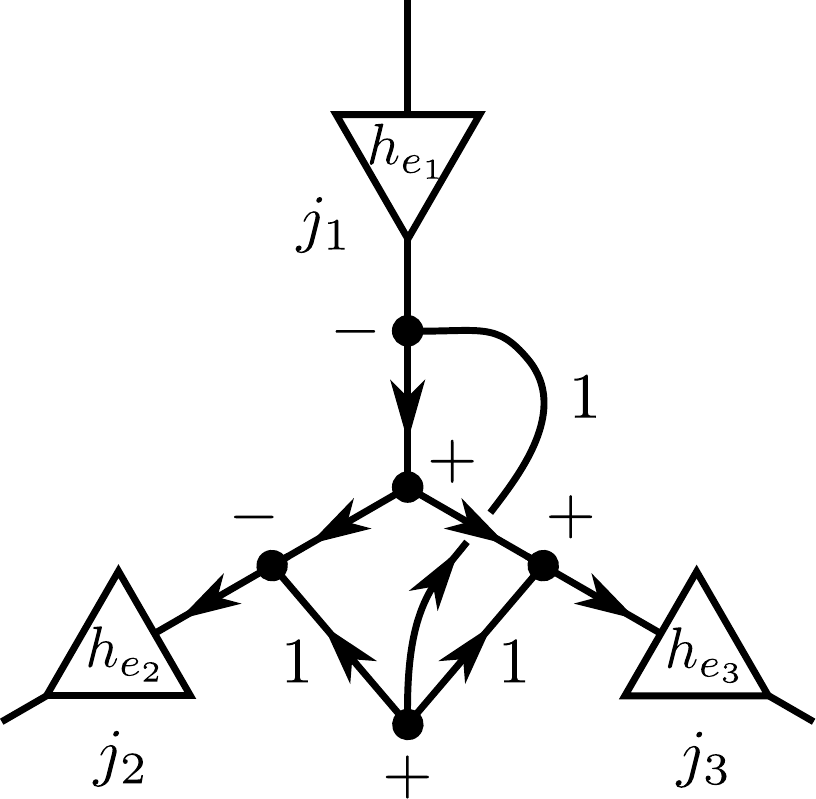}{0.6}
\ee
Compared against \Eq{q*vol-node}, there are three differences on the right-hand side:
\begin{itemize}
\item The arrow on the line carrying spin $j_1$ is reversed;
\item The node with spins $j_1$, $j_1$ and $1$ has the opposite sign;
\item There is no factor of $(-1)$, due to the absence of $(-1)$ in \Eq{J*h target} as opposed to \Eq{J*h source}.
\end{itemize}
By \Eqs{invarrow} and \eqref{3jminus g}, the first two differences contribute respectively the factors $(-1)^{2j_1}$ and $(-1)^{2j_1+1}$. Thus the overall factor in \Eq{q*vol-node_inv} relative to \Eq{q*vol-node} is $(-1)^{4j_1+2} = +1$. This argument shows that the matrix elements of not only the volume operator, but those of any operator which acts on holonomies only through the angular momentum operator $J_i^{(v,e)}$, are not sensitive to the orientation of the edges in the state on which the operator is acting.

\subsection{Euclidean operator}

The Euclidean operator
\be\label{C12}
C_v^{(e_1,e_2)} = \epsilon^{ijk}\Tr\bigl(\tau_k^{(s)}D^{(s)}(h_{\alpha_{12}})\bigr)J_i^{(v,e_1)}J_j^{(v,e_2)}
\ee
is a central ingredient of the Warsaw Hamiltonian for loop quantum gravity \cite{paper1, paper2}. The Euclidean operator acts on a pair of edges $e_1$, $e_2$ connected to the node $v$. In the definition of the operator, $\alpha_{12}$ denotes a closed loop which is tangent to the edges $e_1$ and $e_2$ (but does not overlap with them). As far as the graph of a spin network state is concerned, the action of the Euclidean operator consists of attaching the loop $\alpha_{12}$ (with the holonomy of the loop carrying spin $s$) to the graph of the state on which it is acting, as illustrated by the drawing below:
\be
\RealSymb{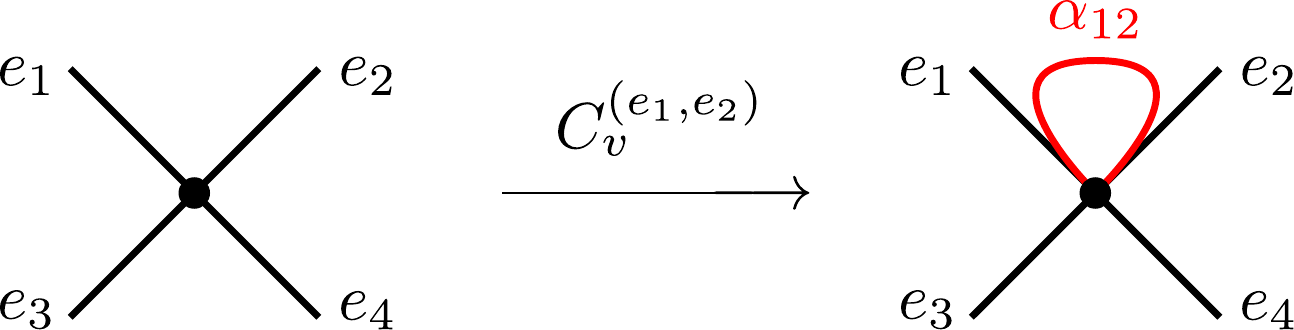}{0.6}
\ee
As an example of computing the action of the Euclidean operator in detail, let us consider the action of $C_v^{(e_1,e_2)}$ on the state
\be\label{E12-state}
\RealSymb{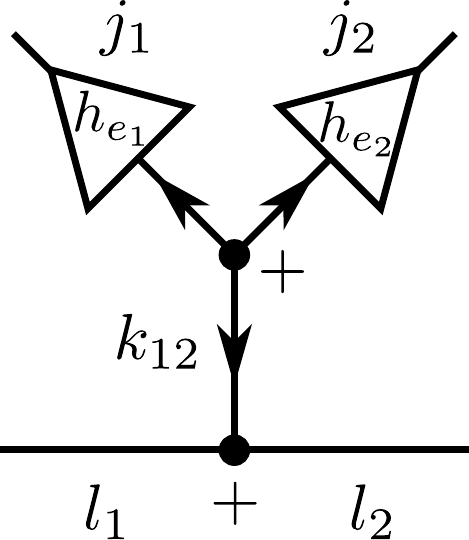}{0.6}
\ee
Here we have again assumed that the edges are oriented outwards from the node; however, the argument given in the previous section shows that the matrix elements of the Euclidean operator, which we will obtain as the result of our calculation, are independent of the orientation of the edges on which the operator acts.

Given our choice of orientation for the edges $e_1$ and $e_2$, the angular momentum operators in the Euclidean operator act on the holonomies of the state \eqref{E12-state} according to \Eq{J*h source}, namely \vspace{-28pt}
\be\label{J source}
J_i^{(v,e)}\;\RealSymb{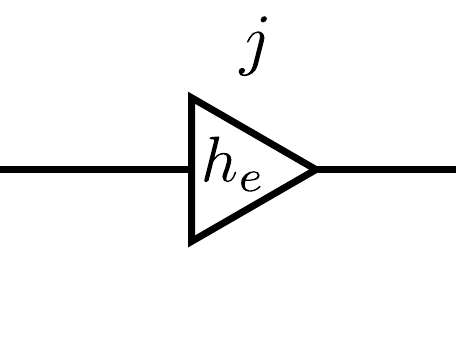}{0.6} \quad = \quad -W_j\;\RealSymb{figA-tauh_e.pdf}{0.6}
\ee
The remaining part of the operator \eqref{C12} may also be expressed in graphical form. Making use of the graphical representations of $\epsilon^{ijk}$ and $\tau_k^{(s)}$ from \Eqs{epsuvw g} and \eqref{tau=C g}, we find
\be\label{epsTr g}
\epsilon^{ijk}\Tr\bigl(\tau_k^{(s)}D^{(s)}(h_{\alpha_{12}})\bigr) \quad = \quad (i\sqrt 6)(iW_s)\;\RealSymb{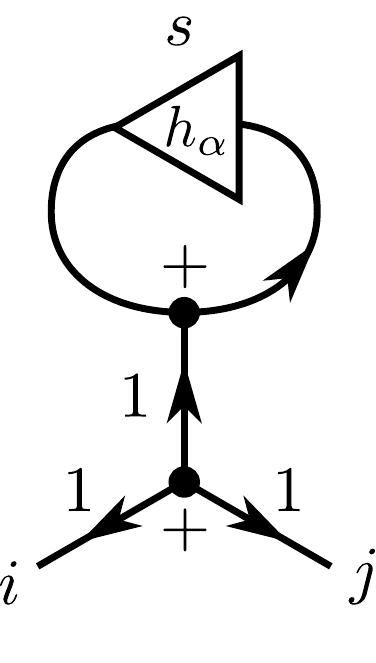}{0.6}
\ee
where the free indices $i$ and $j$ are to be contracted against the indices of the angular momentum operators.

With the help of \Eqs{J source} and \eqref{epsTr g}, we can now write down the action of the operator $\displaystyle C_v^{(e_1,e_2)}$ on the state \eqref{E12-state}:
\be\label{CE*state12}
C_v^{(e_1,e_2)}\;\RealSymb{fig16-state12_E.pdf}{0.6} \quad = \quad -\sqrt 6W_{j_1}W_{j_2}W_s\RealSymb{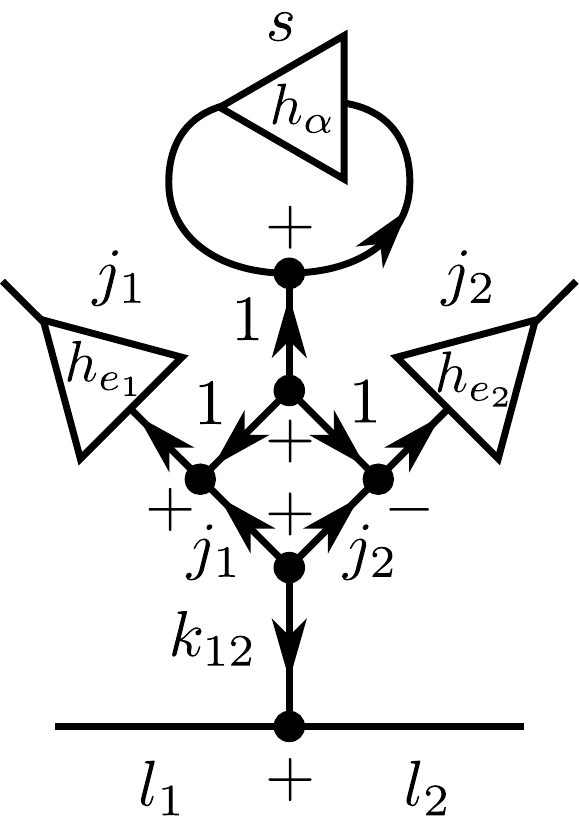}{0.6}
\ee
In order to bring out the matrix elements of the operator with respect to a particular basis, the intertwiner on the right-hand side must be expanded in the desired basis. Focusing on the five-valent intertwiner carrying spins $j_1$, $j_2$, $l_1$, $l_2$ and 1, we choose to expand it as
\be\label{E12-intw-expanded}
\RealSymb{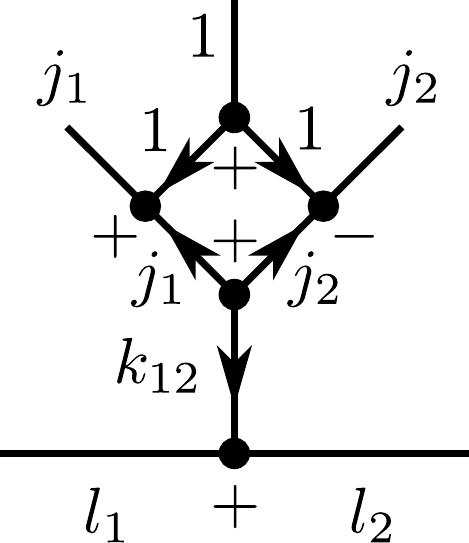}{0.6} \quad = \quad \sum_{x_{12}y} d_{x_{12}}d_y\,E_{12}(k_{12},x_{12},y)\;\RealSymb{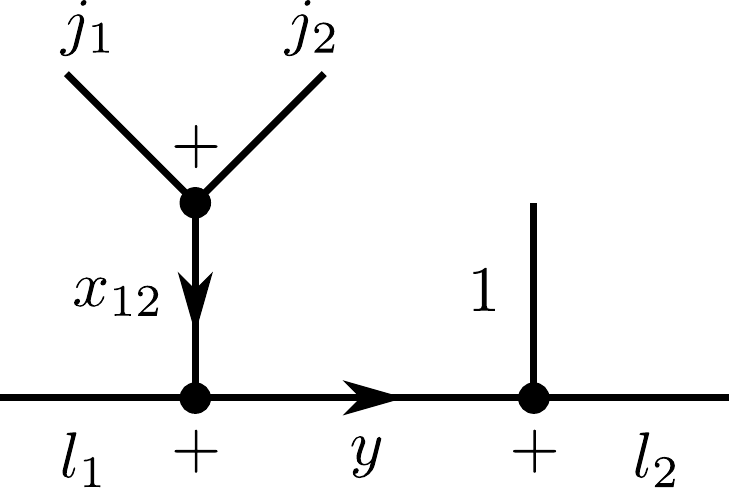}{0.6}
\ee
At the end of the calculation, the loop created by the Euclidean operator,
\be\label{E12-loop}
\RealSymb{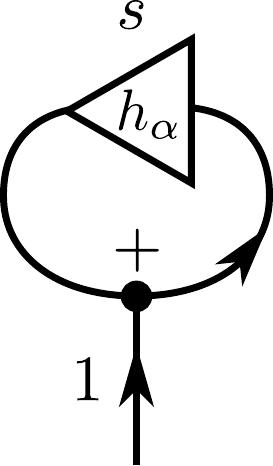}{0.6}
\ee
will be attached to the line carrying spin 1 on the right-hand side of \Eq{E12-intw-expanded}.

The coefficients $E_{12}(k_{12},x_{12},y)$ in \Eq{E12-intw-expanded} are obtained by contracting the intertwiner on the left-hand side of the equation with that on the right-hand side:
\be\label{E_12}
E_{12}(k_{12},x_{12},y) \quad = \quad \RealSymb{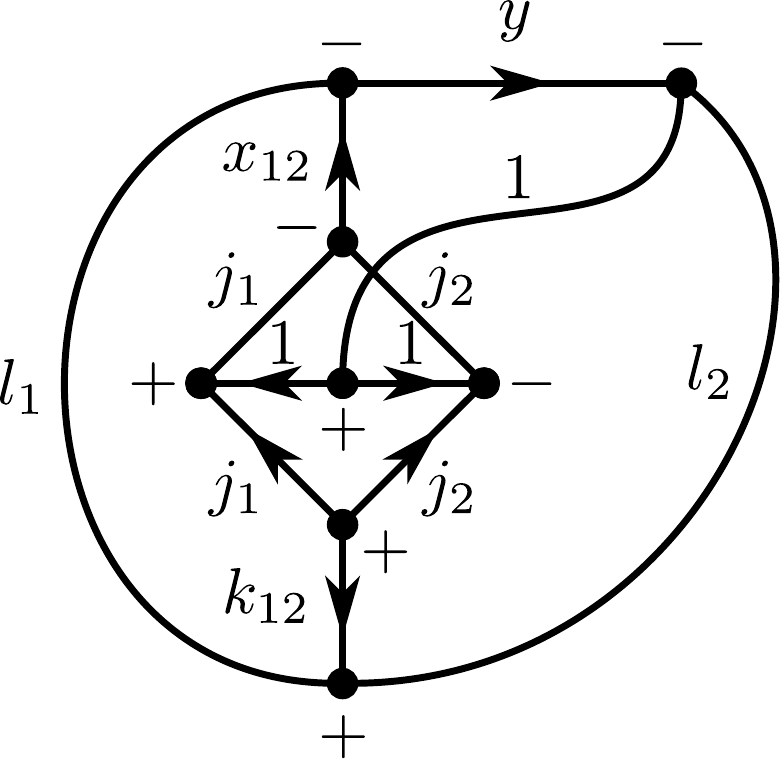}{0.6}
\ee
To deal with this diagram, we appeal to the fundamental theorem of graphical calculus in the form \eqref{thm3'}. The diagram can be separated into two disconnected pieces by cutting three lines according to
\be\label{E12-cut}
\RealSymb{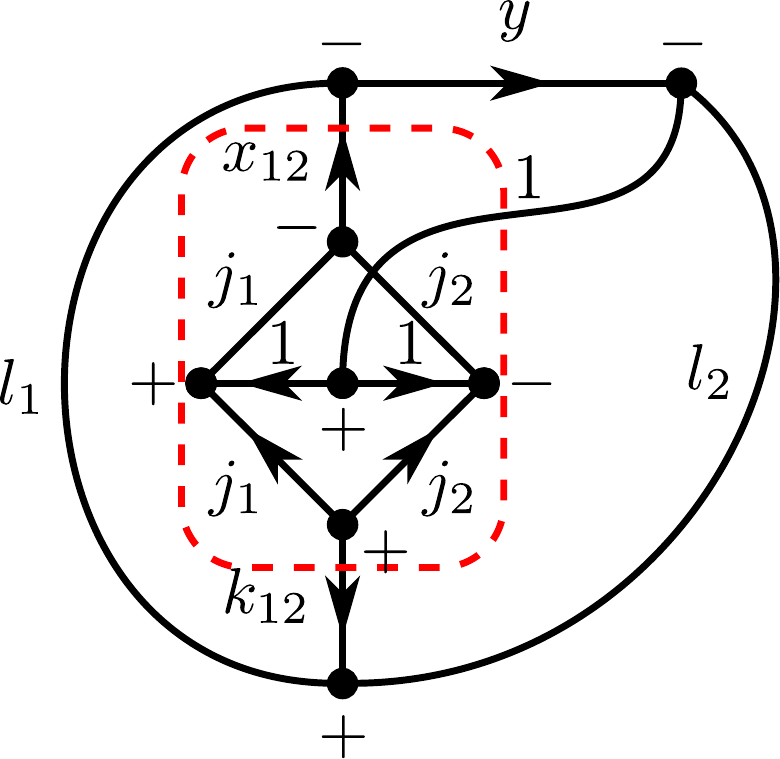}{0.6}
\ee
Therefore the diagram is equal to the product of the two pieces resulting from the cutting. The simpler of the two pieces is \vspace{-10pt}
\be
\RealSymb{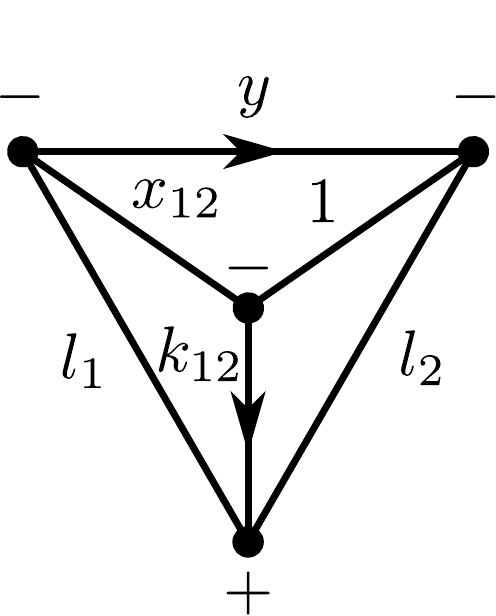}{0.6} \quad = \quad (-1)^{k_{12}+l_1-l_2}\sixj{k_{12}}{x_{12}}{1}{y}{l_2}{l_1},
\ee
where the factor $(-1)^{k_{12}+l_1-l_2}$ arises when \Eqs{invarrow}, \eqref{3jminus g} and \eqref{3jarrows g} are used to adjust the arrows and signs in the diagram so that they agree with \Eq{6j g}. Similarly, by comparing the remaining piece of the diagram \eqref{E12-cut} with \Eq{9j g}, we find
\be
\RealSymb{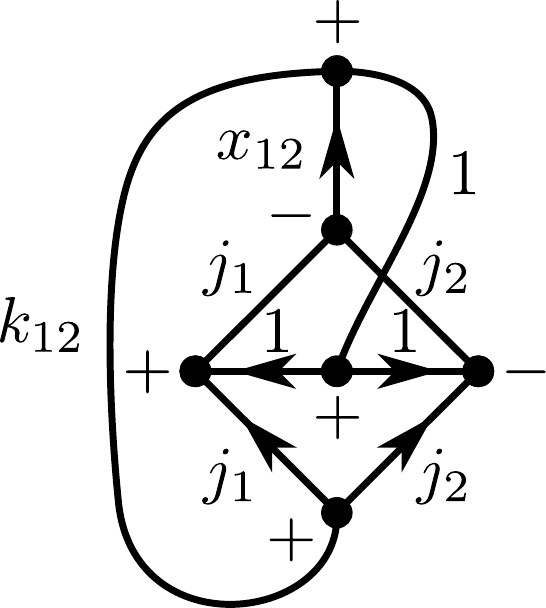}{0.6} \quad = \quad \RealSymb{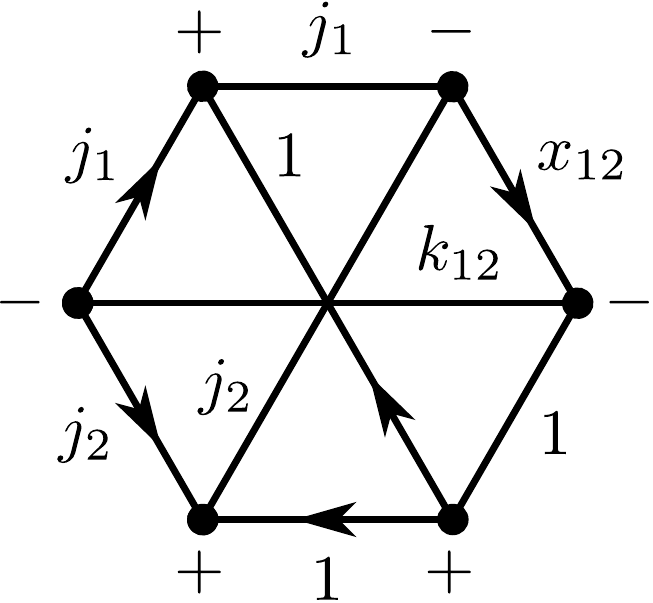}{0.6} \quad = \quad (-1)^{k_{12}-x_{12}}\ninej{j_1}{j_1}{1}{j_2}{j_2}{1}{k_{12}}{x_{12}}{1}.
\ee
Hence we have shown that
\be\label{E_12-result}
E_{12}(k_{12},x_{12},y) = (-1)^{l_1-l_2+x_{12}}\sixj{k_{12}}{x_{12}}{1}{y}{l_2}{l_1}\ninej{j_1}{j_1}{1}{j_2}{j_2}{1}{k_{12}}{x_{12}}{1}.
\ee
We may now read off the result of our calculation from \Eqs{CE*state12} and \eqref{E12-intw-expanded}. We see that the action of the Euclidean operator on the state \eqref{E12-state} is given by
\begin{align}
C_v^{(e_1,e_2)}\;&\RealSymb{fig16-state12_E.pdf}{0.6} \quad = \quad -\sqrt 6W_{j_1}W_{j_2}{W_s}\sum_{x_{12}y}d_{x_{12}}d_y(-1)^{l_1-l_2+x_{12}} \notag \\
&\times\sixj{k_{12}}{x_{12}}{1}{y}{l_2}{l_1}\ninej{j_1}{j_1}{1}{j_2}{j_2}{1}{k_{12}}{x_{12}}{1}\;\RealSymb{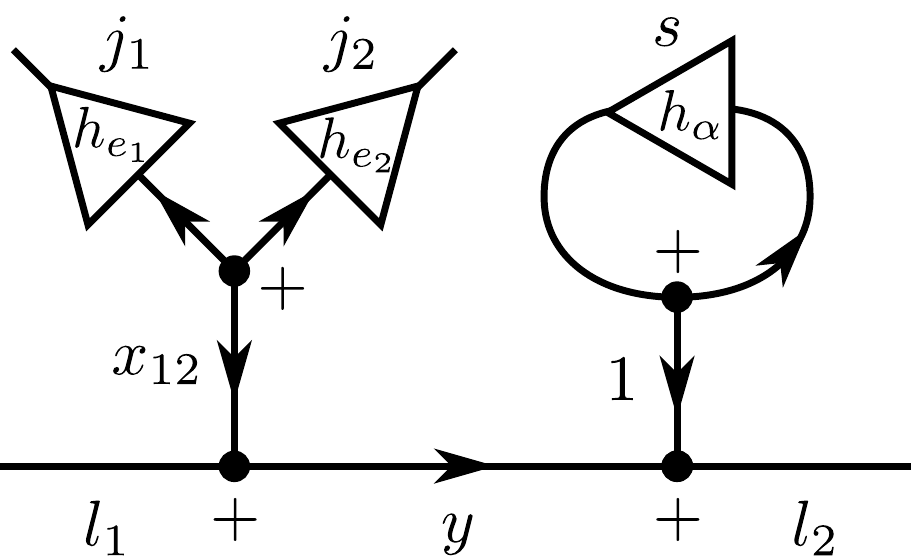}{0.6}\label{CE12-result}
\end{align}

\subsection{Curvature operator}\label{sec:curvature}

The curvature operator $R_v^{(e_1,e_2)}$ \cite{curvature}, just as the Euclidean operator considered in the previous section, acts on a pair of edges $(e_1,e_2)$ belonging to the node $v$. If the edges $e_1$ and $e_2$ are coupled to a definite total spin by the intertwiner at the node, the action of the curvature operator is diagonal:
\be\label{R_v}
R_v^{(e_1,e_2)}\;\RealSymb{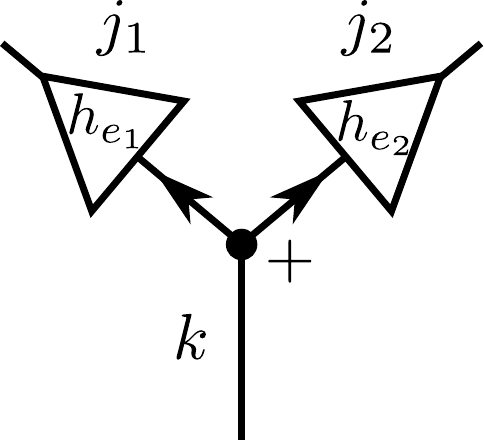}{0.6} \quad = \quad \Lambda(j_1,j_2,k)\;\RealSymb{figR-state12.pdf}{0.6}
\ee
where $\Lambda(j_1,j_2,k)$ is a certain function of the spins $j_1$, $j_2$ and $k$.

The curvature operator is an example of an operator whose action on spin network states cannot be expressed in fully graphical form, by equations analogous to \Eq{q*vol-node} or \Eq{CE*state12} (this is because the operator $R_v^{(e_1,e_2)}$ is a non-polynomial function of the angular momentum operators $J_i^{(v,e)}$). Graphical techniques can nevertheless be useful in computing the action of the operator. For example, let us consider the problem of acting with $R_v^{(e_1,e_2)}$ on the state
\be\label{R12state}
\RealSymb{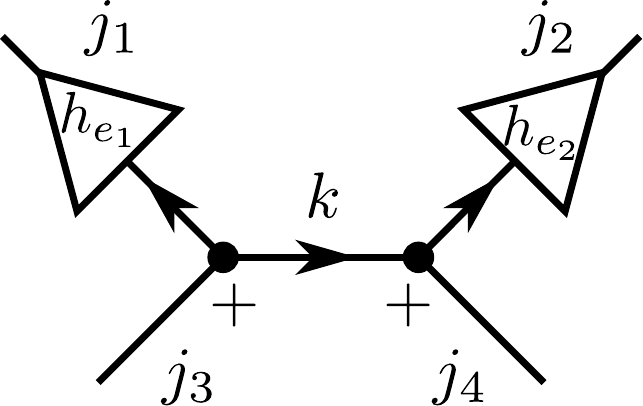}{0.6}
\ee
A straightforward (and essentially non-graphical) approach to the problem would consist of the following three steps:
\begin{itemize}
\item Perform a suitable change of basis in the intertwiner space of the node, expressing the intertwiner at the node in a basis in which the edges $e_1$ and $e_2$ are coupled together.
\item Apply the operator $R_v^{(e_1,e_2)}$, which now acts diagonally on the new intertwiner basis.
\item Reverse the change of basis performed in the first step, transforming the intertwiner back into the original basis.
\end{itemize}
However, there exists also an alternative method\footnote{This method was pointed out to the author by Cong Zhang.}, which can be used to attack the problem graphically, and in which the action of the curvature operator is computed in a single step, without having to perform any changes of basis in the intertwiner space.

\Eq{R_v} shows that the operator $R_v^{(e_1,e_2)}$ can essentially be viewed as an operator acting on the intertwiner space of the node $v$, the action of the operator being given by
\be\label{R12}
R_{12}\;\RealSymb{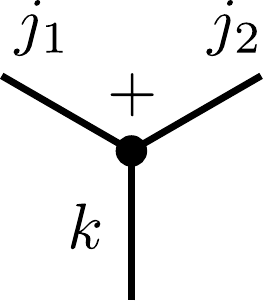}{0.6} \quad = \quad \Lambda(j_1,j_2,k)\;\RealSymb{figR-iota12.pdf}{0.6}
\ee
It then turns out that the operator $R_{12}$ can be expressed in graphical form as
\be\label{R12 g}
R_{12} \; = \; \sum_x d_x\,\Lambda(j_1,j_2,x)\;\RealSymb{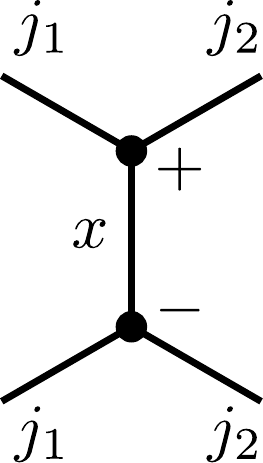}{0.6}
\ee
This graphical operator acts by attaching its two legs onto the two lines of the intertwiner on which it is acting. To check that \eqref{R12 g} is indeed a correct representation of the operator $R_{12}$, it suffices to act with it on an intertwiner in which the spins $j_1$ and $j_2$ are coupled, and observe that \Eq{R12} is reproduced:
\be
R_{12}\;\RealSymb{figR-iota12.pdf}{0.6} \quad = \quad \sum_x d_x\,\Lambda(j_1,j_2,x)\;\RealSymb{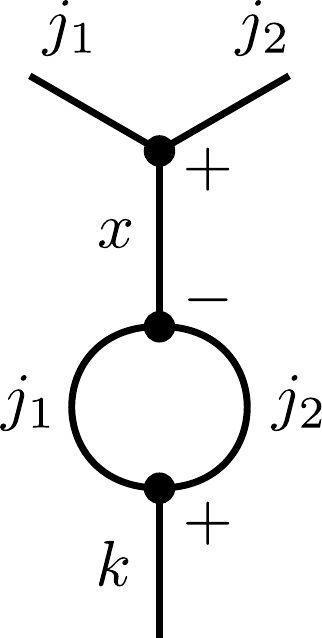}{0.6} \quad = \quad \Lambda(j_1,j_2,k)\;\RealSymb{figR-iota12.pdf}{0.6}
\ee
where we have recalled that
\be
\RealSymb{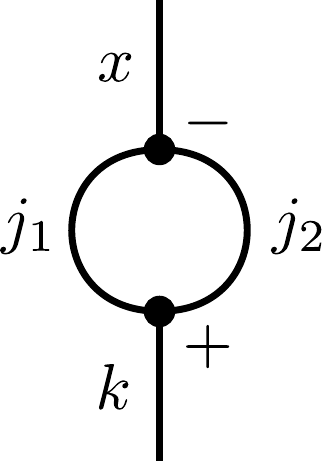}{0.6} \quad = \quad \delta_{kx}\frac{1}{d_k}\quad\RealSymb{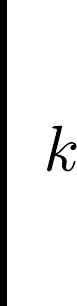}{0.6}
\ee
To illustrate the use of the graphical operator \eqref{R12 g}, let us apply it to the problem of evaluating the action of the curvature operator on the state \eqref{R12state} -- or, equivalently, the action of the operator $R_{12}$ on the intertwiner
\be\label{R12-intw}
\RealSymb{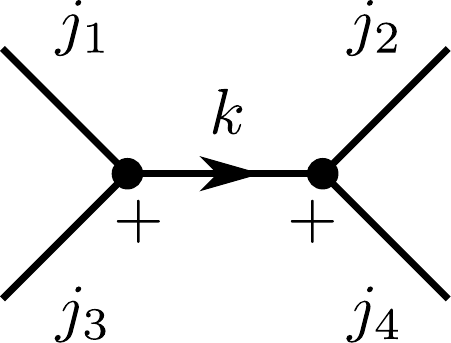}{0.6}
\ee
Attaching the operator \eqref{R12 g} to the intertwiner, we get
\be\label{R12*intw}
R_{12}\;\RealSymb{R12-intw}{0.6} \quad = \quad \sum_x d_x\,\Lambda(j_1,j_2,x)\;\RealSymb{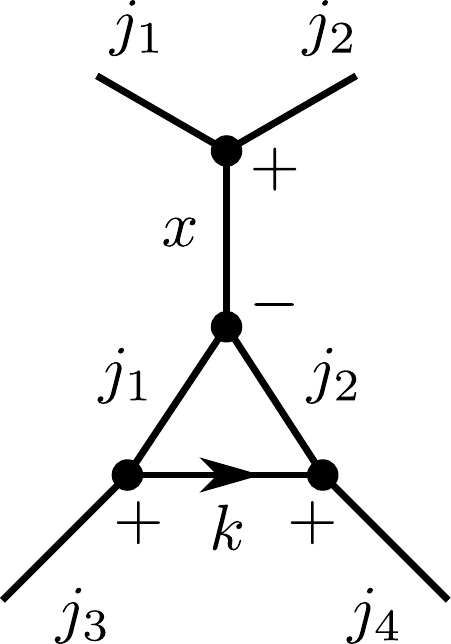}{0.6}
\ee
To obtain the matrix elements of $R_{12}$ in the basis formed by the intertwiners \eqref{R12-intw}, it suffices to expand the intertwiner on the right-hand side of \Eq{R12*intw} in this basis. Using the fundamental theorem in its four-valent version \eqref{thm4}, we have
\be\label{R12-action}
\RealSymb{R13-action}{0.6} \quad = \quad \sum_y d_y\quad\RealSymb{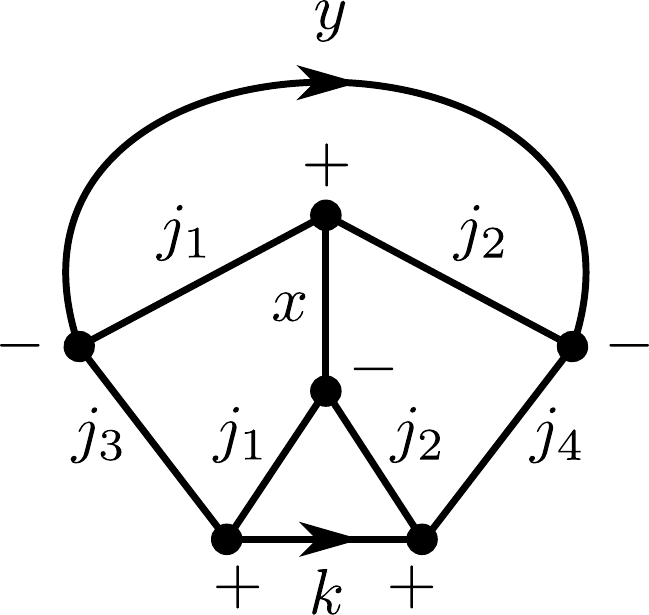}{0.6}\quad\RealSymb{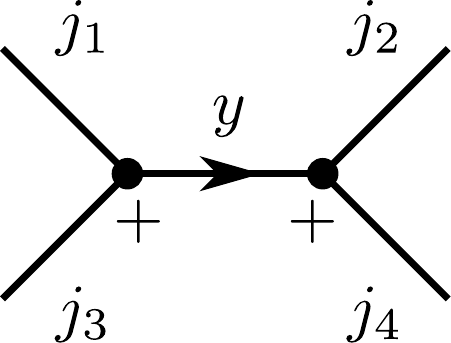}{0.6}
\ee
Here the diagram within the sum should clearly be split into two pieces by cutting it across the three vertical lines. Before performing the cut, we introduce (at no cost of a sign factor) two oppositely oriented arrows on the line carrying spin $x$. This is done so that each of the the two pieces resulting from the cut will be a properly contracted, invariant object. In this way we find 
\newpage
\be\label{R12-cut}
\RealSymb{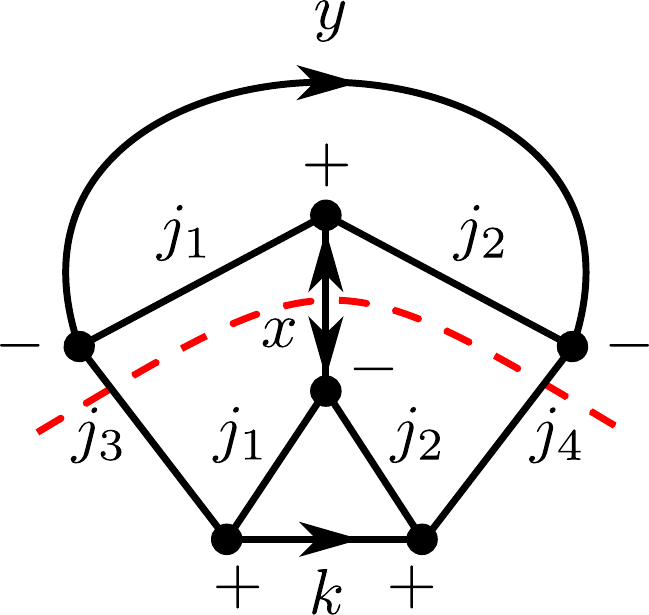}{0.6} \quad = \quad \RealSymb{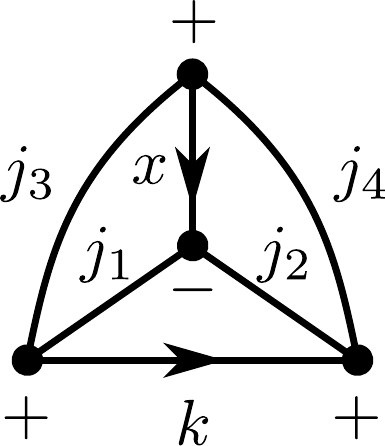}{0.6}\qquad\RealSymb{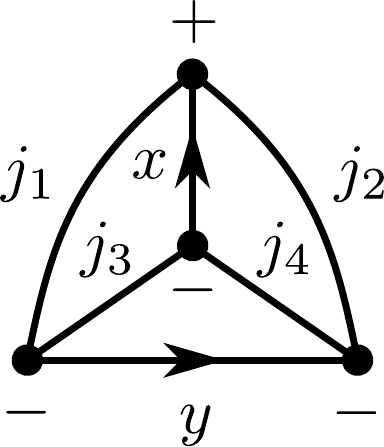}{0.6}
\ee
By comparing the first piece with \Eq{6j g}, we see that
\be\label{R12-piece1}
\RealSymb{R13-6j-1}{0.6} \quad = \quad (-1)^{j_1+j_2+x}\;\;\RealSymb{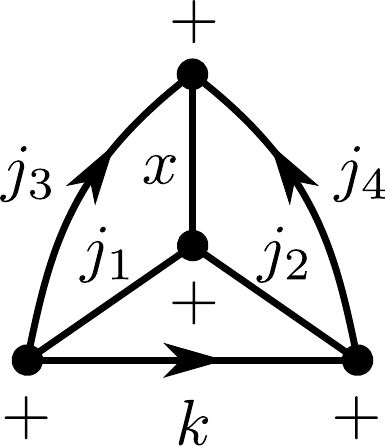}{0.6} \quad = \; (-1)^{j_1+j_2+x}(-1)^{2j_3}\sixj{j_1}{j_2}{x}{j_4}{j_3}{k}.
\ee
Similarly, comparing the second piece with \Eq{6jminus g}, we get
\be\label{R12-piece2}
\RealSymb{R13-6j-2}{0.6} \quad = \quad (-1)^{j_1+j_2+x}(-1)^{2j_1}(-1)^{2y}\sixj{j_1}{j_2}{x}{j_4}{j_3}{y}.
\ee
Hence the overall sign on the right-hand side of \Eq{R12-cut} is $(-1)^{2j_1+2j_3+2y} = +1$.

The calculation has now been completed. Inserting \Eqs{R12-piece1} and \eqref{R12-piece2} back into \Eq{R12-action}, we conclude that the action of the curvature operator on the intertwiner \eqref{R12-intw} is given by
\begin{align}
R_{12}\;&\RealSymb{R12-intw}{0.6} \notag \\
&=\quad\sum_{xy} d_xd_y\,\sixj{j_1}{j_2}{x}{j_4}{j_3}{k}\sixj{j_1}{j_2}{x}{j_4}{j_3}{y}\Lambda(j_1,j_2,x)\;\RealSymb{R12-intw-y}{0.6}
\end{align}
We leave it for the reader to check that the same result is obtained from the non-graphical calculation outlined in the beginning of this section, using \Eq{6j-def g} to interchange the lines carrying spins $j_2$ and $j_3$ in the intertwiner \eqref{R12-intw}. In the example at hand, the graphical solution does not really require any less effort than the non-graphical approach. However, the graphical representation of the curvature operator can be significantly more effective in more complicated cases, especially if a formula for the change of basis required in the non-graphical calculation is not readily available.

\newpage

\subsection*{Exercises}
\addcontentsline{toc}{subsection}{{\hspace{21.4pt} Exercises}}

\begin{enumerate}[leftmargin=*]

\item The scalar product between two states based on a graph $\Gamma$ is defined as
\[
\braket{\Psi_\Gamma}{\Phi_\Gamma} = \int dh_{e_1}\cdots dh_{e_N}\,\overline{\Psi_\Gamma(h_{e_1},\dots,h_{e_N})}\Phi_\Gamma(h_{e_1},\dots,h_{e_N}),
\]
where each $dh_e$ is the $SU(2)$ Haar measure. Show that the scalar product between two spin network states of the form \eqref{spinnetwork} is
\[
\braket{\Psi_{\Gamma,\{j_e\},\{\iota_v\}}}{\Psi_{\Gamma,\{j_e'\},\{\iota_v'\}}} = \biggl(\prod_{e\in\Gamma}\frac{1}{d_{j_e}}\delta_{j_e,j_e'}\biggr)\biggl(\prod_{v\in\Gamma} \braket{\iota_v}{\iota_v'}\biggr).
\]
How can the definition of the scalar product be extended to states which are not based on the same graph? Formulate a set of assumptions which the graphs must satisfy so that the states $\Psi_{\Gamma,\{j_e\},\{\iota_v\}}$ and $\Psi_{\Gamma',\{j_e'\},\{\iota_v'\}}$, based on two different graphs, are guaranteed to be orthogonal to each other. (See \eg \cite{status} or \cite{thesis}.)

\item Check that the operator $J_i^{(v,e)}$ satisfies the commutation relation
\[
\bigl[J_i^{(v,e)},J_j^{(v',e')}\bigr] = \delta_{vv'}\delta_{ee'}i\epsilon\downup{ij}{k}J_k^{(v,e)}.
\]

\item Calculate the commutator $\bigl[\D{j}{m}{n}{h_e},J_i^{(v,e)}\bigr]$, when the point $v$ is
\begin{itemize}
\item[(a)] the beginning point of the edge $e$;
\item[(b)] the endpoint of the edge $e$.
\end{itemize}

\item Show that the generic spin network node
\[
\iota\updown{m_1\cdots m_M}{m_1'\cdots m_N'}\D{j_1}{n_1}{m_1}{h_{e_1}}\cdots\D{j_M}{n_M}{m_M}{h_{e_M}}\D{j_1'}{m_1'}{n_1'}{h_{e_1'}}\cdots\D{j_N'}{m_N'}{n_N'}{h_{e_N'}},
\]
which contains $M$ outgoing edges $e_1,\dots,e_M$ and $N$ incoming edges $e_1',\dots,e_N'$, is annihilated by the Gauss operator
\[
G_i^{(v)} = \sum_{\substack{\text{edges}\\ \text{at $v$}}} J_i^{(v,e)}.
\]

\item Verify that the angle operator
\[
\Delta_v^{(e_1,e_2)} = J_i^{(v,e_1)}J_{\phantom{i}}^{(v,e_2)i}
\]
and the volume operator
\[
q_v^{(e_1,e_2,e_3)} = \epsilon^{ijk}J_i^{(v,e_1)}J_j^{(v,e_2)}J_k^{(v,e_3)}
\]
commute with the Gauss operator $G_i^{(v)}$ defined in the previous problem.

\item Show, without making use of the graphical formalism, that the action of the volume operator
\[
q_v^{(e_1,e_2,e_3)} = \epsilon^{ijk}J_i^{(v,e_1)}J_j^{(v,e_2)}J_k^{(v,e_3)}
\]
on a three-valent spin network node gives zero. (Use the gauge invariance condition $J_i^{(v,e_1)} + J_i^{(v,e_2)} + J_i^{(v,e_3)} = 0$ and the commutation relations of the angular momentum operator.)

\item Consider the following spin network state, whose graph consists of $N$ closed loops $\alpha_1,\dots,\alpha_N$ all intersecting each other a single node $v$:
\[
\RealSymb{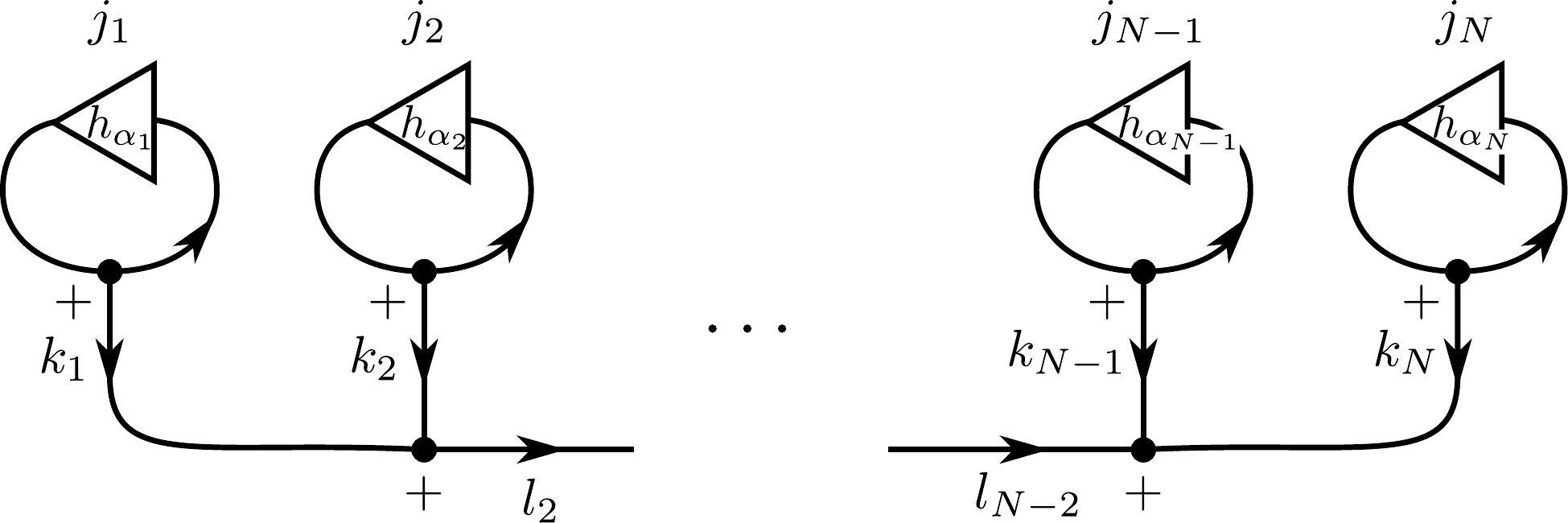}{0.6}
\]
Find how the state behaves under a transformation which reverses the orientation of all the loops, \ie $h_{\alpha_i}\to h_{\alpha_i}^{-1}$ for every $\alpha_i$.

\item Calculate the action of the area operator \cite{area, RSvolume}
\[
A_v^{(e)} = J_i^{(v,e)}J_{\phantom{i}}^{(v,e)i}
\]
on the holonomy $\D{j}{m}{n}{h_e}$, assuming that the node $v$ is the beginning point of the edge $e$.

\item If the surface whose area we wish to measure contains a node of a spin network state, the area operator associated to the surface is given by
\[
A_v \; = \sum_{e_I,e_J\;\text{at}\;v} \delta(e_I,e_J)J_i^{(v,e_I)}J_{\phantom{i}}^{(v,e_J)i},
\]
where $\delta(e_I,e_J)$ equals $0$ if either $e_I$ or $e_J$ intersects the surface tangentially at the node, and otherwise equals $+1$ if $e_I$ and $e_J$ lie on the same side of the surface, or $-1$ if they lie on opposite sides of it. Find the eigenvalues \mbox{of the operator $A_v$.} ({\em Hint:} It is useful to divide the edges belonging to the node into three classes -- ''above'', ''below'' and ''tangential'' -- as shown in the drawing, and express $A_v$ in terms of the operators
\[
J_i^{(v,a)} \; = \sum_{\text{$e$ above $S$}} J_i^{(v,e)}, \qquad J_i^{(v,b)} \; = \sum_{\text{$e$ below $S$}} J_i^{(v,e)}, \qquad J_i^{(v,t)} \; = \sum_{\substack{\text{$e$ tangential} \\ \text{to $S$}}} J_i^{(v,e)}.)
\]

\[
\RealSymb{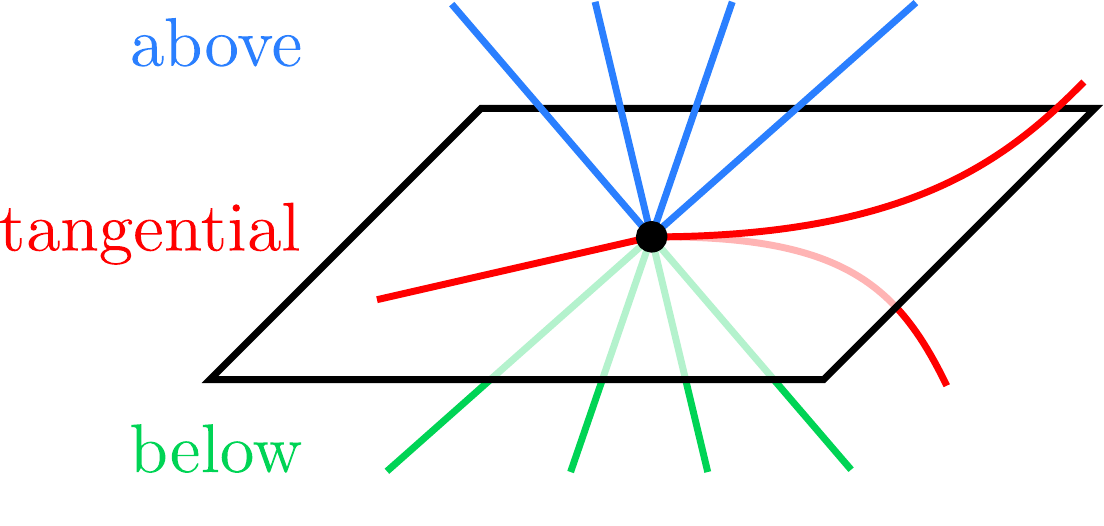}{0.6}
\]

\newpage

\item Calculate the action of the angle operator \cite{Major}
\[
\Delta_v^{(e_1,e_2)} = J_i^{(v,e_1)}J_{\phantom{i}}^{(v,e_2)i}
\]
on the three-valent node
\[
\RealSymb{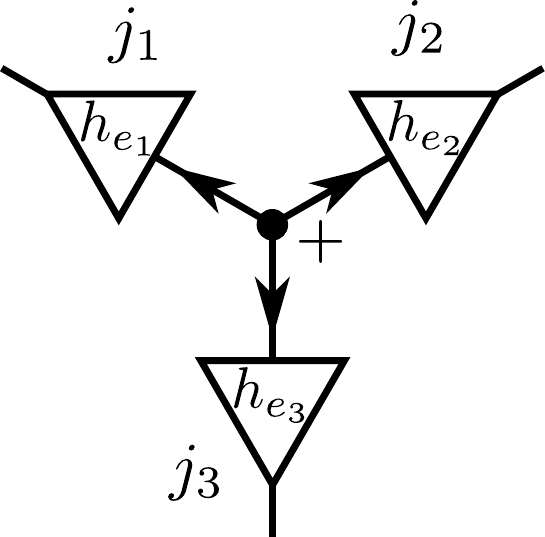}{0.6}
\]
\begin{itemize}
\item[(a)] Graphically;
\item[(b)] Non-graphically, using the condition $J_i^{(v,e_1)} + J_i^{(v,e_2)} + J_i^{(v,e_3)} = 0$.
\end{itemize}
Verify that your answers to parts (a) and (b) are consistent with each other.

\item Calculate the action of the volume operator $q_v^{(e_1,e_2,e_3)}$ on the four-valent node
\[
\RealSymb{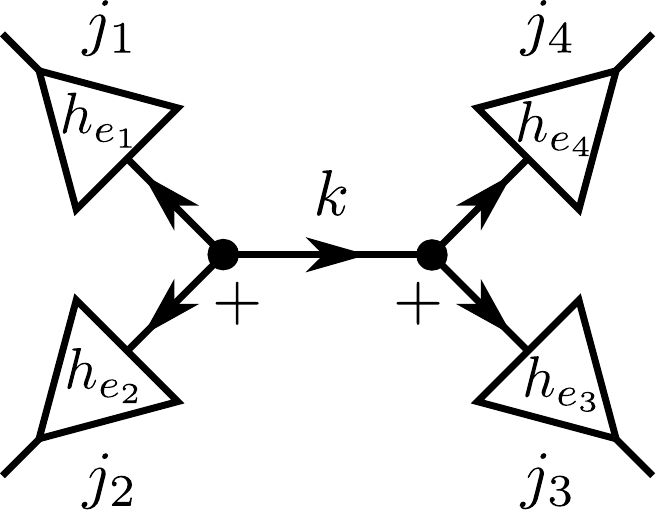}{0.6}
\]

\item Calculate the action of the volume operator
\[
q_v^{(e_I,e_J,e_K)} = \epsilon^{ijk}J_i^{(v,e_I)}J_j^{(v,e_J)}J_k^{(v,e_K)}
\]
on the state
\[
\RealSymb{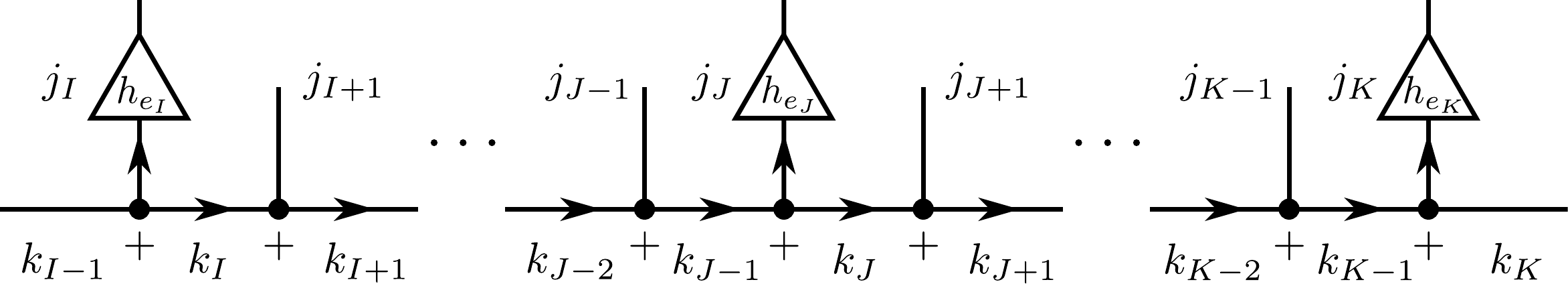}{0.6}
\]

\item Calculate the action of the Euclidean operator
\[
C_v^{(e_1,e_3)} = \epsilon^{ijk}\Tr\bigl(\tau_k^{(s)}D^{(s)}(h_{\alpha_{13}})\bigr)J_i^{(v,e_1)}J_j^{(v,e_3)}
\]
on the state
\[
\RealSymb{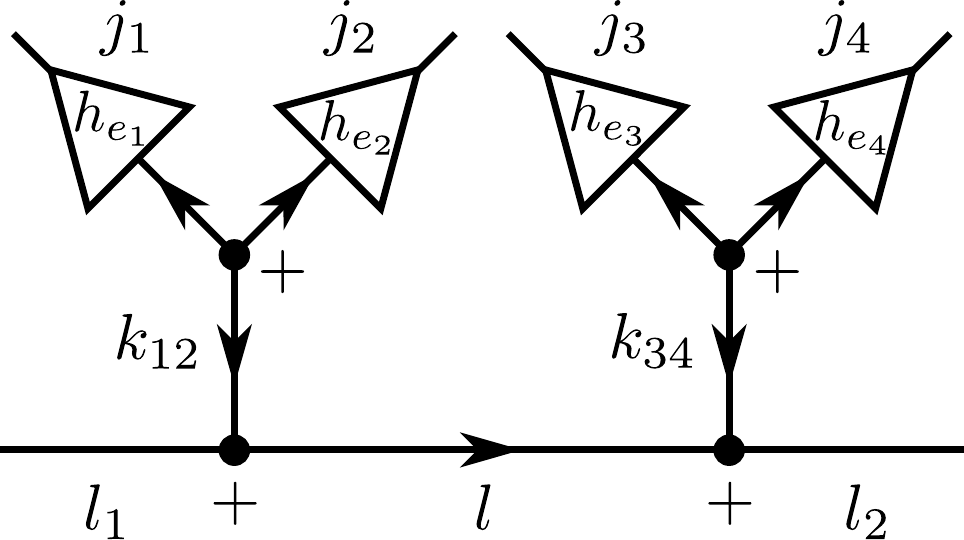}{0.6}
\]
Express the result in terms of states of the form
\[
\RealSymb{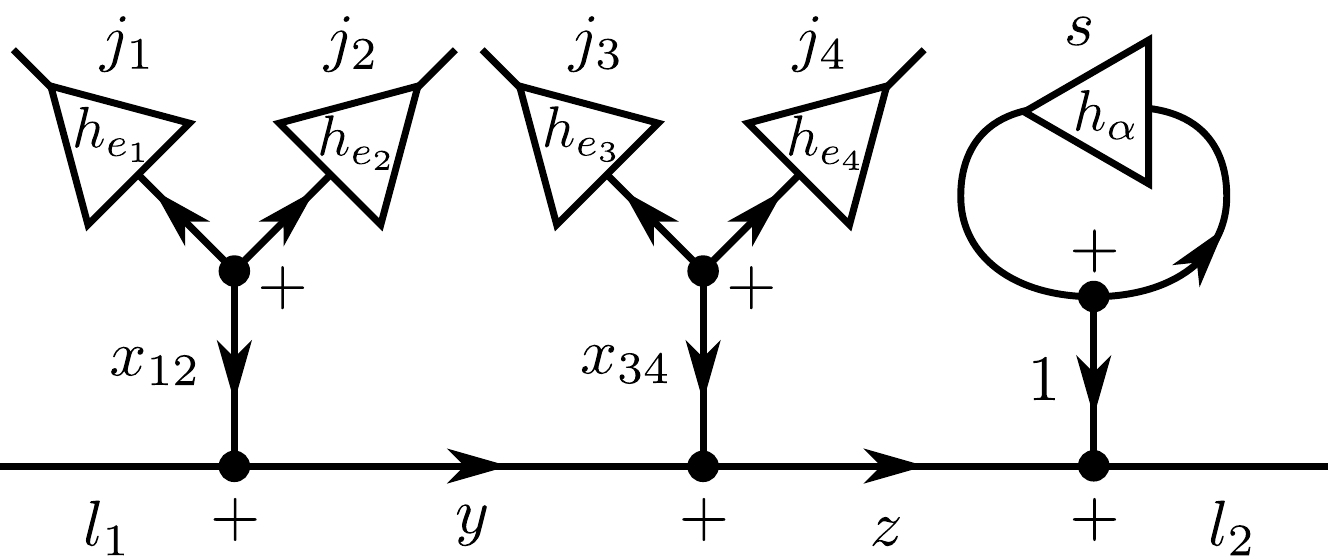}{0.6}
\]

\item Calculate the action of the curvature operator $R_v^{(e_1,e_3)}$ on the state
\[
\RealSymb{fig16-state13_v2.pdf}{0.6}
\]
\begin{itemize}
\item[(a)] Using the graphical representation of the curvature operator;
\item[(b)] By transforming the state into a basis in which $R_v^{(e_1,e_3)}$ acts diagonally -- recall \Eq{iota6change} -- and reversing the change of basis after the operator has acted.
\end{itemize}
Verify that your answers to parts (a) and (b) are consistent with each other.

\item The Euclidean part of Thiemann's Hamiltonian (see \eg \cite{Thiemann} or \cite{QSD}) is essentially the operator
\[
H_v^{(e_1,e_2,e_3)} = \Tr\Bigl(D^{(s)}(h_{\alpha_{12}})D^{(s)}(h_{s_3}^{-1})V_vD^{(s)}(h_{s_3})\Bigr).
\]
The notation is explained by the drawing below: $s_I$ denotes a segment of the edge $e_I$, and $\alpha_{IJ}$ is the loop $s_J^{-1}\circ p_{JI}\circ s_I$, where $p_{JI}$ is an edge connecting the endpoints of $s_I$ and $s_J$. Furthermore, $V_v$ is the actual volume operator (not the operator $q_v$) acting on the node $v$.
\[
\RealSymb{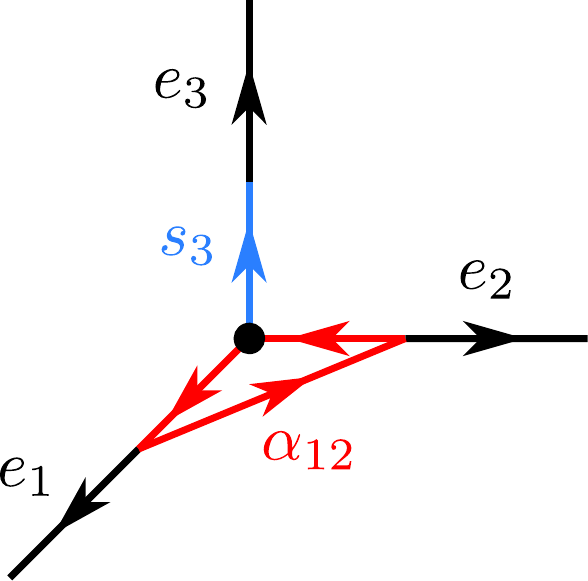}{0.6}
\]
Calculate the action of $H_v^{(e_1,e_2,e_3)}$ on the three-valent node
\[
\RealSymb{state123.pdf}{0.6}
\]
Assume that the action of the volume operator on a three-valent, non-gauge invariant node has the generic form
\[
V_v\;\RealSymb{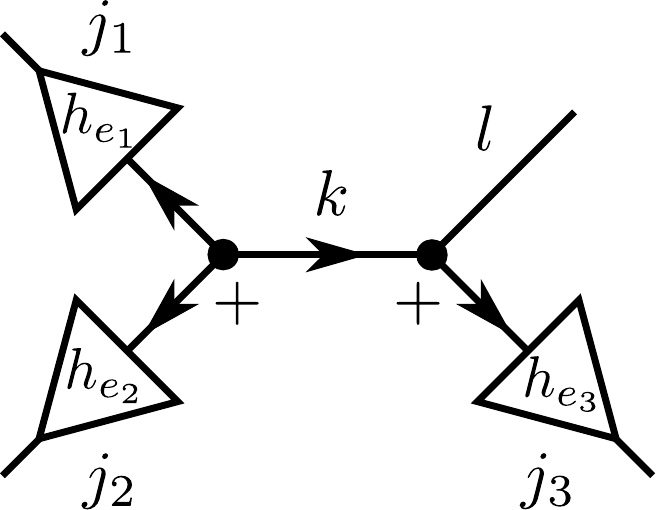}{0.6} \quad = \quad \sum_x \; V^{(j_1j_2j_3l)}(x,k) \; \RealSymb{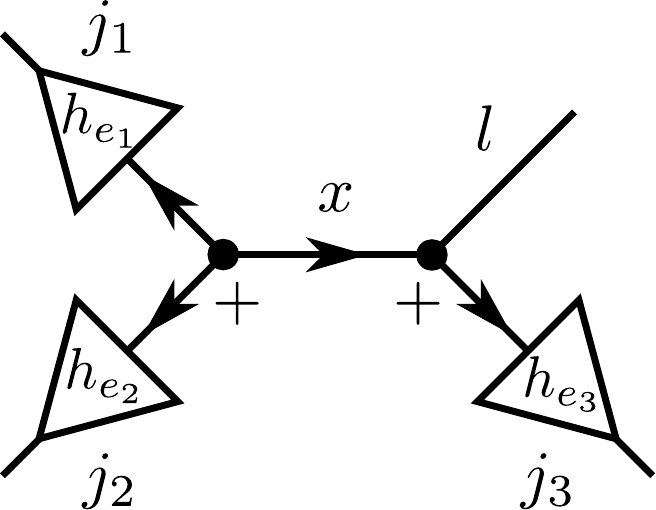}{0.6}
\]
Are the matrix elements of $H_v^{(e_1,e_2,e_3)}$ independent of the orientation of the edges $e_1$, $e_2$ and $e_3$?

(This calculation is discussed using graphical techniques \eg in \cite{AlesciLiegenerZipfel} and \cite{AlesciThiemannZipfel}.)

\end{enumerate}

\newpage

\renewcommand{\refname}{{\Large Bibliography}}

\newpage
\thispagestyle{empty}

\newgeometry{hmargin={25mm,25mm},vmargin={22mm,0mm}}

\begin{footnotesize}

\begin{minipage}{0.06\textwidth}
\end{minipage}
{\begin{minipage}{0.33\textwidth}
\[
\threej{j_1}{j_2}{j_3}{m_1}{m_2}{m_3}\quad = \quad\RealSymb{figA-3j.pdf}{0.49}
\]
\end{minipage}}
{\begin{minipage}{0.68\textwidth}
\[
\epsilon^{(j)}_{mn}\quad = \quad\RealSymb{figA-epsilon-indices.pdf}{0.49}
\]
\vspace{-6pt}
\[
\RealSymb{figA-invepsilon.pdf}{0.49}\quad = \quad(-1)^{2j}\;\RealSymb{figA-epsilon.pdf}{0.49}
\]
\vspace{-12pt}
\[
\RealSymb{figA-epseps2.pdf}{0.49}\quad = \quad\RealSymb{figA-delta.pdf}{0.49}
\]
\end{minipage}}
\vspace{6pt}
\[
\RealSymb{figA-3j.pdf}{0.49} \quad = \quad (-1)^{j_1+j_2+j_3}\RealSymb{figA-3j-minus.pdf}{0.49} \quad = \quad \RealSymb{figA-3j-arrows.pdf}{0.49}\vspace{-12pt}
\]

\[
\CG{j_1j_2}{j}{m_1m_2}{m} \; = \; (-1)^{j_1-j_2-j}\sqrt{d_j}\;\RealSymb{figA-clebsch.pdf}{0.49} \qquad\qquad \Tau{j}{i}{m}{n} \; = \; iW_j\,\RealSymb{figA-tau.pdf}{0.49}
\]

\[
\RealSymb{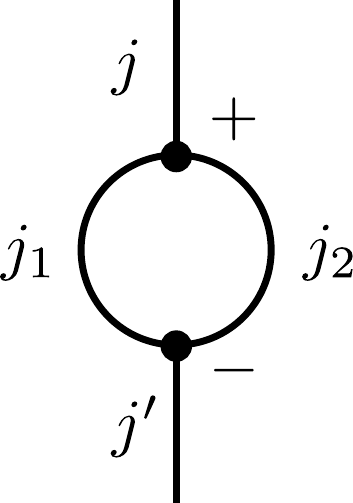}{0.49} \quad = \quad\delta_{jj'}\frac{1}{d_j}\RealSymb{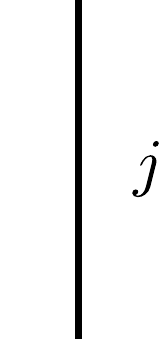}{0.49} \qquad \qquad \qquad \qquad \RealSymb{figA-3j-zero.pdf}{0.49} \quad = \quad\delta_{jj'}\frac{1}{\sqrt{d_j}}\quad\RealSymb{figA-eps-vertical.pdf}{0.49}
\]

\[
\sixj{j_1}{j_2}{j_3}{k_1}{k_2}{k_3} \quad = \; \RealSymb{figA-6j.pdf}{0.49} \qquad\qquad \ninej{j_1}{j_2}{j_3}{k_1}{k_2}{k_3}{l_1}{l_2}{l_3} \quad = \quad \RealSymb{figA-9j-plus.pdf}{0.49}
\]
\vspace{6pt}
\[
\RealSymb{figA-iota412.pdf}{0.49} \quad = \quad \sum_l d_l(-1)^{j_2+j_3+k+l}\sixj{j_1}{j_2}{k}{j_4}{j_3}{l}\;\RealSymb{figA-iota413.pdf}{0.49}
\]

{\begin{minipage}{0.38\textwidth}
\[
\D{j}{m}{n}{g} \; = \; \RealSymb{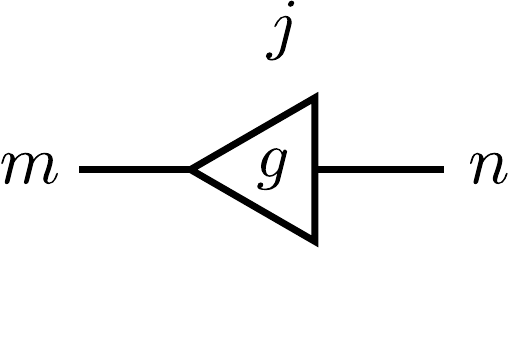}{0.49}
\]
\[
\RealSymb{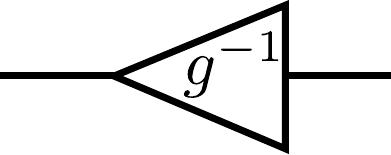}{0.49} \quad = \quad \RealSymb{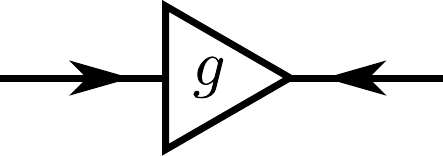}{0.49}
\]
\end{minipage}}
\begin{minipage}{0.05\textwidth}
\end{minipage}
{\begin{minipage}{0.56\textwidth}
\vspace{16pt}
\[
\RealSymb{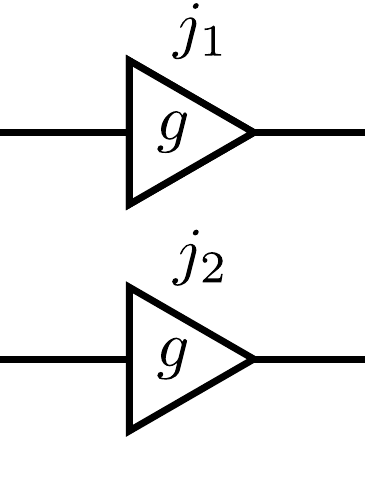}{0.49} \quad = \quad \sum_j d_j\;\RealSymb{figA-CDC.pdf}{0.49}
\]
\end{minipage}}

\vspace{-20pt}
\begin{align*}
J_i^{(v,e)}\;\RealSymb{figA-h_e-source.pdf}{0.49} \quad &= \quad -W_j\;\RealSymb{figA-tauh_e.pdf}{0.49} \\
\intertext{\vspace{-36pt}}
J_i^{(v,e)}\;\RealSymb{figA-h_e-target.pdf}{0.49} \quad &= \quad W_j\;\RealSymb{figA-h_etau.pdf}{0.49}
\end{align*}

\end{footnotesize}

\end{document}